\documentclass[longauth]{aa}

\usepackage{graphicx}
\usepackage{txfonts}
\usepackage{hyperref}
\hypersetup{
    colorlinks=true,
    linkcolor=blue,   
    urlcolor=blue,
    citecolor=blue
    }
\usepackage[modulo,switch]{lineno} 
\usepackage{multirow} 
\usepackage{adjustbox} 

\usepackage{subcaption}
\usepackage{float}
\usepackage{orcidlink}
\usepackage{ulem}

\begin{document} 


   \title{Unveiling the correlation between the UV 2175\,\r{A} dust attenuation bump, PAH emission, and gas-phase metallicity at cosmic noon}


    \author{L. Arriscado
          \inst{1}\fnmsep\inst{2}\orcidlink{0009-0001-8226-4637}
          \and
          I. Shivaei\inst{1}\orcidlink{0000-0003-4702-7561}
          \and
          L. A. Boogaard\inst{3}\orcidlink{0000-0002-3952-8588}
          \and
          A. Mizener\inst{4}\orcidlink{0000-0003-2722-6600}
          \and
          D. Calzetti\inst{4}\orcidlink{0000-0002-5189-8004}
          \and
          P. G. P\'erez-Gonz\'alez\inst{1}\orcidlink{0000-0003-4528-5639}
          \and
          F. Rodr\'iguez Montero\inst{5}\orcidlink{0000-0001-6535-1766}
          \and
          G. H. Rieke\inst{6}\orcidlink{0000-0003-2303-6519}
          \and
          Y. Zhu\inst{7}\orcidlink{0000-0003-3307-7525}
          \and
          A. Battisti\inst{8}\fnmsep\inst{9}\fnmsep\inst{10}\orcidlink{0000-0003-4569-2285}
          \and
          A. J. Bunker\inst{11}\orcidlink{0000-0002-8651-9879}
          \and
          E. Curtis-Lake\inst{12}\orcidlink{0000-0002-9551-0534}
          \and
          T. D\'iaz-Santos\inst{13}\fnmsep\inst{14}\orcidlink{0000-0003-0699-6083}
          \and
          Z. Ji\inst{7}\orcidlink{0000-0001-7673-2257}
          \and
          P. Rinaldi\inst{15}\fnmsep\inst{7}\orcidlink{0000-0002-5104-8245}
          \and
          B. Rodr\'iguez del Pino\inst{1}\orcidlink{0000-0001-5171-3930}
          \and
          S. Tacchella\inst{16}\fnmsep\inst{17}\orcidlink{0000-0002-8224-4505}
          \and
          A. Varo-O'Ferrall\inst{1}\fnmsep\inst{2}\orcidlink{0009-0005-7278-1541}
          \and
          C. C. Williams\inst{18}\orcidlink{0000-0003-2919-7495}
          \and
          J. Witstok\inst{19}\fnmsep\inst{20}\orcidlink{0000-0002-7595-121X}
          }

   \institute{Centro de Astrobiolog\'ia (CAB), CSIC-INTA, Carretera de Ajalvir km 4, Torrej\'on de Ardoz 28850, Madrid, Spain
              \email{larriscado@cab.inta-csic.es}
         \and
             Departamento de F\'isica de la Tierra y Astrof\'isica, Facultad de Ciencias F\'isicas, Universidad Complutense de Madrid, E-28040, Madrid, Spain 
         \and
             Leiden Observatory, Leiden University, PO Box 9513, 2300 RA Leiden, The Netherlands 
         \and
             Department of Astronomy, University of Massachusetts Amherst MA 01003, USA 
         \and 
            Kavli Institute for Cosmological Physics, University of Chicago, Chicago IL 60637, USA 
         \and
            Steward Observatory and Dept of Planetary Sciences, University of Arizona 933 N. Cherry Avenue Tucson AZ 85721, USA 
        \and
             Steward Observatory, University of Arizona, 933 North Cherry Avenue, Tucson, AZ 85721, USA 
        \and
            Research School of Astronomy and Astrophysics, Australian National University, Cotter Road, Weston Creek, ACT 2611, Australia 
        \and
            International Centre for Radio Astronomy Research (ICRAR), University of Western Australia, M468, 35 Stirling Highway, Crawley, WA 6009, Australia 
        \and
            ARC Centre of Excellence for All Sky Astrophysics in 3 Dimensions (ASTRO 3D), Australia 
        \and
            Department of Physics, University of Oxford, Denys Wilkinson Building, Keble Road, Oxford OX1 3RH, UK 
        \and
            Centre for Astrophysics Research, Department of Physics, Astronomy and Mathematics, University of Hertfordshire, Hatfield AL10 9AB, UK 
        \and
            Institute of Astrophysics, Foundation for Research and Technology-Hellas (FORTH), Heraklion, 70013, Greece 
        \and
            Department of Physics, University of Crete, Heraklion, 71003, Greece 
        \and
            Space Telescope Science Institute, 3700 San Martin Drive, Baltimore, Maryland 21218, USA 
        \and
            Kavli Institute for Cosmology, University of Cambridge, Madingley Road, Cambridge, CB3 0HA, UK 
        \and
            Cavendish Laboratory, University of Cambridge, 19 JJ Thomson Avenue, Cambridge, CB3 0HE, UK 
        \and
            NSF National Optical-Infrared Astronomy Research Laboratory, 950 North Cherry Avenue, Tucson, AZ 85719, USA 
        \and
            Cosmic Dawn Center (DAWN), Copenhagen, Denmark 
        \and    
            Niels Bohr Institute, University of Copenhagen, Jagtvej 128, DK-2200, Copenhagen, Denmark} 

   \date{Received June 16, 2026; accepted October 16, 2026}

 
    \abstract
   {The UV 2175\,\AA\ attenuation bump is one of the most prominent yet poorly understood features of galaxy attenuation and extinction curves. It is observed from the local universe up to $z\sim 7$, but its physical origin remains uncertain. Its main carrier candidate is Polycyclic Aromatic Hydrocarbon (PAH) molecules, which are more prevalent in high-metallicity environments.}
   {We present the first systematic study of the relationship between PAH abundance and the UV attenuation bump in 49 galaxies at cosmic noon ($1.5<z<2.7$), and examine its dependence on stellar mass ($M_*$), star formation rate (SFR), and $V$-band optical depth ($\tau_V$). For 19 galaxies, we also explore the role of gas-phase metallicity via oxygen abundance measurements.}
   {We measure the UV bump amplitude using VLT/MUSE spectroscopy from the MUSE/HUDF survey. Galaxy properties, including the PAH-to-dust mass fraction $q_{\rm PAH}$, are derived via SED fitting with \texttt{Prospector}, combining \textit{JWST}/NIRCam and \textit{JWST}/MIRI photometry ($\lambda_{\rm rest}\sim0.1-8\,\mu$m at $z=2$). Oxygen abundances are obtained from strong-line diagnostics using Keck/MOSFIRE and \textit{JWST}/NIRSpec data.}
   {We find a positive correlation between the UV bump amplitude $E_{\rm bump}$ and $q_{\rm PAH}$ (Spearman $\rho=0.60$); at low $q_{\rm PAH}$, UV bumps are rarely detected, supporting PAHs as the primary carriers. Both $E_{\rm bump}$ and $q_{\rm PAH}$ correlate with $M_*$, oxygen abundance, and $\tau_V$, and anti-correlate with specific SFR (sSFR), indicating that PAHs and the UV bump are suppressed in low-metallicity, dust-poor, and/or intensely star-forming environments. This is consistent with the destruction of PAHs in regions with limited dust shielding and/or strong radiation fields produced by young stellar populations. A principal component analysis of $E_{\rm bump}$, $q_{\rm PAH}$, $\tau_V$, sSFR, and oxygen abundance explains 97\% of the data variance, supporting previous findings that reported correlations with $M_*$ reflect an underlying correlation with metallicity. Finally, variations in dust geometry can explain the wide range of UV bump amplitudes ($\sim0-0.5$\,mag) at high $q_{\rm PAH}$.}
    {}


   \titlerunning{UV 2175\,\r{A} bump, PAH emission, and metallicity at cosmic noon}
   \authorrunning{L. Arriscado}
   \maketitle
%
\section{Introduction}\label{sect01}

    Dust is a key component of galaxies. Although it comprises less than 1\% of the interstellar medium (ISM) mass, dust strongly influences the chemical composition and thermodynamics of the gas \citep{draine_interstellar_2003}. It also plays a role in regulating star formation \citep{ galliano_interstellar_2018, schneider_formation_2024}. Moreover, dust grains play a critical role in shaping the spectral energy distributions (SEDs) of galaxies by absorbing and scattering ultraviolet (UV) and optical light and re-radiating this energy in the infrared (IR) \citep{conroy_modeling_2013, galliano_interstellar_2018, salim_dust_2020}; these mechanisms give rise to dust extinction and dust attenuation\footnote{Extinction refers to the absorption and scattering of light (originating from a single source) by a foreground screen of dust, where scattering occurs out of the line of sight. Attenuation refers to the combined effects of absorption and scattering (both into and out of the line of sight) in systems with multiple light sources and complex dust–star geometries.}. In particular, dust extinction depends on the size distribution of grains and their chemical composition \citep[e.g.,][]{weingartner_dust_2001, jones_evolution_2013, hensley_astrodustpah_2023}, while dust attenuation depends also on the dust-star geometry \citep[e.g.,][]{witt_multiple_2000, narayanan_theory_2018}. 
    
    Dust has a substantial impact on observational studies. A proper understanding of dust attenuation curves across all wavelengths is therefore essential in order to accurately identify galaxies and model SEDs, and to correctly derive widely used physical properties such as star formation rates (SFRs) and intrinsic UV luminosity functions \citep[e.g.,][]{meurer_dust_1999, conroy_modeling_2013}. Moreover, the bulk of star formation at $z<4$ occurs in dust-obscured environments \citep{zavala_evolution_2021}; recent ALPINE \citep{le_fevre_alpine-alma_2020, bethermin_alpine-alma_2020, faisst_alpinealma_2020} results indicate that dust remains important even at higher redshifts, with a flattening of the cosmic SFR density beyond $z \sim 4$ \citep{gruppioni_alpine-alma_2020}. Thus, dust significantly affects the detection and classification of galaxies, and can cause heavily obscured systems, such as merging galaxies, to be overlooked (e.g., \citealt{langan_alma_2024}).

    Beyond their radiative effects, dust grains serve as key coolants of the ISM, helping to trigger gravitational collapse and star formation. By shielding dense cloud cores from UV photons, dust enables the buildup of key molecules required for star formation \citep{peroux_cosmic_2020, schneider_formation_2024}; in particular, the formation of H$_2$ primarily ocurrs on the surfaces of dust grains \citep[e.g.,][]{hollenbach_molecule_1979}. 
    
    The dominant dust seed producers are asymptotic giant branch (AGB) stars and supernovae (SNe) \citep[e.g.,][]{draine_physics_2011, gall_production_2011, micelotta_dust_2018}, tightly linking dust content to stellar evolution. Observationally, dust attenuation has been shown to correlate with stellar mass (e.g., \citealt{shivaei_mosdef_2020, shapley_mosfire_2022, woodrum_jades_2025, maheson_big_2025}), SFR (e.g., \citealt{salim_dust_2020, maheson_big_2025}), the UV-optical slope of the dust attenuation curve (e.g., \citealt{salim_dust_2020, shivaei_diversity_2025, markov_unveiling_2025}), and galaxy inclination \citep[e.g.,][]{chevallard_insights_2013}, among others.
    
    One of the most prominent features of dust extinction and attenuation curves is the 2175 \AA\ UV bump \citep{stecher_interstellar_1965}. This broad absorption feature is present along many lines of sight in the Milky Way and the Large Magellanic Cloud, but is weak or absent in the Small Magellanic Cloud (\citealt{salim_dust_2020}). The bump has also been detected up to $z\sim7$ \citep[e.g.,][]{witstok_carbonaceous_2023, markov_dust_2023, ormerod_detection_2025}. The UV bump strength has been found to correlate with several galaxy properties, including stellar mass \citep[e.g.,][]{shivaei_uv_2022}, metallicity \citep[e.g.,][]{shivaei_mosdef_2020, markov_unveiling_2025}, slope of the dust attenuation curve \citep[e.g.,][]{kriek_dust_2013}, specific star formation rate \citep[sSFR; e.g.,][]{battisti_constraining_2025}, and galaxy inclination \citep[e.g.,][]{wild_empirical_2011, battisti_characterizing_2017}.

    Despite its prominence and connection with other properties, the physical origin of the UV bump remains uncertain. The prevailing view is that the bump originates from graphitic carbon \citep{draine_optical_1984}. Other possibilities include UV-processed hydrogenated amorphous carbon (HAC) particles \citep{mennella_new_1998} and hyperfullerene carbon particles \citep{wada_carbonaceous_1999}. However, these candidates cannot fully explain variations in bump width \citep{draine_graphite_1993}, and the feature is widely thought to arise predominantly, if not entirely, from a mixture of polycyclic aromatic hydrocarbon (PAH) molecules \citep{tielens_interstellar_2008, li_spitzers_2020, lin_polycyclic_2025}.

    PAH molecules are complex carbonaceous particles that absorb individual UV photons and re-emit them by fluorescence in the mid-IR \citep{allamandola_polycyclic_1985, draine_interstellar_2003, tielens_interstellar_2008}. They are primarily produced in AGB stars \citep{galliano_interstellar_2018, li_spitzers_2020}, but may also form in supernova ejecta \citep{otaki_effective_2026}, Wolf–Rayet winds \citep{marchenko_search_2017, endo_detection_2022}, or intermediate-luminosity red transients \citep{rose_investigating_2025}. PAH emission itself correlates with SFR \citep[e.g.,][]{calzetti_calibration_2007, calzetti_polycyclic_2011, gregg_calibration_2026, alberts_calibrating_2026}, stellar age \citep[e.g.,][]{galliano_stellar_2008, shivaei_mosdef_2017}, stellar mass \citep[e.g.,][]{shivaei_new_2024}, and metallicity \citep[e.g.,][]{engelbracht_metallicity_2005, draine_dust_2007, aniano_modeling_2020}, from the local universe to cosmic noon.

    Despite decades of study, the origin of the UV bump and its connection to PAH emission remain poorly understood. Most existing work linking these phenomena is either theoretical (e.g., \citealt{stecher_graphite_1965,li_infrared_2001,papoular_polycrystalline_2009}) or laboratory-based (e.g., \citealt{joblin_contribution_1992,bradley_astronomical_2005,steglich_electronic_2010}). The few observational studies that have explored this relationship generally suggest a positive connection between PAH emission and the UV bump, although often with substantial scatter and limited sample sizes. These studies are restricted to the local universe \citep[e.g.,][]{blasberger_observational_2017, gordon_expanded_2024, battisti_constraining_2025} and to cosmic noon \citep[$1<z<3$;][]{noll_gmass_2009,shivaei_uv_2022}, where analyses have so far relied on stacked data rather than individual galaxies. Cosmic noon  represents the most distant epoch at which PAH emission can be robustly traced in individual galaxies with current facilities. It therefore provides a unique window beyond the local Universe to study the link between PAHs and the UV bump, while still allowing both to be directly observed: PAH features in the (observed) mid-IR with \textsl{JWST}/MIRI, and the UV attenuation bump in the (observed) optical with instruments such as VLT/MUSE. The capabilities of the James Webb Space Telescope \citep[\textsl{JWST};][]{gardner_james_2023}, combined with ground-based spectroscopy from Keck and the VLT, now enable more sensitive multi-wavelength studies of statistically significant samples of individual galaxies at this epoch.

    In this work, we present the first simultaneous study of the relationship between the UV attenuation bump strength, PAH emission, and gas-phase metallicity (through oxygen abundance) in star-forming galaxies (SFGs). We use multi-wavelength datasets to probe these components in a sample of 49 galaxies at cosmic noon: VLT/MUSE spectra to measure the UV bump strength, \textit{JWST}/MIRI photometry to probe the PAH emission, and Keck/MOSFIRE and \textit{JWST}/NIRSpec spectra to constrain the gas-phase metallicity using strong-line\footnote{Here, ``strong-line'' refers to methods that estimate gas-phase metallicities using bright emission-line ratios and empirical calibrations, as opposed to $T_e$-based methods that rely on weaker auroral emission lines.} diagnostics. Sect.~\ref{sect02} describes the data and samples used. Sect.~\ref{sect03} details the methods for measuring UV bump strength, PAH mass fraction, oxygen abundance, and other galaxy properties, as well as the correlations among these quantities. Sect.~\ref{sect05} discusses the implications of our results, and Sect.~\ref{sect06} summarizes our conclusions.

    Throughout this paper, all magnitudes are in the AB system. We adopt a \citet{chabrier_galactic_2003} initial mass function (IMF), and assume a $\Lambda$CDM cosmology with $H_0=67.74\,\mathrm{km\,s^{-1}\,Mpc^{-1}}$, $\Omega_{\mathrm{m}} = 0.3075$, and $\Omega_{\Lambda} = 0.6925$ \citep{planck_collaboration_planck_2016}.

\section{Sample and data}\label{sect02}

\subsection{Spectroscopic data}\label{sect02:spec}

    We use spectroscopic data from the optical VLT/MUSE spectrometer \citep{bacon_muse_2010} to trace the UV 2175\,\r{A} bump. Oxygen abundances are derived from near-IR (tracing rest-frame optical emission) Keck/MOSFIRE (\citealt{mclean_design_2010}, \citealt{mclean_mosfire_2012}) observations and complemented with emission-line flux measurements from two \textsl{JWST}/NIRSpec \citep{jakobsen_near-infrared_2022} surveys. The sample selection and spectroscopic data are described below.

\paragraph{MUSE} The parent sample is drawn from \cite{shivaei_uv_2022} and we refer to that paper for details. In brief, the sample consists of 120 galaxies at $1.5<z<2.7$ from the MUSE \textsl{Hubble} Ultra Deep Field (HUDF) Survey (\citealt{bacon_muse_2017}, \citealt{inami_muse_2017}). The MUSE spectral range (4750-9350\,\r{A}) covers the rest-frame range 1900-2527\,\r{A}, including the UV bump.
    
Sources are selected to have secure redshifts (CONFID $\geq 2$; \citealt{inami_muse_2017}), determined from at least one robust spectral feature\footnote{Emission and absorption features are identified in 1D spectra and confirmed with narrowband images, with CONFID assigned only upon confirmation. The final sample comprises 13 CONFID=2 (1 spectral feature) and 36 CONFID=3 (2+ spectral features) galaxies.}. Redshifts are verified using NIRSpec spectroscopic redshifts (see below). The sample spans a wide range of star formation rates, $\log_{10}(\mathrm{SFR}/M_\odot\,\mathrm{yr}^{-1}) \sim -1$ to 2, and stellar masses, $\log_{10}(M_\ast/M_\odot) \sim 7.5$ to $11.5$, placing them broadly along the star-forming main sequence.

\paragraph{MOSFIRE} We observe 26 galaxies (from the MUSE parent sample) with Keck/MOSFIRE in the $H$ (1.46–1.81\,$\mu$m, $R=3660$) and $K$ (1.93–2.45\,$\mu$m, $R=3620$) bands (programs 2022B and 2024B; PIs: D.~Calzetti and I.~Shivaei). We derive oxygen abundances for 6 galaxies in the final sample (see Sect.~\ref{sect03:met} for details on the metallicity estimation). These spectra cover key rest-frame optical emission lines (H$\beta$, [O\,\textsc{iii}]$\lambda 4960$, [O\,\textsc{iii}]$\lambda 5008$, H$\alpha$, [N\,\textsc{ii}]$\lambda 6585$). Details on the observations, data reduction, and flux calibration are provided in Appendix~\ref{app:mosfire}.

Briefly, the data is reduced using the \texttt{PypeIt} pipeline \citep{prochaska_pypeit_2020}, with minor modifications to object tracing and detection to account for faint sources and small positional drifts. Individual exposures are quality-selected, coadded, and flux-calibrated. One-dimensional spectra are optimally extracted following \cite{horne_optimal_1986}. Since we rely on emission-line ratios of nearby lines, we do not perform slit-loss corrections.

\paragraph{NIRSpec} We complement the Keck data with spectroscopic observations from the SMILES DR2 \citep{zhu_smiles_2026} and JADES DR4 \citep{curtis-lake_jades_2025, scholtz_jades_2025} NIRSpec catalogs. The data is briefly described below and we refer to the corresponding papers for details of data reduction, flux measurement, and slit-loss corrections.

SMILES observations are obtained in MOS mode using medium-resolution gratings, covering $0.97$–$3.17\,\mu$m ($R\sim1000$). SMILES NIRSpec data is reduced using a customized pipeline \citep{zhu_smiles_2026} based on the \textit{JWST} Science Calibration Pipeline \citep{bushouse_jwst_2026}; emission lines are measured using the \texttt{GELATO} pipeline \citep{hviding_theskyentistgelato_2022}. JADES observations provide broader wavelength coverage ($0.6$–$5.3\,\mu$m, $R=500$–1500), and have been reduced with the NIRSpec GTO pipeline, as described in detail in \citet{scholtz_jades_2025}. Emission-line fluxes are measured using \texttt{QubeSpec}\footnote{\url{https://github.com/honzascholtz/Qubespec}}, assuming single-component Gaussian profiles. For both surveys, we adopt the publicly available emission-line measurements without re-fitting the spectra. These include all key emission lines for metallicity diagnostics at $1.5<z<2.7$.

\subsection{Photometric data}\label{sect02:phot}

    We use photometric data from \textsl{JWST}/NIRCam \citep{rieke_performance_2023} and \textsl{JWST}/MIRI \citep{wright_mid-infrared_2023} to perform SED fitting from $0.4$ to $25\,\mu$m ($\lambda_{\rm rest}\sim0.1-8\,\mu$m at $z=2$) with \texttt{Prospector} \citep[][Sect.~\ref{sect03:prosp}]{johnson_stellar_2021}. In addition, we use the GOODS-S morphology catalog presented by \citet{genin_dawn_2025}, obtained from the DAWN \textsl{JWST} Archive \citep[DJA;][]{valentino_atlas_2023}, which compiles NIRCam photometry from major public extragalactic surveys (including JADES). This allows us to obtain morphological parameters (semimajor and semiminor axes) as well as rest-frame $U$, $V$, and $J$ colors for our galaxies.

\paragraph{NIRCam and HST} We use NIRCam photometry from the JADES DR5 catalog ($0.8-5\,\mu$m; \citealt{johnson_jwst_2026, robertson_jwst_2026}). The available filters are F090W, F115W, F150W, F182M, F200W, F210M, F277W, F335M, F356W, F410M, F430M, F444W, F460M, and F480M. We also use \textit{Hubble} Space Telescope (\textsl{HST}) photometry ($0.4-1.7\,\mu$m) included in the JADES DR5 catalog \citep{illingworth_hubble_2017, whitaker_hubble_2019}, obtained with the ACS/WFC instrument (filters F435W, F606W, F775W, F814W, F850LP). HST and JADES photometry are extracted using the same Kron aperture, defined on the detection image\footnote{JADES detection image in GOODS-S is constructed by combining filters F277W, F335M, F356W, F410M, F430M, F444W, F460M, and F480M.} with a Kron parameter of 2.5. Photometric uncertainties are floored at 5\% to account for systematic errors not captured by the JADES data reduction pipelines.

\paragraph{MIRI} The \textsl{JWST}/NIRCam and \textsl{HST} data is complemented with \textsl{JWST}/MIRI photometry from the SMILES DR1 catalog ($5-25\,\mu$m; \citealt{alberts_smiles_2024, rieke_smiles_2024}), which provides data in all available MIRI filters: F560W, F770W, F1000W, F1280W, F1500W, F1800W, F2100W, and F2550W. We use Kron aperture photometry as provided by the catalog, measured using a $2.5\times$ scaled Kron aperture defined on the F560W+F770W detection image.

\paragraph{DJA morphology catalog} Morphological information is taken from the DJA catalog \citep{genin_dawn_2025}, where sources were modeled with \texttt{SourceXtractor++} \citep{bertin_sourcextractor_2020, kummel_working_2020}, fitting a 2D surface brightness profile simultaneously across the NIRCam bands F090W, F115W, F150W, F200W, F277W, F356W, and F444W. 
We adopt the bulge+disk decomposition fitting provided in \citet{genin_dawn_2025}.
Morphological parameters were constrained to be identical across all bands, yielding weighted-average values over the $0.8-5\,\mu$m range. We use the semimajor and semiminor axes, along with the UVJ colors provided by this catalog, to compute morphological parameters.

\subsection{Far-IR/sub-mm data and constraints on $q_{\rm PAH}$}\label{sect02:fir-data}

We incorporate \textit{Herschel} and ALMA data into the photometry for SED fitting to better constrain the dust emission and improve inferences on PAH mass fractions. 13 galaxies in our sample are part of the ASPECS survey \citep{walter_alma_2016}, a deep ALMA band 3 and 6 survey in HUDF, with ALMA band 6 (1.1\,mm) detections. We adopt the ALMA and \textit{Herschel} photometry for this sample from \cite{boogaard_alma_2020}.
An additional four galaxies of the sample have ALMA continuum detections from the compilation of various programs in the ECOGAL ALMA data mining project \citep{lee_ecogal_2025}.
Inclusion of far-IR data does not significantly change the inferred $q_{\rm PAH}$ values, which remain consistent within uncertainties, but provides improved constraints on the dust emission component of the SED. A comparison of $q_{\rm PAH}$ values derived with and without ALMA data, as well as an example SED fit, are shown in Appendix~\ref{app:alma}.

\subsection{Sample selection and final sample}\label{sect02:sample}

\begin{figure}
   \resizebox{\hsize}{!}
            {\includegraphics[]{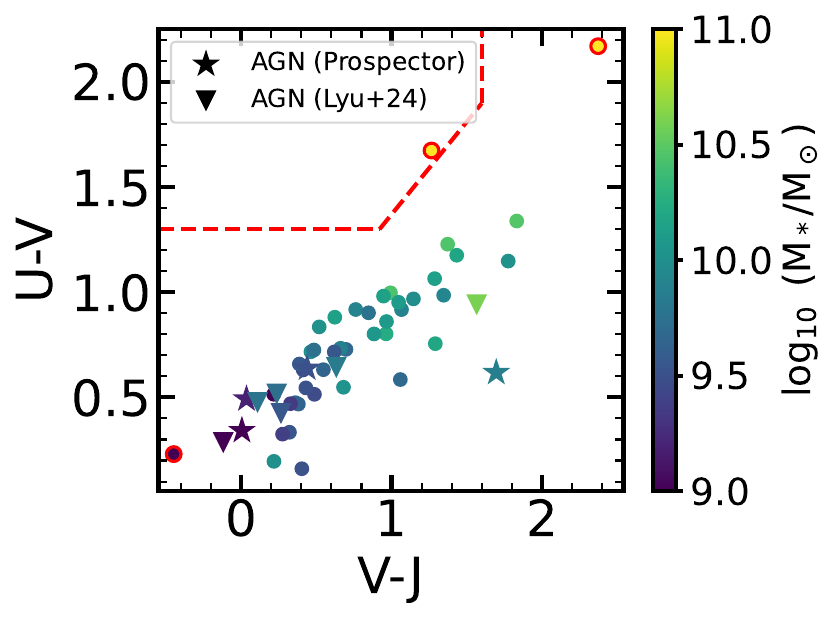}}
      \caption{UVJ diagram for the 52 galaxies remaining after the first two selection steps. The red dashed line indicates the quiescent galaxy region defined by \citet{williams_detection_2009}. Star symbols indicate AGN selected based on \texttt{Prospector} results, and triangles indicate AGN from the \cite{lyu_active_2024} catalog (see ``AGN fraction'' paragraph in Sect.~\ref{sect02:sample}). Galaxies discarded in subsequent selection steps and not included in the final sample are highlighted with red contours around the data points.}
      \label{fig03}
\end{figure}

\begin{figure*}
   \resizebox{\hsize}{!}
            {\includegraphics[]{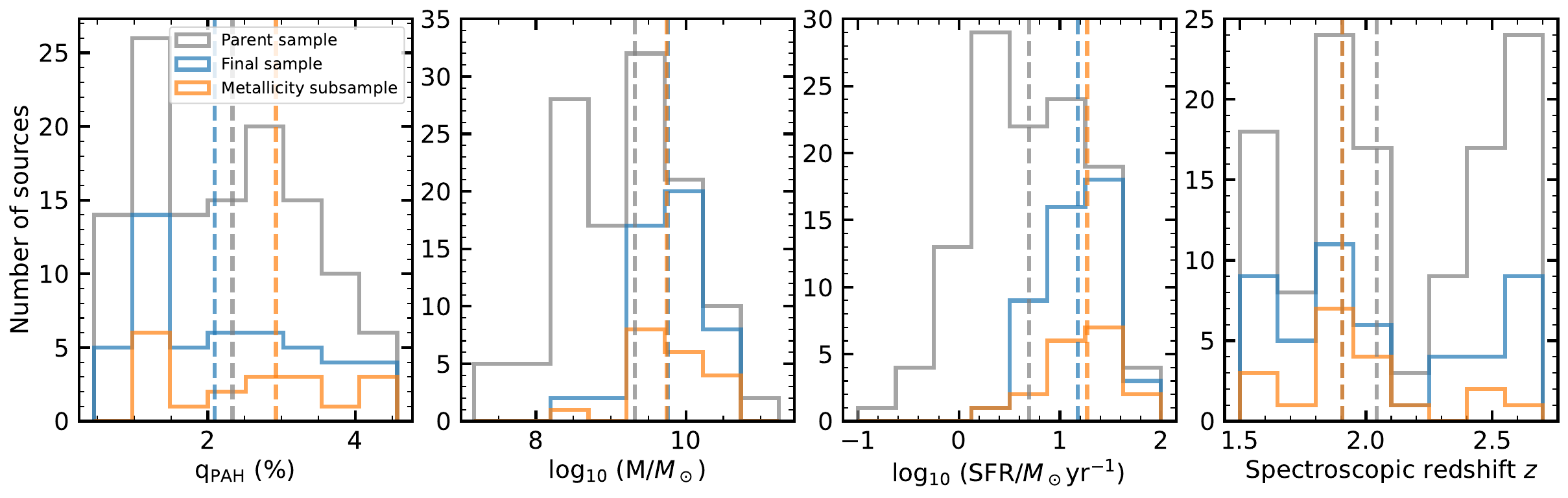}}
      \caption{Distributions of $q_{\rm PAH}$ (left), stellar mass (center left), SFR (center right), and spectroscopic redshift (right) for the parent sample (120 galaxies, gray), the final sample (49 galaxies, blue), and the metallicity subsample (19 galaxies, orange), defined by the criteria in Sect.~\ref{sect02:sample}. The dashed vertical lines indicate the median values per sample/subsample. The differences in the stellar mass and SFR distributions between the parent sample (gray) and the final sample (blue), visible as a lack of low-mass galaxies in the final sample, arise from the selection step in which we removed galaxies with fewer than two detections ($>3\sigma$) in MIRI filters longward of F1280W.}
      \label{fig04}
\end{figure*}

    Sample selection is performed in four steps. First, we start with the MUSE parent sample of 120 galaxies at $1.5<z<2.7$ from \cite{shivaei_uv_2022}, described in Sect.~\ref{sect02:spec}. We discard all sources with fewer than two detections ($>3\sigma$) in MIRI filters longward of F1280W\footnote{Since this spectral region is dominated by PAH emission, it must be well constrained to provide robust $q_{\rm PAH}$ measurements.}. This reduces the sample to 55 galaxies. Second, we exclude galaxies with poor SED fits, based on the reduced $\chi^2$ values from \texttt{Prospector} (Sect.~\ref{sect03:prosp} and Tab.~\ref{tab:prospector} for a summary of the parameters and priors used in the SED fitting). We require reduced $\chi^2 < 10$, removing three additional sources and leaving 52. Third, we classify galaxies as star-forming or quiescent based on their positions in the UVJ diagram (Fig.~\ref{fig03}), following the criteria of \citet{williams_detection_2009} for $1<z<2$ galaxies. We identify one quiescent galaxy and one extremely dusty SFG ($U-V>2$, $V-J>2$), and remove both, leaving 50 sources. The fact that most galaxies lie in the star-forming region is expected, as the sources in the MUSE parent sample are selected based on rest-frame UV continuum magnitude and blind emission line detections, with secure redshifts determined from at least one robust spectral feature (Sect.~\ref{sect02:spec}). This selection naturally excludes many dusty and quiescent galaxies (see \citealt{inami_muse_2017} for more details). Finally, we visually inspect the \texttt{Prospector} fits and discard one additional poor fit, for which the modeled SED fails to reproduce the mid-IR photometry. Our final robust sample contains 49 galaxies with reliable $q_{\rm PAH}$ measurements. The sample distributions are shown in Fig.~\ref{fig04}.

\paragraph{AGN fraction} Additionally, we flag (but do not remove) sources with a significant AGN contribution, based on the \texttt{Prospector} parameter $f_{\rm AGN}$, which quantifies the fraction of the $4-20~\mu$m luminosity contributed by the AGN. Four sources with $f_{\rm AGN}>0.2$ are flagged and shown as star symbols in the relevant figures. No X-ray or radio AGN are identified in the sample. 
We compare our AGN identification to the six AGN candidates reported by \cite{lyu_active_2024}, identified using a different implementation of \texttt{Prospector} and mid-IR data. These six galaxies are shown as triangles in Fig.~\ref{fig03}. Repeating the full analysis performed in this paper using the \cite{lyu_active_2024} AGN classification (instead of our AGN classification) yields results consistent within the uncertainties, with the AGN candidates from both methods occupying similar regions of parameter space. We retain our AGN classification throughout the paper for consistency with the adopted modeling framework.

\paragraph{Metallicity subsample} Of the 49 galaxies in the final sample, 19 have available measurements of oxygen abundance, defining the subsample used for metallicity analysis. Of these, four are from the MOSFIRE observations and 15 are drawn from the NIRSpec catalogs.

\paragraph{Sample completeness} We assess the impact of our selection criteria by comparing the distributions of $q_{\rm PAH}$, stellar mass, SFR, and redshift between the MUSE parent sample, the final sample, and the metallicity subsample. In Fig.~\ref{fig04}, we observe a large overlap between the samples, and although the median values are different, they are within scatter. As mentioned above, the MUSE selection criteria naturally exclude dusty and quiescent galaxies. Within this MUSE-selected population, the final sample is shifted toward higher stellar masses and SFRs relative to the parent sample, with two-sample KS tests\footnote{We use the two-sided two-sample Kolmogorov-Smirnov (KS) test, which tests the null hypothesis that two samples are drawn from the same underlying continuous distribution. We compute the test using the \texttt{ks\_2samp} function with the ``auto'' method from the \texttt{scipy} package.} indicating statistically significant differences in both distributions ($p=1.0\times10^{-6}$ and $5.0\times10^{-6}$, respectively). The final sample has higher median stellar mass and SFR than the MUSE parent sample, although the distributions mostly overlap (see Fig.~\ref{fig04}). This shift is likely driven by the requirement of robust MIRI detections in at least two filters longward of F1280W, which preferentially selects more massive and actively star-forming galaxies. In contrast, the $q_{\rm PAH}$ and redshift distributions are statistically consistent between the parent and final samples ($p=0.95$ and 0.82, respectively). No significant differences are found between the final sample and metallicity subsample for any of these properties ($p>0.6$). Thus, our samples are biased toward more massive and higher-SFR galaxies relative to the MUSE parent population, although the differences are modest in terms of their median values, while the $q_{\rm PAH}$ and redshift distributions remain statistically consistent.


\section{Methods}\label{sect03}

\subsection{UV bump fitting}\label{sect03:uvbump}

    The UV 2175\AA\ bump catalog is adopted from \cite{shivaei_uv_2022}. We briefly describe here the methodology and refer to \cite{shivaei_uv_2022} for more details. The UV attenuation bump $A_{\lambda,\mathrm{bump}}$ is defined as the excess attenuation relative to a smooth continuum without a bump, $A_{\lambda,\mathrm{bump}}=A_{\lambda,\mathrm{tot}}-A_{\lambda,\mathrm{cont}}$, with the bump amplitude $E_{\rm bump}$ being the excess attenuation measured at 2175\,\r{A}, $E_{\rm bump}=A_{2175,\mathrm{bump}}$. Assuming the same reddening $E(B-V)$ for all attenuation, we can write the attenuation bump as $A_{\lambda,\mathrm{bump}}=\kappa_{\lambda,\mathrm{bump}}\ E(B-V)$, with $\kappa_{\lambda,\mathrm{bump}}=\kappa_{\lambda,\mathrm{tot}}-\kappa_{\lambda,\mathrm{cont}}$ being the attenuation ratio obtained by integrating a Drude profile as described in \cite{fitzpatrick_analysis_2007}:

    \begin{equation}
        \kappa_{\lambda,\mathrm{bump}}=c_3\frac{(1/\lambda)^2}{((1/\lambda)^2-(1/\lambda_0)^2)^2+(1/\lambda)^2\gamma^2},
        \label{eq:01}
    \end{equation}
    where $c_3$ is a parameter defining a Lorentzian-like 2175\,\r{A} bump, $\lambda_0$ is the central wavelength at 2175\,\r{A}, and $\gamma$ is the broadening parameter. The bump amplitude is defined as $E_{\rm bump}=E(B-V)\frac{c_3}{\gamma^2}$. 
    
    The smooth UV stellar continuum is first fit with a power-law function using line-free wavelength windows from \cite{calzetti_dust_1994}, after which the observed spectrum is divided by this model to isolate the bump attenuation. A Drude function centered at 2175\,\r{A} is then fitted across the $1950-2330$\,\r{A} range, allowing the FWHM to vary between 200 and 350\,\r{A}, following the results of \citet{noll_gmass_2009}. Final values of $E_{\rm bump}$, along with their uncertainties, are derived through bootstrapping. We also investigate some of the results for the normalized $E_{\rm bump}$, using $A_V$ and $E(B-V)_{\rm neb}$, in App.~\ref{app:02}.

\subsection{Emission line fitting and metallicity estimation}\label{sect03:met}

Emission line measurements for the MOSFIRE spectra are obtained through Gaussian fitting; full details of the fitting procedure are provided in Appendix~\ref{app:mosfire:lines}. In brief, emission lines are modeled individually (with joint fits for H$\alpha$+[N\,\textsc{ii}]), and flux uncertainties are estimated via Monte Carlo sampling. Emission line measurements for the JADES and SMILES samples are taken directly from the public catalogs described in Sect.~\ref{sect02:spec}.

For all samples, only lines with integrated $\mathrm{SNR} > 2.5$ are retained for further analysis. Oxygen abundances are derived using the empirical calibrations from \citet{sanders_aurora_2026}, based on available line ratios, including O3 = [O\,\textsc{iii}]$\lambda5008$/H$\beta$, N2 = [N\,\textsc{ii}]$\lambda6585$/H$\alpha$, and O3N2 = O3/N2. For the galaxies in our final sample, metallicity estimates are obtained for four sources from MOSFIRE, 12 from JADES, and three from SMILES. Throughout this work, metallicities are expressed as oxygen abundances, $12+\log_{10}(\mathrm{O/H})$.

\subsection{SED fitting and derivation of $q_{\rm PAH}$}\label{sect03:prosp}

We use the Python-based SED fitting code \texttt{Prospector} \citep{johnson_stellar_2021} to derive galaxy physical properties, such as $q_{\rm PAH}$, stellar mass, and SFR. Photometric data from \textit{HST}, \textit{JWST} NIRCam and MIRI, and when available ALMA and \textit{Herschel}, serve as input to \texttt{Prospector} (Sect.~\ref{sect02:phot}), providing wavelength coverage from the UV to the mid-infrared ($0.4-25$\,$\mu$m, corresponding to rest-frame $0.1-8$\,$\mu$m at $z \sim 2$). An example SED fit is shown in Fig.~\ref{fig05}. Redshifts are fixed to the spectroscopic redshifts from the MUSE sample (Sect.~\ref{sect02:spec}). All parameters and priors used in \texttt{Prospector} are summarized in Tab.~\ref{tab:prospector} (see App.~\ref{app:prosp} for full details).

Briefly, stellar populations are modeled using the Flexible Stellar Population Synthesis (FSPS; \citealt{conroy_propagation_2009, conroy_propagation_2010-1}) code, using the MESA Isochrones and Stellar Tracks \citep[MIST;][]{dotter_mesa_2016} and the MILES stellar library \citep{falcon-barroso_updated_2011}. The attenuation curve follows the prescription of \citet{noll_gmass_2009} and \citet{kriek_dust_2013}. Nebular emission is included using photoionization models from \texttt{Cloudy} \citep{byler_nebular_2017}, which depend on the stellar populations. Dust emission is modeled following \citet{draine_infrared_2007}. We adopt a parametric star formation history (SFH) consisting of a delayed-$\tau$ model of the form $t\,e^{-t/\tau}$\footnote{We adopt a parametric SFH, as non-parametric approaches typically require higher spectral resolution data than available from photometric observations; e.g. \citet{conroy_modeling_2013}.}, and assume a \citet{chabrier_galactic_2003} initial mass function (IMF). Metallicity, dust, and SFH parameters are allowed to vary within broad priors. We include a two-parameter AGN dust emission model \citep{leja_hot_2018}, and intergalactic medium (IGM) attenuation is applied following the prescription of \citet{madau_radiative_1995}.  

For the offset from the star-forming main sequence ($\Delta$MS), we adopt the same definition as in \cite{chastenet_resolved_2024}: $\Delta \rm MS=\log_{10}(sSFR)-\log_{10}(sSFR_{MS})$\footnote{SFGs are typically defined as those within $|\Delta \mathrm{MS}| \leq 0.3$.}, where $\log_{10}(\rm sSFR_{MS})$ denotes the specific star formation rate of the star-forming main sequence at the redshift of each galaxy.

\subsection{Correlations}\label{sect04}

\begin{figure*}[ht]
    \centering
    \begin{subfigure}[ht]{0.49\hsize}
        \centering
        \includegraphics[width=\linewidth]{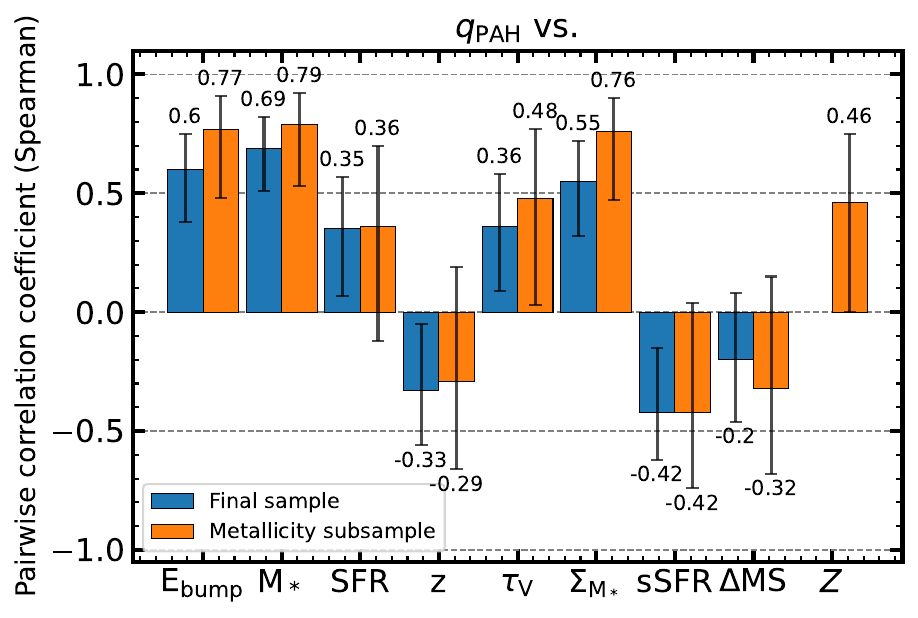}
        \label{fig06a}
    \end{subfigure}
    \hfill
    \begin{subfigure}[ht]{0.49\hsize}
        \centering
        \includegraphics[width=\linewidth]{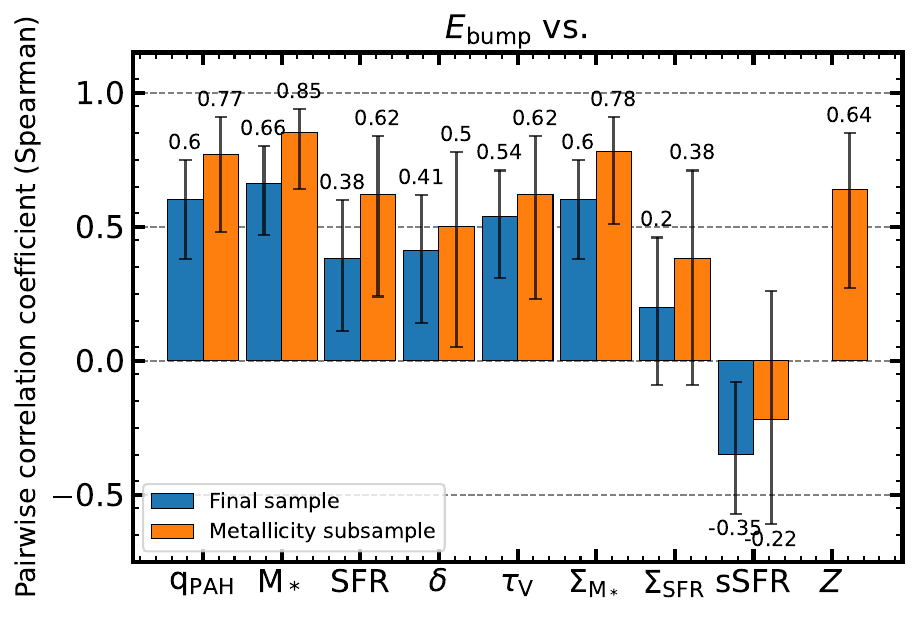}
        \label{fig06b}
    \end{subfigure}
    \caption{Pairwise Spearman correlation coefficients between $q_{\rm PAH}$ (left) and $E_{\rm bump}$ (right) and various galaxy parameters: stellar mass ($M_*$), star formation rate (SFR), redshift ($z$), slope of the attenuation curve ($\delta$), optical depth at 5500\,\AA\ ($\tau_V$), stellar surface mass density ($\Sigma_{M_\ast}$), surface SFR density ($\Sigma_{\rm SFR}$), specific star formation rate (sSFR), offset from the star-forming main sequence ($\Delta \rm MS$), and oxygen abundance ($Z$). We include AGN in this analysis. Only parameters with a Spearman rank correlation coefficient of $|\rho| > 0.3$ in either of the two samples are shown. Blue bars indicate pairwise correlation coefficients for galaxies in the final sample (49 galaxies), while orange bars represent the metallicity subsample (19 galaxies; Sect.~\ref{sect02:sample}). Error bars show the 95\% confidence intervals of the correlations.}
    \label{fig06}
\end{figure*}

    We use the Python package \texttt{pingouin} \citep{vallat_pingouin_2018} to compute pairwise Spearman rank correlation coefficients for the dataset, assuming non-directional hypotheses. For each correlation coefficient, we report the corresponding $p-$value (Figs.~\ref{app_fig01} and \ref{app_fig02}) as well as the 95\% confidence interval (Fig.~\ref{fig06}). 

    In Fig.~\ref{fig06}, we display the pairwise Spearman rank correlation coefficients between $q_{\rm PAH}$ (left) and all galaxy parameters exhibiting a correlation coefficient $|\rho|\geq 0.3$. The same analysis is performed for $E_{\rm bump}$ (right). To assess the correlations, we follow the criteria reported in \cite{kuckartz_statistik_2013} where $|\rho|<0.1$ signifies no correlation, $|\rho|=0.1-0.3$ low correlation, $|\rho|=0.3-0.5$ moderate correlation, $|\rho|=0.5-0.7$ high correlation, and $|\rho|>0.7$ very high correlation. For each correlation coefficient, we also report the corresponding 95\% confidence interval. The inferred correlation coefficients are broadly consistent between the final sample and the metallicity subsample, although the smaller size of the latter results in larger uncertainties.

    We refer the reader to Figs.~\ref{app_fig01} and \ref{app_fig02} for detailed correlation matrices (heatmaps) of the pairwise Spearman rank correlation coefficients and the corresponding $p-$values for all parameters, including those that show weak or no correlations ($|\rho|<0.3$). Fig.~\ref{app_fig01} shows the results for the final sample (49 galaxies; Sect.~\ref{sect02:sample}), while Fig.~\ref{app_fig02} presents the results for the metallicity subsample (19 galaxies; Sect.~\ref{sect02:sample}).

    In addition, we perform a principal component analysis (PCA) on the same dataset, using the \texttt{decomposition} module of the \texttt{scikit-learn} Python package \citep{pedregosa_scikit-learn_2011}. When we use PCA throughout this paper, the data is previously normalized to ensure a correct analysis.

    Given the relatively small sample sizes, particularly for the metallicity subsample ($N=19$), we additionally assess the significance of the main correlations using permutation tests. For each correlation, we randomly permute one variable while keeping the other fixed and recompute Spearman's $\rho$ over $10^5$ permutations, thereby constructing an empirical null distribution under the hypothesis of no association. The two-sided permutation $p$-value is calculated as the fraction of permutations for which $|\rho_{\rm perm}|\geq|\rho_{\rm obs}|$. The resulting null distributions are shown in App.~\ref{app:permutation}.

\section{Results and discussion}\label{sect05}

In this section, we investigate the relationships between the 2175\,\AA\ UV bump, PAH emission, and other galaxy properties at cosmic noon, extending work done at $z\sim0$ (see Sect.~\ref{sect01} for references), using our final sample of 49 galaxies and metallicity subsample of 19 galaxies. In Sect.~\ref{sect05:01}, we examine the connection between the UV bump and PAH emission, and how these relate to stellar mass, metallicity, and dust optical depth. In Sects.~\ref{sect05:02} and \ref{sect05:03}, we analyze how $q_{\rm PAH}$ and $E_{\rm bump}$ vary independently with different physical parameters, such as stellar mass, SFR, redshift, oxygen abundance, and dust optical depth.

\subsection{PAH fraction and UV attenuation bump relation}\label{sect05:01}

\begin{figure*}
   \resizebox{\hsize}{!}
            {\includegraphics[]{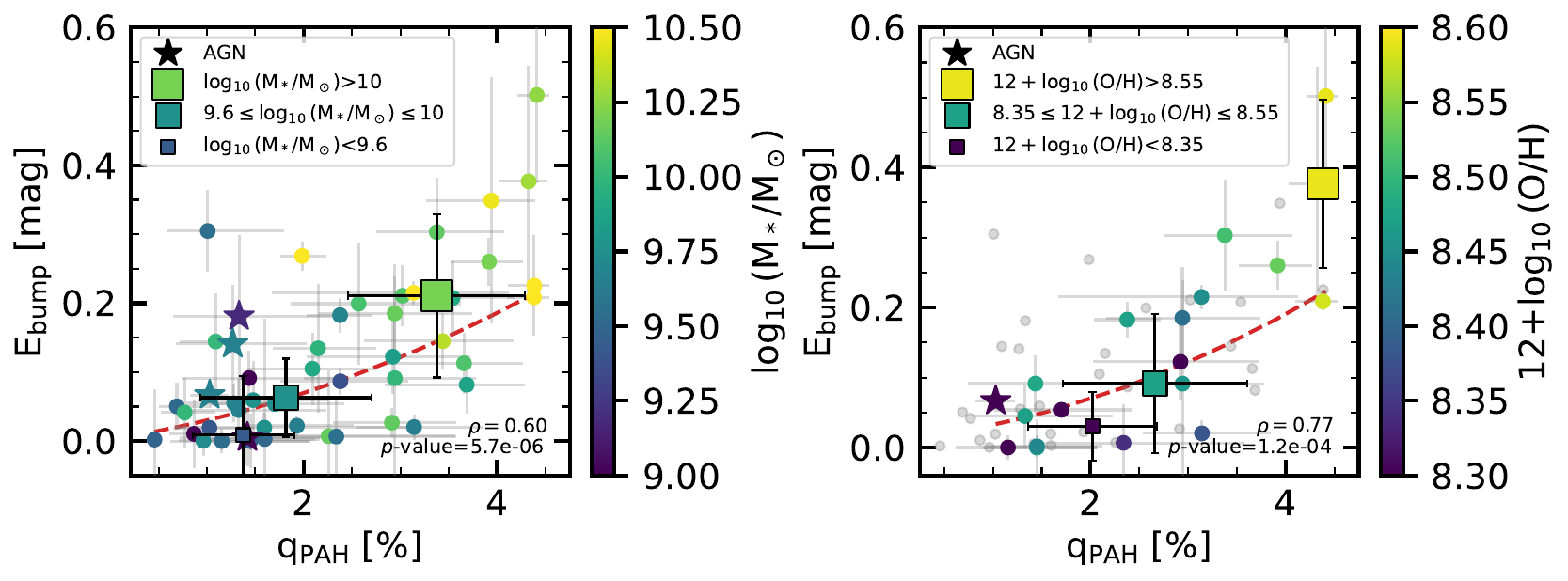}}
      \caption{2175\,\AA\ UV attenuation bump amplitude ($E_{\rm bump}$) as a function of the PAH-to-total dust mass fraction ($q_{\rm PAH}$). Points are color-coded by stellar mass (left; 49 galaxies) and oxygen abundance (right; 19 galaxies). AGN are shown as stars and are included in the analysis. In the right panel, gray points indicate galaxies without metallicity measurements and are shown for reference only. Squares show medians and standard deviations for subsamples built according to mass and metallicity, as shown in the legend and Tab.~\ref{tab02}.  We fit the data points with a  weighted quadratic model of the form $ax^2+bx+c$, shown as red dashed lines; the coefficients are $a=(6.11\pm7.99)\times10^{-3}$, $b=(2.14\pm4.11)\times10^{-2}$, and $c=(2.55\pm43.58)\times10^{-3}$ for the final sample (left panel) and $a=(7.91\pm12.31)\times10^{-3}$, $b=(1.31\pm6.74)\times10^{-2}$, and $c=(1.16\pm7.60)\times10^{-2}$ for the metallicity subsample (right panel). The grey vertical and horizontal lines represent the error bars on the data points in the y and x variables, respectively. Spearman rank correlation coefficient and corresponding $p-$values between $E_{\rm bump}$ and $q_{\rm PAH}$ are reported in each panel. Overall, the figure shows a clear positive correlation between $E_{\rm bump}$ and $q_{\rm PAH}$, with a possible non-linear trend at high $q_{\rm PAH}$.}
      \label{fig07}
\end{figure*}

Fig.~\ref{fig07} shows the UV 2175\,\AA\ bump amplitude ($E_{\rm bump}$) as a function of the PAH mass to total dust mass fraction ($q_{\rm PAH}$), color-coded by stellar mass (left panel) and oxygen abundance (right panel), the latter obtained from spectroscopic measurements as described in Sect.~\ref{sect03:met}. We find a clear positive correlation between $E_{\rm bump}$ and $q_{\rm PAH}$, with a Spearman rank correlation coefficient of $\rho = 0.60$ ($p=5.7\cdot10^{-6}$) for the final sample and $\rho = 0.77$ for the metallicity subsample ($p=1.2\cdot10^{-4}$). These values are consistent within their 95\% confidence intervals (Fig.~\ref{fig06}). The $E_{\rm bump}-q_{\rm PAH}$ correlation is robust to a permutation test, with $p_{\rm perm}=10^{-5}$ and $p_{\rm perm}=2.5\times 10^{-4}$ for the final sample and metallicity subsample, respectively (App.~\ref{app:permutation}). Similar results have been shown previously using smaller samples and/or stacks of cosmic noon galaxies \citep{shivaei_uv_2022, noll_gmass_2009}\footnote{Overall, we find that the AGN-host subset of galaxies follows the same general trends as the general galaxy population. Inspection of the outlier with $E_{\rm bump}>0.2$\,mag and $q_{\rm PAH}<1.5\,\%$ reveals that it shows an SED with substantial uncertainties in the mid-infrared, from where we extract the $q_{\rm PAH}$ values, manifested as large error bars along the $x$-axis.}.

Figs.~\ref{fig06} and~\ref{fig07} show strong correlations between $E_{\rm bump}$ and stellar mass ($\rho=0.66$, $p=2.2\cdot 10^{-7}$) and between $q_{\rm PAH}$ and stellar mass ($\rho=0.69$, $p=3.4\cdot 10^{-8}$) for the final sample. A trend among all three parameters is visible in Fig.~\ref{fig07} (left panel), with median values and standard deviations within each bin reported in Tab.~\ref{tab02}\footnote{The metallicity bins are chosen to distinguish galaxies closer to solar metallicity (high-metallicity bin), galaxies below the metallicity transition associated with the increase in PAH abundance reported in this and previous studies \citep[e.g., ][see also Sect.~\ref{sect05:02}]{engelbracht_metallicity_2005,draine_dust_2007,shivaei_mosdef_2017}, and intermediate-metallicity galaxies. The stellar-mass bins were chosen to approximately correspond to these metallicity regimes based on the MZR from \cite{sanders_co_2023}.
}. Similarly, we also find correlations between $E_{\rm bump}$ and oxygen abundance ($\rho=0.64$, $p=3.0\cdot 10^{-3}$) and between $q_{\rm PAH}$ and oxygen abundance ($\rho=0.46$, $p=4.9\cdot 10^{-2}$) for the metallicity subsample. A similar trend is visible in Fig.~\ref{fig07} (right panel), with median values and standard deviations within each bin also reported in Tab.~\ref{tab02}. The $E_{\rm bump}-$metallicity correlation remains statistically significant under the permutation test ($\rho=0.64$, $p_{\rm perm}=3.6\times 10^{-3}$), whereas the $q_{\rm PAH}-$metallicity correlation is only marginally inconsistent with the null hypothesis ($\rho=0.46$, $p_{\rm perm}=5.0\times 10^{-2}$; App.~\ref{app:permutation}). In both panels, we observe an apparent non-linear correlation between $E_{\rm bump}$ and $q_{\rm PAH}$, characterized by a plateau at lower $q_{\rm PAH}$ values and a sharp increase for $q_{\rm PAH}\geq 3\,\%$. This behaviour is similar to the trends observed between PAH emission and metallicity in studies of local galaxies (e.g., \citealt{engelbracht_metallicity_2005, draine_dust_2007}). The results shown in Fig.~\ref{fig07} are consistent with a scenario in which low-mass and low-metallicity galaxies have a low abundance of PAH molecules (as found by, e.g., \citealt{engelbracht_metallicity_2005, draine_dust_2007, marble_aromatic_2010, shivaei_mosdef_2017}; see Sect.~\ref{sect05:02} for more details), resulting in weak or absent UV bump features. These results support the hypothesis that PAHs are the primary carriers of the UV bump. The larger scatter in $E_{\rm bump}$ at high $q_{\rm PAH}$ values suggests that, in this regime, additional mechanisms are at play (e.g. patchy geometry), as the presence of PAHs in the ISM does not necessarily imply that they manifest as absorption features in the attenuation curve, as discussed below.

\begin{figure*}
   \centering
   \includegraphics[width=\hsize]{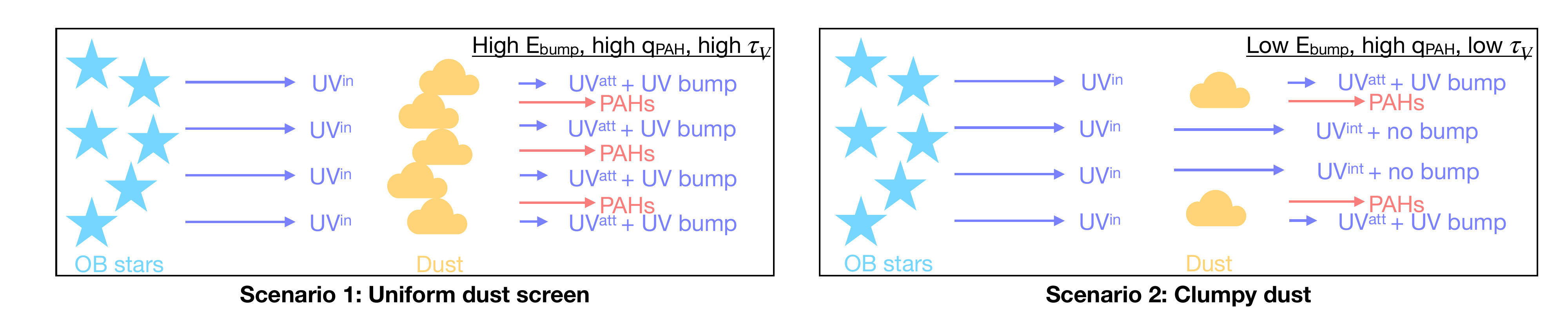}
      \caption{Diagram illustrating two different scenarios that could explain the variation in $E_{\rm bump}$ at high $q_{\rm PAH}$. In the first scenario, the young stellar population is obscured by a uniform dust screen; in this case, galaxies with high PAH abundances will exhibit strong UV bumps and high optical depths. In the second scenario, clumpy dust geometry gives rise to both obscured and unobscured young stellar populations. The UV-optical emission is dominated by the unobscured component, resulting in weak or absent bumps and low $\tau_V$. In contrast, the mid-IR emission, which traces PAHs, is dominated by the dusty regions, leading to high PAH abundances.}
         \label{fig10}
\end{figure*}

One possible explanation for the large scatter in $E_{\rm bump}$ at high $q_{\rm PAH}$ ($\geq 3\%$) is due to clumpy dust geometry \citep{witt_multiple_1996}. In this scenario, at the high $q_{\rm PAH}$ regime, galaxies with low $E_{\rm bump}$ have their (luminosity-weighted) UV-optical emission diluted by the unobscured young stellar populations. This would result in a very weak or absent bump and also a low dust optical depth ($\tau_V$). In contrast, their (luminosity-weighted) mid-IR emission, which traces the PAH emission (and, in particular, the measured $q_{\rm PAH}$), is dominated by the dusty regions, leading to high $q_{\rm PAH}$. This effect could therefore lead to galaxies with high $q_{\rm PAH}$ and low $E_{\rm bump}$ (see the diagram in Fig.~\ref{fig10}).

A similar interpretation is supported by \citet{massa_relations_2022}, who showed that the scatter between the UV bump amplitude and PAH emission can be significantly reduced when controlling for additional variables. By selecting stars located beyond the Galactic dust layer, they ensured that extinction and emission arise from the same grain population, and they further constrained the sample by requiring agreement (within $\pm10\%$) between dust column densities independently derived from dust emission and dust extinction. Under these conditions, the relation between the UV bump and PAH emission becomes much tighter, suggesting that part of the observed scatter arises from variations in line-of-sight dust column density and, in particular, dust-star geometry.

\begin{table*}
\centering
\begin{tabular}{lccccc}
\hline
Bin & Number of sources & $E_{\rm bump}$ [mag] & $q_{\rm PAH}$ [\%] & $\log_{10}(M_*/M_\odot)$ & $12+\log_{10}(\rm O/H)$ \\
\hline
Low stellar mass & 14 & $0.01 \pm 0.09$ & $1.38 \pm 0.52$ & $9.43 \pm 0.35$ & -- \\
Intermediate stellar mass & 18 & $0.06 \pm 0.06$ & $1.82 \pm 0.88$ & $9.75 \pm 0.13$ & -- \\
High stellar mass & 17 & $0.21 \pm 0.12$ & $3.38 \pm 0.91$ & $10.19 \pm 0.30$ & -- \\
\hline\hline
Low metallicity & 4 & $0.03 \pm 0.05$ & $2.02 \pm 0.66$ & -- & $8.28 \pm 0.03$ \\
Intermediate metallicity & 13 & $0.09 \pm 0.10$ & $2.66 \pm 0.94$ & -- & $8.47 \pm 0.04$ \\
High metallicity & 2 & $0.38 \pm 0.12$ & $4.38 \pm 0.04$ & -- & $8.59 \pm 0.01$ \\
\hline
\end{tabular}
\caption{Median values of $E_{\rm bump}$, $q_{\rm PAH}$, stellar mass, and oxygen abundance in stellar mass and metallicity bins (Fig.~\ref{fig07}).}
\label{tab02}
\end{table*}

To quantitatively investigate the origin of the large scatter in $E_{\rm bump}$ at high $q_{\rm PAH}$, we perform a PCA including stellar mass and oxygen abundance. The PCA on $E_{\rm bump}$, $q_{\rm PAH}$, and stellar mass shows that PC1, PC2, and PC3 account for $\sim56\%$, $\sim32\%$, and $\sim12\%$ of the variance, respectively. We therefore do not find evidence for a strong dimensionality reduction among these parameters, indicating that the UV bump amplitude cannot be explained solely by a combination of PAH mass fraction and stellar mass.
In contrast, the PCA on $E_{\rm bump}$, $q_{\rm PAH}$, and oxygen abundance yields contributions of $\sim66\%$, $\sim28\%$, and $\sim6\%$ for PC1–PC3, respectively, suggesting a more compact parameter space when metallicity is used instead of stellar mass. Extending the analysis to $E_{\rm bump}$, $q_{\rm PAH}$, stellar mass, and oxygen abundance results in a fourth principal component contributing only $\sim4\%$, consistent with the metallicity-only case.
Given the well-known correlation between stellar mass and metallicity via the mass–metallicity relation (MZR; e.g. \citealt{tremonti_origin_2004}), this supports a scenario in which the observed trends of $E_{\rm bump}$ and $q_{\rm PAH}$ with stellar mass are at least partly driven by their underlying dependence on gas-phase metallicity, as suggested by previous studies \citep[e.g.,][]{shivaei_uv_2022}.

As shown in Fig.~\ref{fig06}, both $E_{\rm bump}$ and $q_{\rm PAH}$ correlate with the optical depth $\tau_V$ (and, by definition, with $A_V$), such that galaxies with higher dust attenuation tend to have both larger PAH mass fractions and stronger UV bumps (Sects.~\ref{sect05:02} and \ref{sect05:03}). \citet{massa_relations_2022} also explored how $\tau_V$ affects the observed correlation between the UV bump and PAH emission, showing that controlling for dust column density significantly reduces the scatter between both features. Additionally, we find that both $E_{\rm bump}$ and $q_{\rm PAH}$ anti-correlate with sSFR, consistent with previous studies \citep[e.g.,][]{sutter_fraction_2024, battisti_constraining_2025}. This trend suggests that galaxies undergoing intense star formation tend to exhibit weaker UV bumps and lower PAH fractions, likely reflecting the harsher UV radiation fields associated with young stellar populations, which can efficiently destroy small carbonaceous grains.
Motivated by these trends, we extend the PCA to include $\tau_V$ and sSFR, which are linked to the buildup and destruction of small dust grains (PAHs; see also Sect.~\ref{sect05:02}). 
While metallicity yields a more compact parameter space than stellar mass, a strong dimensionality reduction is not achieved with these parameters alone. Including $\tau_V$ and sSFR, the first four principal components account for 97\% of the variance in the metallicity subsample.

\begin{figure}
   \centering
   \includegraphics[width=\hsize]{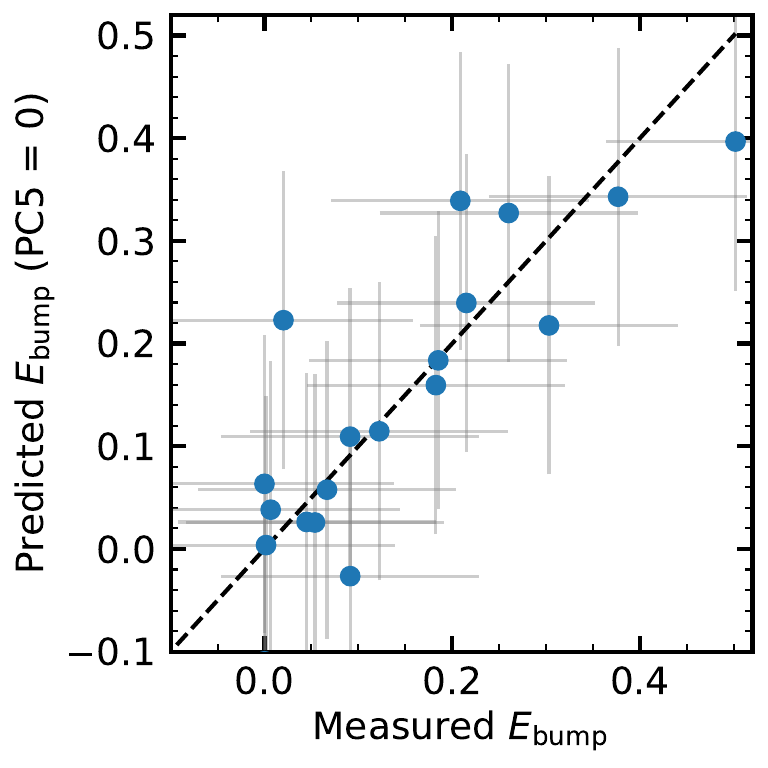}
      \caption{Comparison between measured $E_{\rm bump}$ and values reconstructed from the PCA-derived relation for the metallicity subsample (Eq.~\ref{eq:pca}) using $E_{\rm bump}$, $q_{\rm PAH}$, $12+\log_{10}(\mathrm{O/H})$, sSFR, and $\tau_V$, obtained by setting PC5 (accounting for $\sim3\%$ of the total variance) to zero. Although PCA was performed on standardized variables, the resulting relation is represented in physical units for clarity, following Eq.~\ref{eq:pca}. The grey vertical and horizontal lines represent the error bars on the data points in the y and x variables, respectively. Overall, the PCA-derived $E_{\rm bump}$ shows good agreement with the measured values, supporting the validity of the inferred multivariate relation.}
         \label{fig08}
\end{figure}

For the PCA performed on the metallicity subsample, setting PC5 $=0$ yields an approximate linear relation among the variables, derived in standardized space and expressed in physical units as:
\begin{equation}
\begin{split}
E_{\rm bump} [\rm mag] = 0.11\,q_{\rm PAH}[\%] + 0.58\,(12+\log_{10}(\rm O/H)) \\ - 0.03\,\log_{10}(\rm sSFR [yr^{-1}]) - 0.10\,\tau_V -5.33.
\label{eq:pca}
\end{split}
\end{equation} 
Given the relatively small sample size of the 5-parameter case (19 galaxies), the inferred relation may not be fully stable and should therefore be interpreted with some caution. A comparison between measured $E_{\rm bump}$ and values reconstructed from the PCA-derived relation (Eq.~\ref{eq:pca}) is shown in Fig.~\ref{fig08}, illustrating the degree of agreement between the empirical data and the PC5-based constraint, with a root mean square error (RMSE) of $\sim0.08$\,mag\footnote{This same analysis performed on different UV bump parametrizations gives 96\% (97\%) of the variance explained by PC1-PC4 for $E_{\rm bump}/A_V$ ($E_{\rm bump}/E(B-V)$), and a RMSE between the predicted and measured UV bump of $\sim 33.2$ ($\sim 7.81$).}. This result indicates strong collinearity among the parameters and demonstrates that incorporating dust column density and sSFR substantially improves dimensionality reduction compared to the analysis at the beginning of this section, which considered only stellar mass and gas-phase metallicity. Overall, these findings reinforce the central role of dust (including dust destruction mechanisms) and gas-phase metallicity in shaping the observed UV bump and its connection to PAH emission.

\begin{figure}
   \centering
   \includegraphics[width=\hsize]{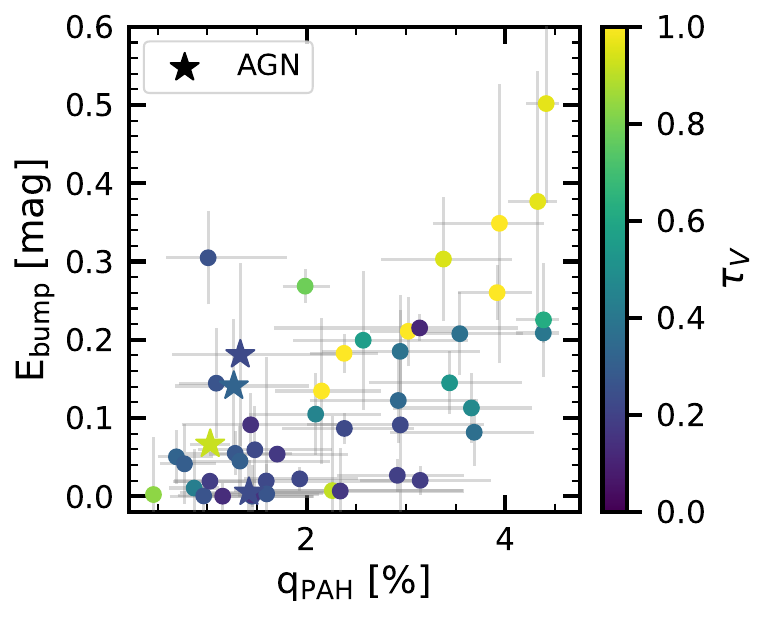}
      \caption{2175\,\AA\ UV attenuation bump amplitude ($E_{\rm bump}$) as a function of PAH-to-total dust mass fraction ($q_{\rm PAH}$), color-coded by dust optical depth (45 galaxies). The grey vertical and horizontal lines represent the error bars on the data points in the y and x variables, respectively. In the high-$q_{\rm PAH}$ regime ($q_{\rm PAH}\geq 3\,\%$), the scatter in the bump amplitude appears to be driven by dust optical depth.}
         \label{fig09}
\end{figure}

Additionally, in Fig.~\ref{fig09}, we show $E_{\rm bump}$ as a function of $q_{\rm PAH}$, color-coded by dust optical depth ($\tau_V$). We find that, in the high-$q_{\rm PAH}$ regime ($q_{\rm PAH}\geq 3\,\%$), the large scatter in $E_{\rm bump}$ correlates with $\tau_V$, such that dustier galaxies exhibit larger apparent UV bump amplitudes\footnote{For consistency with literature that uses parameters such as $E_{\rm bump}/A_V$ or $E_{\rm bump}/E(B-V)$, we also show the correlations for those parameters in App.~\ref{app:02}. The results do not change for the $E_{\rm bump}/E(B-V)$ parametrization: the correlations between $E_{\rm bump}$, $q_{\rm PAH}$, $M_*$, and $12+\log_{10}(\mathrm{O/H})$ persist. However, normalizing $E_{\rm bump}$ by $A_V$ weakens these correlations. This might be due to the fact that $E_{\rm bump}/A_V$ is inversely related to $R_V$ and the attenuation curve shape. Therefore, there might be other factors, such as dust grain properties, that affect the correlations using $E_{\rm bump}/A_V$.}. Indeed, we perform a correlation analysis between $E_{\rm bump}$ and $\tau_V$ for the subsample with $q_{\rm PAH}\geq 3\,\%$ and find $\rho=0.70$ ($p=5.63\cdot10^{-3}$). This behaviour is consistent with the results found by \citet{massa_relations_2022} and with the scenario described above, in which galaxies with high PAH fractions can show weak attenuation bumps if they have a non-uniform (clumpy) dust–star geometry. In this case, the luminosity-weighted UV light arises from low-dust regions, while the luminosity-weighted mid-IR emission is dominated by dustier regions.  We illustrate this scenario schematically in Fig.~\ref{fig10}.

Overall, these results indicate that the relationship between PAH emission and the UV bump is mediated by a broader set of galaxy properties. We further explore the connections between $q_{\rm PAH}/E_{\rm bump}$ and other galaxy properties in Sects.~\ref{sect05:02} and \ref{sect05:03}.

\subsection{PAH emission and galaxy properties}\label{sect05:02}

\begin{figure*}
   \centering
   \includegraphics[width=\hsize]{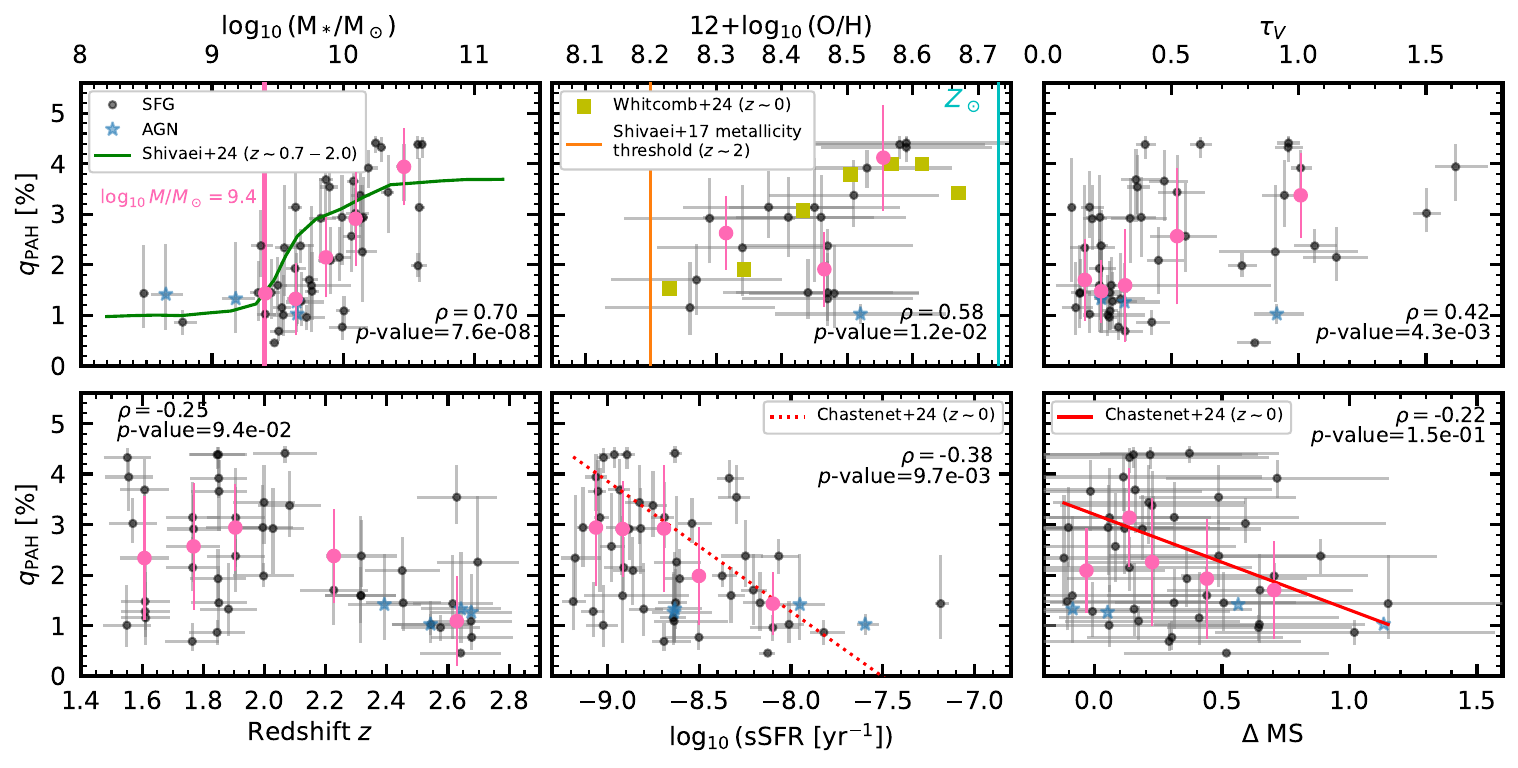}
      \caption{$q_{\rm PAH}$ as a function of various galaxy properties. AGN are included in the plots but not in the analysis. \textsl{Top left:} Stellar mass, including the median for cosmic noon ($z\sim0.7-2.0$) galaxies from \citet{shivaei_new_2024}. \textsl{Top center:} Gas-phase metallicity, with the relation from \citet{whitcomb_metallicity_2024} ($z\sim0$), the solar metallicity value (cyan; \citealt{asplund_chemical_2009}), and the metallicity threshold inferred from the O3N2 ratio in \citet{shivaei_mosdef_2017} ($z\sim2$). \textsl{Top right:} Optical depth. \textsl{Bottom left:} Redshift. \textsl{Bottom center:} sSFR, including the corrected fits from \citet{chastenet_resolved_2024} ($z\sim0$). The correction factor $k=3.5$ is applied to the fit following the form $y=ax+b+k$, where $y=q_{\rm PAH}~[\%]$, $x=\log_{10}(\rm sSFR/yr^{-1})$, and $a=-2.58$ and $b=-22.86$ are the coefficients given in \citet{chastenet_resolved_2024}. $k$ lies within the uncertainties on $b$ reported in that work. \textsl{Bottom right:} Offset from the star-forming main sequence, also compared to the relations from \citet{chastenet_resolved_2024}. Blue stars indicate AGN-host galaxies, and black dots SFGs. We report Spearman rank correlation coefficients and corresponding $p-$values for the sample of SFGs (excluding AGN) in each panel. Median values in bins of the $x-$axis variables are shown in pink, with a minimum bin size of nine sources for the final sample and six sources for the metallicity subsample (see Tab.~\ref{tab03}). The solid pink lines show the best-fit relations for the SFG sample (see Tab.~\ref{tab03} for fitting coefficients). The grey vertical and horizontal lines represent the error bars on the data points in the y and x variables, respectively. The Spearman coefficients are largely consistent with the results obtained from the whole sample (SFGs+AGN; Fig.\ref{fig06}). Overall, PAH abundances are mostly correlated with stellar mass, gas-phase metallicity, dust optical depth, and sSFR.}
         \label{fig11}
\end{figure*}

In Fig.~\ref{fig11}, we examine the dependence of the PAH-to-total dust mass fraction, $q_{\rm PAH}$, on the galaxy properties with which it shows the strongest correlations, as identified in Sect.~\ref{sect05:01}: stellar mass, gas-phase metallicity, optical depth, redshift, and sSFR. We also include the relation with the offset from the star-forming main sequence, $\Delta$MS, which has been reported to correlate with PAH emission in the local universe \citep[e.g.,][]{chastenet_resolved_2024}. For this analysis, we remove AGN from the sample, resulting in a sample size of 45 galaxies for the final sample, and 18 galaxies for the metallicity subsample.

We find that the relation between $q_{\rm PAH}$ and stellar mass is in good agreement with that observed in \cite{shivaei_new_2024} at cosmic noon (top-left panel of Fig.~\ref{fig11}). We identify the transition between low- and high-$q_{\rm PAH}$ regimes at $\log_{10}(M/M_\odot)\sim9.4$, consistent with the value reported by \cite{shivaei_new_2024} ($\sim9.4$, $z\sim0.7-2.0$). As noted in previous studies \citep[e.g.,][]{shivaei_uv_2022}, the trend between PAH abundance and stellar mass is likely driven by the MZR (see below). This is further supported by the PCA performed in Sect.~\ref{sect05:01} (Fig.~\ref{fig08} and Eq.~\ref{eq:pca}), which shows that the scatter between the parameters is more effectively reduced when using metallicity rather than stellar mass. Indeed, PAH emission has been shown to correlate strongly with metallicity both locally \citep{engelbracht_metallicity_2005,draine_dust_2007,remy-ruyer_linking_2015,shim_metallicitypah_2023} and at higher redshifts \citep{shivaei_mosdef_2017}. Consistently, we find that galaxies in lower-metallicity regimes exhibit lower $q_{\rm PAH}$ values, in agreement with scenarios proposing either enhanced PAH destruction or reduced PAH formation efficiencies in metal-poor environments \citep[e.g.][]{draine_dust_2007, marble_aromatic_2010, shivaei_mosdef_2017}.

Our measurements follow a similar $q_{\rm PAH}$-metallicity trend as observed within local galaxies by \citet{whitcomb_metallicity_2024} (top-middle panel of Fig.~\ref{fig11}). In that work, the authors analyze a sample of three spatially resolved galaxies and use SED fitting from 5-160 $\mu$m to derive $q_{\rm PAH}$, combining it with metallicity gradients from \citet{rogers_chaos_2021}. They find that regions with higher metallicity have median PAH abundances of  $q_{\rm PAH}\sim 3.44\%$, while the PAH abundance decreases quickly below a metallicity threshold of $\sim 0.5~Z_\odot$ (which, considering the canonical value for solar oxygen abundance of $12+\log_{10}\rm(O/H)=8.69$ from \citealt{asplund_chemical_2009}, corresponds to an oxygen abundance of $\sim 8.4$). This trend was previously reported in \cite{draine_dust_2007}, among others. In \citet{shivaei_mosdef_2017}, a sample of 476 galaxies at cosmic noon ($1.37\leq z\leq2.61$) was observed using \textit{Spitzer} observations tracing the $7.7~\mu$m PAH feature; gas-phase metallicities were derived through optical emission lines. The metallicity threshold separating low- and high-metallicity regimes identified in that work appears broadly consistent with our data; however, the limited size of our metallicity subsample and the lack of galaxies in the low-metallicity regime prevent a robust comparison. We identify a tentative transition in $q_{\rm PAH}$ at an oxygen abundance of $\sim8.3$, consistent with values reported in other studies (i.e., $\sim8.3$ in \citealt{engelbracht_metallicity_2005} for local galaxies, $\sim8.1$ in \citealt{draine_dust_2007} for local galaxies, $\sim8.2$ in \citealt{shivaei_mosdef_2017} for cosmic noon galaxies). The metallicity and stellar mass thresholds identified in this work are consistent with each other when adopting the MZR from \cite{sanders_co_2023} at $z=2$. 

Further insight into the role of dust is provided by the relation between $q_{\rm PAH}$ and dust optical depth (top-right panel of Fig.~\ref{fig11}). The low-$\tau_V$, low-$q_{\rm PAH}$ regime is consistent with metal-poor systems, where reduced dust shielding facilitates PAH destruction by intense radiation fields \citep[e.g.,][]{madden_ism_2006, hunt_spitzer_2010, xie_new_2019}. Low $q_{\rm PAH}$ may also be linked to low $\tau_V$ because the processes responsible for dust formation and growth in galaxies, such as grain coagulation in the ISM and grain production in AGB winds and SNe ejecta, also drive the formation of PAHs. Conversely, galaxies with high $q_{\rm PAH}$ span a wide range of optical depths. In high-metallicity systems with low $\tau_V$, PAHs may survive owing to softer radiation fields, whereas in high-$\tau_V$ systems, enhanced dust shielding likely promotes PAH survival and growth \citep[e.g.,][]{madden_ism_2006, egorov_phangsjwst_2023, shivaei_tight_2024}. Additionally, different dust-star geometries can lead to a wide distribution of $\tau_V$ at a given $q_{\rm PAH}$ (Fig.~\ref{fig10} and Sect.~\ref{sect05:01}).

A tentative anti-correlation between redshift and $q_{\rm PAH}$ is visible in the bottom-left panel of Fig.~\ref{fig11}, although this trend is not statistically significant ($p\sim0.1$) and should therefore be interpreted with caution. The relation between $q_{\rm PAH}$ and sSFR (bottom-middle panel of Fig.~\ref{fig11}) shows a mild anti-correlation, in agreement with results for local galaxies \citep[e.g.,][]{chastenet_resolved_2024, battisti_constraining_2025}. Galaxies with higher sSFR host younger stellar populations and more intense radiation fields, which can contribute to PAH destruction, particularly for small grains \citep{salgado_orion_2016, chastenet_polycyclic_2019, rigopoulou_properties_2021}. Indeed, previous studies of nearby SFGs have found reduced PAH emission in regions of intense star formation \citep[e.g.,][]{egorov_phangsjwst_2023, sutter_fraction_2024}. However, this trend appears to be driven by two outliers at $\log_{10}(\mathrm{sSFR},[\mathrm{yr}^{-1}]) > -8.0$. After excluding these sources, the correlation weakens to $\rho = -0.33$ with a $p-$value of 0.03. Although this remains close to the conventional significance threshold, it suggests that the observed anti-correlation may be sensitive to these outliers.

Finally, although $\Delta$MS has been shown to anti-correlate with PAH emission in other studies \citep[e.g.,][]{chastenet_resolved_2024, dale_pah_2025}, we find only a weak trend that is not statistically relevant ($p-$value $=0.15$) and has a large scatter (bottom-right panel of Fig.~\ref{fig11}), broadly consistent with the local relations reported by \citet{chastenet_resolved_2024}. In Tab.~\ref{tab03}, we report the binned median values of $q_{\rm PAH}$ and the galaxy properties discussed above (see Fig.~\ref{fig11}).


\begin{table}
\centering
\begin{tabular}{lccc}
\hline
$x-$axis variable & Median $x$ & Median $q_{\rm PAH}$ & $\sigma_{q_{\rm PAH}}$ \\
\hline
$\log_{10}(M_*/M_\odot)$ & 9.41 & 0.01 & 0.04 \\
 & 9.64 & 0.02 & 0.10 \\
 & 9.87 & 0.09 & 0.05 \\
 & 10.09 & 0.14 & 0.09 \\
 & 10.46 & 0.26 & 0.10 \\
\hline
 $12+\log_{10}(O/H)$ & 8.31 & 0.04 & 0.07 \\
 & 8.46 & 0.07 & 0.08 \\
 & 8.55 & 0.28 & 0.13 \\
\hline
 $\tau_V$ & 0.16 & 0.02 & 0.07 \\
 & 0.23 & 0.06 & 0.09 \\
 & 0.32 & 0.05 & 0.07 \\
 & 0.52 & 0.14 & 0.09 \\
 & 1.01 & 0.26 & 0.14 \\
\hline
$z$ & -0.56 & 0.05 & 0.07 \\
 & -0.43 & 0.04 & 0.06 \\
 & -0.29 & 0.06 & 0.10 \\
 & -0.09 & 0.21 & 0.15 \\
 & 0.08 & 0.19 & 0.11 \\
\hline
$\log_{10}(\mathrm{sSFR}/\mathrm{yr}^{-1})$ & -9.06 & 2.94 & 1.15 \\
 & -8.92 & 2.91 & 0.94 \\
 & -8.69 & 2.92 & 1.26 \\
 & -8.50 & 1.98 & 0.97 \\
 & -8.10 & 1.44 & 0.62 \\
\hline
$\Delta\mathrm{MS}$ & -0.03 & 2.09 & 0.84 \\
 & 0.14 & 3.14 & 0.97 \\
 & 0.23 & 2.26 & 1.26 \\
 & 0.44 & 1.93 & 1.19 \\
 & 0.70 & 1.70 & 0.97 \\
\hline
\end{tabular}
\caption{Median values of $q_{\rm PAH}$ [\%] and $x-$axis variables in bins of the galaxy properties shown in Fig.~\ref{fig11}. 
Column 4 shows the standard deviation of $q_{\rm PAH}$ within each bin.}
\label{tab03}
\end{table}

Interestingly, the correlations with surface densities are weaker than those with global quantities (Fig.~\ref{fig06}). In particular, the correlations with stellar mass and SFR are stronger than those with their corresponding surface densities, for both $q_{\rm PAH}$ and $E_{\rm bump}$, suggesting that the trends with surface densities are primarily driven by the underlying integrated quantities. Alternatively, this may indicate that the spatial regions over which stellar mass and SFR are measured do not fully coincide with those dominating the PAH emission and UV bump, pointing to a potential role of dust–star geometry in shaping these correlations. Overall, we find that the low-$q_{\rm PAH}$ regime is associated with low-mass, low-metallicity, and low-optical depth galaxies, whereas high-$q_{\rm PAH}$ values are found in low-sSFR galaxies. These results suggest that dust is linked to the survival, growth, and destruction of PAHs in galaxies.

\subsection{UV bump and galaxy properties}\label{sect05:03}

\begin{figure*}
   \centering
   \includegraphics[width=\hsize]{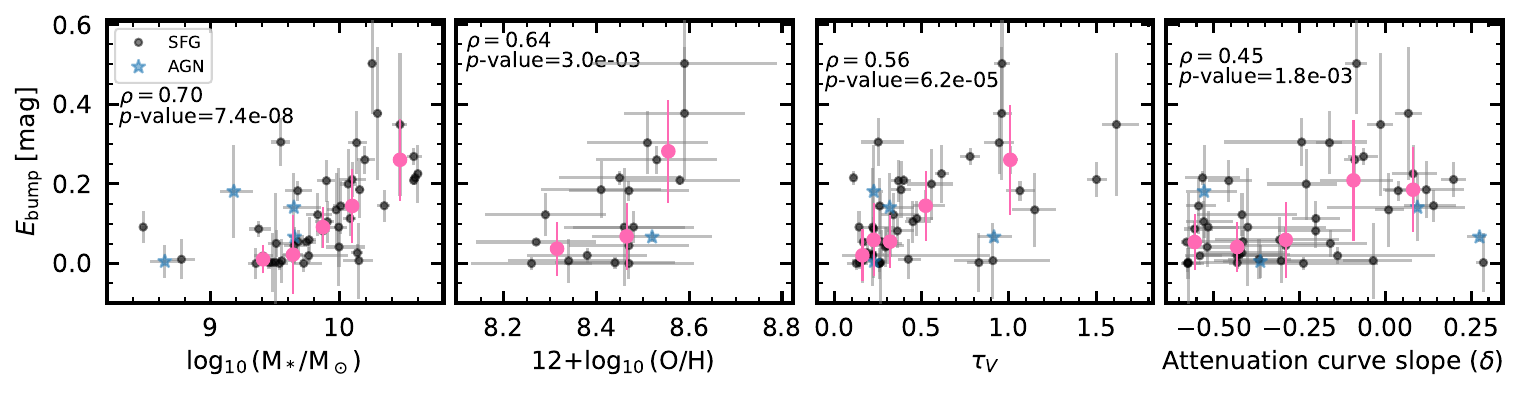}
      \caption{$E_{\rm bump}$ as a function of various galaxy properties: stellar mass (left), oxygen abundance (center-left), optical depth (center-right), and slope of the attenuation curve (right). Blue stars indicate AGN-host galaxies, and black dots SFGs. AGN are included in the plots but not in the analysis. We report Spearman rank correlation coefficients and corresponding $p-$values for the sample of SFGs (excluding AGN) in each panel. Median values in bins of the $x-$axis variables are shown in pink, with a minimum bin size of nine sources for the final sample and six sources for the metallicity subsample (see Tab.~\ref{tab04}). The grey vertical and horizontal lines represent the error bars on the data points in the y and x variables, respectively. The Spearman coefficients are largely consistent with the results obtained from the whole sample (SFGs+AGN; Fig.~\ref{fig06}). Overall, the UV bump amplitude is mostly correlated with stellar mass, gas-phase metallicity, dust optical depth, and slope of the dust attenuation curve.}
         \label{fig12}
\end{figure*}

Fig.~\ref{fig12} shows the apparent UV bump amplitude, $E_{\rm bump}$, as a function of different global galaxy properties: stellar mass, oxygen abundance, dust optical depth, and slope of the attenuation curve. While the UV bump and oxygen abundance are derived from rest-frame UV and optical spectra, respectively, the other parameters are derived from SED fitting to photometry. For this analysis, we remove AGN from the sample, resulting in a sample size of 45 galaxies for the final sample, and 18 galaxies for the metallicity subsample.

Previous studies have reported correlations between $E_{\rm bump}$ and stellar mass \citep{kashino_2175_2021, shivaei_uv_2022}, as well as gas-phase metallicity either from spectroscopy \citep{shivaei_mosdef_2020} or SED fitting \citep{markov_unveiling_2025}. In this work, based on independently derived UV bump, mass, and metallicity parameters, we find a strong correlation between $E_{\rm bump}$ and stellar mass ($\rho=0.7$, $p-$value$=7.4\cdot 10^{-8}$; left panel of Fig.~\ref{fig12}). We also find a correlation with metallicity ($\rho=0.64$, $p=3\cdot 10^{-3}$; second panel of Fig.~\ref{fig12}). 
The two correlations are likely linked to each other owing to the mass-metallicity relation. They clearly show that at low metallicities, the UV bump is weak: strong UV attenuation bumps only appear in metal-rich massive systems. The scatter of $E_{\rm bump}$ at high metallicities can be attributed to clumpy dust (Sect.~\ref{sect05:01}).

Correlations with optical depth or $V$-band attenuation in the literature have been less conclusive, likely due to differences in the UV bump parametrization, as well as the influence of dust clumpiness. For example, \cite{markov_unveiling_2025} analyze a sample of 173 galaxies at $z\sim 2-12$ using NIRSpec prism data and perform SED fitting with a modified version of the \texttt{BAGPIPES} code (\citealt{markov_dust_2023} for details), from which they derived dust attenuation curves and UV bump amplitudes. They report an anti-correlation between the bump amplitude, defined using the parametrization of \cite{salim_dust_2020}, and $A_V$. However, here (third panel of Fig.~\ref{fig12}), we find a positive correlation between the bump strength and dust optical depth. In this work, the UV attenuation bump is defined as the apparent (i.e., observed) feature measured directly from the spectra, whereas studies based on SED fitting \citep[e.g.,][]{markov_evolution_2025} typically derive intrinsic bump amplitudes (i.e., observed bump amplitudes normalized by a measure of total dust attenuation or reddening). These intrinsic estimates are subject to uncertainties associated with the assumed dust attenuation curve (see App.~\ref{app:02}).

A connection between UV bump amplitude and the slope of the attenuation curve has been reported in previous works, however, with the opposite trend that we find here. \cite{kriek_dust_2013} construct composite SEDs from $\sim3500$ galaxies at $0.5 < z < 2.0$ (\citealt{kriek_h_2011} for details), using photometry from the NEWFIRM medium-band survey \citep[NMBS;][]{van_dokkum_newfirm_2009}. They find that the bump amplitude anti-correlates with $\delta$, such that steeper (lower $\delta$) attenuation curves exhibit stronger UV bumps. However, this correlation is likely due to the fact that in \citet{kriek_dust_2013} the dust attenuation curve parametrization links the bump amplitude to the slope of the curve (which is different from the original \citealt{noll_gmass_2009} parametrization used in this work). The trend between bump amplitude and $\delta$ found in \citet{kriek_dust_2013} is opposite to our results (right panel of Fig.~\ref{fig12}), that show that galaxies with high $E_{\rm bump}$ tend to be greyer with shallower UV-optical dust attenuation curve slopes (i.e., higher $\delta$). As our measurements of the UV bump amplitude and attenuation curve slope are derived independently, one from rest-UV spectra and the other from SED fitting to photometry, the two parameters are not degenerate. In general, if a galaxy’s {\em extinction} curve includes a UV bump, radiative transfer effects (e.g., variations in optical depth or dust covering fraction) tend to both weaken the bump in {\em attenuation} and produce a shallower attenuation curve slope \citep{witt_multiple_1996}. This would lead to a bump-slope anti-correlation, for cases where dust clumpiness is the dominant factor in determining the shape of the attenuation curve \citep{narayanan_theory_2018}. However, in this work, the presence of strong bumps in galaxies with shallow attenuation curves (Fig.~\ref{fig12} right) indicates that the bump carriers exist in these systems, regardless of the dust clumpiness, while they are absent in the galaxies with the steepest attenuation curves. These trends are consistent with the findings of \cite{shivaei_mosdef_2020}, who reported a steep average attenuation curve without a bump at low metallicities, and a shallow curve with a significant bump at high metallicities at Cosmic Noon. Similar results were found by \cite{battisti_characterizing_2017}, who studied the impact of inclination on the shape of average dust attenuation curves obtained from a sample of $\sim 10{,}000$ SFGs in the local universe. They also reported evidence for a UV bump feature in the shallowest curves at high galaxy inclinations. Our observations are similar to those findings for individual galaxies using robust spectroscopic bump measurements at Cosmic Noon. We note that despite the moderate $E_{\rm bump}-$slope correlation ($\rho=0.45$, $p=1.8\cdot 10^{-3}$), we observe a large scatter in the relation, particularly at high $\delta$ values where galaxies exhibit a wide range of bump amplitudes, from 0 to 0.5\,mag. 

As consistently observed throughout this work, dusty, massive, and metal-rich galaxies exhibit a wide range of UV bump amplitudes, spanning from negligible values up to $\sim0.5$\,mag. In contrast, lower-mass, lower-metallicity systems generally display weak or absent UV bumps. These results suggest a scenario in which less massive galaxies, which are typically less metal-rich and less dusty, preferentially exhibit steeper attenuation curves and weak UV bumps. At low metallicities, although galaxies are usually less dusty, efficient dust growth and processing (such as shattering of dust grains into smaller grains) skews the grain size distribution toward smaller sizes, leading to steeper attenuation curve slopes (lower $\delta$). In contrast, in high-metallicity, dust-rich environments, grain coagulation becomes more efficient, increasing the fraction of large grains \citep{asano_dust_2013, hirashita_remodelling_2019, relano_dust_2022}, and thus producing shallower attenuation curves \citep{weingartner_dust_2001, hou_dust_2019}.  As discussed in Sect.~\ref{sect05:01}, dust clumpiness also affects the UV bump amplitude and may lead to weak or absent bumps, even in the presence of strong PAH emission. In Tab.~\ref{tab04}, we report the binned median values of $E_{\rm bump}$ and the galaxy properties discussed above (see Fig.~\ref{fig12}).

\begin{table}
\centering
\begin{tabular}{lccc}
\hline
$x-$axis variable & Median $x$ & Median $E_{\rm bump}$ & $\sigma(E_{\rm bump})$ \\
\hline
$\log_{10}(M_*/M_\odot)$ & 9.41 & 0.01 & 0.04 \\
 & 9.64 & 0.02 & 0.10 \\
 & 9.87 & 0.09 & 0.05 \\
 & 10.09 & 0.14 & 0.09 \\
 & 10.46 & 0.26 & 0.10 \\
\hline
$12+\log_{10}(O/H)$ & 8.31 & 0.04 & 0.07 \\
 & 8.46 & 0.07 & 0.08 \\
 & 8.55 & 0.28 & 0.13 \\
\hline
$\tau_V$ & 0.16 & 0.02 & 0.07 \\
 & 0.23 & 0.06 & 0.09 \\
 & 0.32 & 0.05 & 0.07 \\
 & 0.52 & 0.14 & 0.09 \\
 & 1.01 & 0.26 & 0.14 \\
\hline
$\delta$ & -0.56 & 0.05 & 0.07 \\
 & -0.43 & 0.04 & 0.06 \\
 & -0.29 & 0.06 & 0.10 \\
 & -0.09 & 0.21 & 0.15 \\
 & 0.08 & 0.19 & 0.11 \\
\hline
\end{tabular}
\caption{Median values of $E_{\rm bump}$ [mag] and $x-$axis variables in bins of the galaxy properties shown in Fig.~\ref{fig12}. Column 4 shows the standard deviation of $E_{\rm bump}$ within each bin.}
\label{tab04}
\end{table}

\section{Summary and conclusions}\label{sect06}

In this work, we present the first systematic analysis of the relationship between the UV attenuation bump and the PAH-to-total dust mass fraction, $q_{\rm PAH}$, in a statistically significant sample of individual galaxies, and examine how this connection varies with gas-phase metallicity (through oxygen abundance) and other global galaxy properties. We analyze a sample of 49 galaxies at cosmic noon ($z\sim 1$--3) and derive their physical properties, including $q_{\rm PAH}$, from SED modeling spanning the rest-frame UV to the mid-IR. The SEDs are fitted using the \texttt{Prospector} code to photometry from \textit{JWST}/NIRCam and MIRI, \textit{HST}, and, where available, ALMA and \textit{Herschel}. VLT/MUSE spectra are used to measure the spectrosopic redshifts and the UV bump amplitude, $E_{\rm bump}$. Oxygen abundances are derived from oxygen and nitrogen emission line ratios (O3N2, N2, and O3). The required line fluxes are measured from observations with Keck/MOSFIRE in bands K and H, covering the rest-frame optical for our redshift range, and from the \textit{JWST}/NIRSpec observations.

Our main results can be summarized as follows:
\begin{itemize}
    \item We find a clear correlation between the apparent UV bump amplitude, $E_{\rm bump}$, and the PAH mass fraction, $q_{\rm PAH}$ (Fig.~\ref{fig07}, $\rho=0.60$), extending works performed in the local Universe or with stacking at cosmic noon. In particular, we find that galaxies with low PAH abundances have weak to absent bumps. These results support a scenario in which PAH molecules are the primary carriers of the 2175\,\AA\ UV bump. Moreover, we observe a large range of $E_{\rm bump}$ ($\sim 0 - 0.5$\,mag) at high $q_{\rm PAH}$ ($>3$\,\%) that appears to be driven by dust optical depth (Fig.~\ref{fig09}). The absence of a bump even in high $q_{\rm PAH}$ regimes can be explained by non-uniform (clumpy) dust-star geometry, where the mid-IR emission arises from dusty regions, while the UV light is dominated by the low-dust regions (Fig.~\ref{fig10}).
    
    \item Both $E_{\rm bump}$ and $q_{\rm PAH}$ correlate with stellar mass ($\rho=0.66$ and $\rho=0.69$, respectively), oxygen abundance ($\rho=0.64$ and $\rho=0.46$), and dust optical depth ($\rho=0.54$ and $\rho=0.36$, Fig.~\ref{fig09}). Both show a moderate anti-correlation with sSFR ($\rho=-0.35$ and $\rho=-0.42$, respectively). Several PCAs performed on different parameters show that PAHs are not solely responsible for the strength of the UV attenuation bump, but that a combination of $q_{\rm PAH}$, oxygen abundance, sSFR, and optical depth is necessary to explain $E_{\rm bump}$ (Fig.~\ref{fig08}). This confirms that the UV bump can be suppressed in low-metallicity environments, and/or in galaxies with higher sSFRs, where harder UV radiation fields associated with young stellar populations efficiently destroy PAH molecules, especially in the absence of dust. 
    
    \item The PAH mass fraction shows correlations with mass, metallicity, and $V$-band optical depth; and an anti-correlation with sSFR (Fig.~\ref{fig11}). These results likely reflect the interplay between formation and survival of PAHs. In particular, the processes responsible for dust growth (such as AGB winds and SNe ejecta) also contribute to the formation of PAHs. Moreover, galaxies with low metallicity and low optical depth are more susceptible to PAH destruction in the presence of hard UV radiation, due to reduced dust shielding. Indeed, galaxies with high sSFR will have larger populations of OB-type stars that produce harder radiation fields; they will also have larger SN rates. Both mechanisms may contribute to PAH destruction, particularly at low metallicitites.
    
    \item The UV bump amplitude also correlates with dust attenuation curve slope (Fig.~\ref{fig12}). Variations in optical depth appear to regulate the observed scatter in $E_{\rm bump}$ at high $q_{\rm PAH}$ (Fig.~\ref{fig09}), highlighting how clumpy dust geometry can suppress the apparent UV bump (Fig.~\ref{fig10}). The correlations of the UV bump with metallicity and mass also indicate the absence of UV bump carriers in metal-poor/low-mass galaxies. Interestingly, galaxies with steep attenuation curves, on average, have weak (or absent) bumps.
\end{itemize}

Very few studies have explored the relationship between PAH emission and the UV 2175\,\AA\ attenuation bump, particularly beyond the local Universe and in statistically significant samples of individual galaxies rather than stacked data. In this work, we provide the first analysis of this connection at the level of individual sources, while also incorporating gas-phase metallicity, a key parameter long known to be linked to PAH emission. Our results highlight the importance of jointly considering dust properties and metallicity in shaping these features. Future studies based on larger samples will be essential to confirm and extend the trends identified here and to further constrain the physical mechanisms governing both PAH formation and the origin of the UV bump.

This study demonstrates the power of multi-wavelength analyses combining state-of-the-art facilities such as \textit{JWST}, VLT, and Keck to probe the dust properties of galaxies from the rest-frame UV to the mid-infrared. Future facilities, including the Extremely Large Telescope \citep[ELT;][]{padovani_extremely_2023} and the \textit{PRobe far-Infrared Mission for Astrophysics} \citep[PRIMA;][]{glenn_prima_2025}, will enable these studies to be extended to fainter systems and longer wavelengths, providing further insight into the origin and evolution of dust and PAHs in galaxies \citep[e.g.,][]{arriscado_metallicity-pah_2025}.

\begin{acknowledgements}
     This work is supported by the European Research Council (ERC) under the European Union’s Horizon 2020 research and innovation programme (DistantDust, Grant agreement No. 101117541). LA is supported by the CSIC Program 'Programa JAE (JAE-Pre 2023), by the grant PID2024-158856NA-I00 funded by Spanish Ministerio de Ciencia e Innovación MCIN/AEI/10.13039/501100011033 and by “ERDF A way of making Europe”. 
     IS is supported by the Atracc\'{i}on de Talento Grant No.2022-T1/TIC-20472 of the Comunidad de Madrid, Spain.
     LAB acknowledges support from the Dutch Research Council (NWO) under grant VI.Veni.242.055 (\url{https://doi.org/10.61686/LAJVP77714}).
     PGP-G acknowledges support from grant PID2022-139567NB-I00 funded by Spanish Ministerio de Ciencia, Innovaci\'on y Universidades MCIU/AEI/10.13039/501100011033, FEDER {\it Una manera de hacer Europa}.
     The work of GHR  was supported in part by grant 80NSSC18K0555, from NASA Goddard Space Flight Center to the University of Arizona.
     YZ acknowledges JWST/NIRCam contract to the University of Arizona NAS5-02105.
     AJB acknowledges funding from the “FirstGalaxies” Advanced Grant from the European Research Council (ERC) under the European Union’s Horizon 2020 research and innovation program (Grant agreement No. 789056).
     ECL acknowledges support of an STFC Webb Fellowship (ST/W001438/1).
     ZJ acknowledges JWST/NIRCam contract to the University of Arizona, NAS5-02105.
     BRP acknowledges support from grant PID2024-158856NA-I00 funded by Spanish Ministerio de Ciencia e Innovación MCIN/AEI/10.13039/501100011033 and by “ERDF A way of making Europe”.
     ST acknowledges support by the Royal Society Research Grant G125142.
     The research of CCW is supported by NOIRLab, which is managed by the Association of Universities for Research in Astronomy (AURA) under a cooperative agreement with the National Science Foundation.
     JW gratefully acknowledges support from the Cosmic Dawn Center through the DAWN Fellowship. The Cosmic Dawn Center (DAWN) is funded by the Danish National Research Foundation under grant No. 140.
     \\
     Some of the data presented herein were obtained at the W. M. Keck Observatory, which is operated as a scientific partnership among the California Institute of Technology, the University of California and the National Aeronautics and Space Administration. The Observatory was made possible by the generous financial support of the W. M. Keck Foundation. The authors recognize and acknowledge the very significant cultural role and reverence that the summit of Maunakea has always had within the Native Hawaiian community. We are most fortunate to have the opportunity to conduct observations from this mountain.
     \\
     This work is based in part on observations made with the NASA/ESA/CSA James Webb Space Telescope. The data was obtained from the Mikulski Archive for Space Telescopes at the Space Telescope Science Institute, which is operated by the Association of Universities for Research in Astronomy, Inc., under NASA contract NAS 5-03127 for JWST.
     JADES DR4 includes NIRSpec observations from JWST programs 1180, 1181, 1210, 1286, 1287 and 3215.
     JADES DR5 includes NIRCam data from JWST programs 1176, 1180, 1181, 1210, 1264, 1283, 1286, 1287, 1895, 1963, 2079, 2198, 2514, 2516, 2674, 3215, 3577, 3990, 4540, 4762, 5398, 5997, 6434, 6511, and 6541.
     JADES DR5 includes MIRI data from JWST programs 1180, 1181, and 1207.
     The authors acknowledge the teams of programs 1895, 1963, 2079, 2514, 3215, 3577, 3990, 6434, and 6541 for developing their observing program with a zero-exclusive-access period.
     SMILES observations are associated with JWST program 1207. The specific observations analyzed can be accessed via \url{https://archive.stsci.edu/hlsp/smiles} under doi:10.17909/et3f-zd57.
     \\
     The analysis presented in this paper was made possible by several Python packages, including \texttt{Astropy} \citep{the_astropy_collaboration_astropy_2013, the_astropy_collaboration_astropy_2018, the_astropy_collaboration_astropy_2022}, \texttt{lmfit} \citep{newville_lmfit_2025}, \texttt{Matplotlib} \citep{hunter_matplotlib_2007}, \texttt{Numpy} \citep{harris_array_2020}, \texttt{Pandas} \citep{the_pandas_development_team_pandas-devpandas_2026, mckinney_data_2010}, \texttt{pingouin} \citep{vallat_pingouin_2018}, and \texttt{sklearn} \citep{pedregosa_scikit-learn_2011}.
\end{acknowledgements}

%
%





\bibliographystyle{aa} 
\bibliography{bibliography} 

@article{madau_radiative_1995,
	title = {Radiative {Transfer} in a {Clumpy} {Universe}: {The} {Colors} of {High}-{Redshift} {Galaxies}},
	volume = {441},
	issn = {0004-637X},
	shorttitle = {Radiative {Transfer} in a {Clumpy} {Universe}},
	url = {https://ui.adsabs.harvard.edu/abs/1995ApJ...441...18M},
	doi = {10.1086/175332},
	urldate = {2025-10-28},
	journal = {ApJ},
	publisher = {IOP},
	author = {Madau, Piero},
	month = mar,
	year = {1995},
	pages = {18},
}

@article{leja_hot_2018,
	title = {Hot {Dust} in {Panchromatic} {SED} {Fitting}: {Identification} of {Active} {Galactic} {Nuclei} and {Improved} {Galaxy} {Properties}},
	volume = {854},
	issn = {0004-637X},
	shorttitle = {Hot {Dust} in {Panchromatic} {SED} {Fitting}},
	url = {https://doi.org/10.3847/1538-4357/aaa8db},
	doi = {10.3847/1538-4357/aaa8db},
	language = {en},
	number = {1},
	urldate = {2025-10-28},
	journal = {ApJ},
	publisher = {The American Astronomical Society},
	author = {Leja, Joel and Johnson, Benjamin D. and Conroy, Charlie and Dokkum, Pieter van},
	month = feb,
	year = {2018},
	pages = {62},
}

@article{draine_infrared_2007,
	title = {Infrared {Emission} from {Interstellar} {Dust}. {IV}. {The} {Silicate}-{Graphite}-{PAH} {Model} in the {Post}-{Spitzer} {Era}},
	volume = {657},
	issn = {0004-637X},
	url = {https://ui.adsabs.harvard.edu/abs/2007ApJ...657..810D},
	doi = {10.1086/511055},
	urldate = {2025-10-28},
	journal = {ApJ},
	publisher = {IOP},
	author = {Draine, B. T. and Li, Aigen},
	month = mar,
	year = {2007},
	pages = {810--837},
}

@article{byler_nebular_2017,
	title = {Nebular {Continuum} and {Line} {Emission} in {Stellar} {Population} {Synthesis} {Models}},
	volume = {840},
	issn = {0004-637X},
	url = {https://ui.adsabs.harvard.edu/abs/2017ApJ...840...44B},
	doi = {10.3847/1538-4357/aa6c66},
	urldate = {2025-10-28},
	journal = {ApJ},
	publisher = {IOP},
	author = {Byler, Nell and Dalcanton, Julianne J. and Conroy, Charlie and Johnson, Benjamin D.},
	month = may,
	year = {2017},
	pages = {44},
}

@article{kriek_dust_2013,
	title = {The {Dust} {Attenuation} {Law} in {Distant} {Galaxies}: {Evidence} for {Variation} with {Spectral} {Type}},
	volume = {775},
	issn = {0004-637X},
	shorttitle = {The {Dust} {Attenuation} {Law} in {Distant} {Galaxies}},
	url = {https://ui.adsabs.harvard.edu/abs/2013ApJ...775L..16K},
	doi = {10.1088/2041-8205/775/1/L16},
	urldate = {2025-10-28},
	journal = {ApJ},
	publisher = {IOP},
	author = {Kriek, Mariska and Conroy, Charlie},
	month = sep,
	year = {2013},
	pages = {L16},
}

@article{pedregosa_scikit-learn_2011,
	title = {Scikit-learn: {Machine} {Learning} in {Python}},
	volume = {12},
	issn = {1533-7928},
	shorttitle = {Scikit-learn},
	url = {http://jmlr.org/papers/v12/pedregosa11a.html},
	number = {85},
	urldate = {2025-10-27},
	journal = {Journal of Machine Learning Research},
	author = {Pedregosa, Fabian and Varoquaux, Gaël and Gramfort, Alexandre and Michel, Vincent and Thirion, Bertrand and Grisel, Olivier and Blondel, Mathieu and Prettenhofer, Peter and Weiss, Ron and Dubourg, Vincent and Vanderplas, Jake and Passos, Alexandre and Cournapeau, David and Brucher, Matthieu and Perrot, Matthieu and Duchesnay, Édouard},
	year = {2011},
	pages = {2825--2830},
}

@article{vallat_pingouin_2018,
	title = {Pingouin: statistics in {Python}},
	volume = {3},
	issn = {2475-9066},
	shorttitle = {Pingouin},
	url = {https://joss.theoj.org/papers/10.21105/joss.01026},
	doi = {10.21105/joss.01026},
	language = {en},
	number = {31},
	urldate = {2025-10-27},
	journal = {Journal of Open Source Software},
	author = {Vallat, Raphael},
	month = nov,
	year = {2018},
	pages = {1026},
}

@article{calzetti_dust_1994,
	title = {Dust {Extinction} of the {Stellar} {Continua} in {Starburst} {Galaxies}: {The} {Ultraviolet} and {Optical} {Extinction} {Law}},
	volume = {429},
	issn = {0004-637X},
	shorttitle = {Dust {Extinction} of the {Stellar} {Continua} in {Starburst} {Galaxies}},
	url = {https://ui.adsabs.harvard.edu/abs/1994ApJ...429..582C},
	doi = {10.1086/174346},
	urldate = {2025-09-15},
	journal = {ApJ},
	author = {Calzetti, Daniela and Kinney, Anne L. and Storchi-Bergmann, Thaisa},
	month = jul,
	year = {1994},
	pages = {582},
}

@article{rieke_performance_2023,
	title = {Performance of {NIRCam} on {JWST} in {Flight}},
	volume = {135},
	issn = {0004-6280},
	url = {https://ui.adsabs.harvard.edu/abs/2023PASP..135b8001R},
	doi = {10.1088/1538-3873/acac53},
	urldate = {2025-08-20},
	journal = {PASP},
	author = {Rieke, Marcia J. and Kelly, Douglas M. and Misselt, Karl and Stansberry, John and Boyer, Martha and Beatty, Thomas and Egami, Eiichi and Florian, Michael and Greene, Thomas P. and Hainline, Kevin and Leisenring, Jarron and Roellig, Thomas and Schlawin, Everett and Sun, Fengwu and Tinnin, Lee and Williams, Christina C. and Willmer, Christopher N. A. and Wilson, Debra and Clark, Charles R. and Rohrbach, Scott and Brooks, Brian and Canipe, Alicia and Correnti, Matteo and DiFelice, Audrey and Gennaro, Mario and Girard, Julien H. and Hartig, George and Hilbert, Bryan and Koekemoer, Anton M. and Nikolov, Nikolay K. and Pirzkal, Norbert and Rest, Armin and Robberto, Massimo and Sunnquist, Ben and Telfer, Randal and Wu, Chi Rai and Ferry, Malcolm and Lewis, Dan and Baum, Stefi and Beichman, Charles and Doyon, René and Dressler, Alan and Eisenstein, Daniel J. and Ferrarese, Laura and Hodapp, Klaus and Horner, Scott and Jaffe, Daniel T. and Johnstone, Doug and Krist, John and Martin, Peter and McCarthy, Donald W. and Meyer, Michael and Rieke, George H. and Trauger, John and Young, Erick T.},
	month = feb,
	year = {2023},
	pages = {028001},
}

@misc{hviding_theskyentistgelato_2022,
	title = {{TheSkyentist}/{GELATO}: {GELATO} v2.5.2},
	shorttitle = {{TheSkyentist}/{GELATO}},
	url = {https://zenodo.org/records/7439084},
	doi = {10.5281/zenodo.7439084},
	urldate = {2025-08-19},
	publisher = {Zenodo},
	author = {Hviding, Raphael Erik},
	month = dec,
	year = {2022},
}

@article{jakobsen_near-infrared_2022,
	title = {The {Near}-{Infrared} {Spectrograph} ({NIRSpec}) on the \textit{{James} {Webb}} {Space} {Telescope}: {I}. {Overview} of the instrument and its capabilities},
	volume = {661},
	copyright = {https://www.edpsciences.org/en/authors/copyright-and-licensing},
	issn = {0004-6361, 1432-0746},
	shorttitle = {The {Near}-{Infrared} {Spectrograph} ({NIRSpec}) on the \textit{{James} {Webb}} {Space} {Telescope}},
	url = {https://www.aanda.org/10.1051/0004-6361/202142663},
	doi = {10.1051/0004-6361/202142663},
	language = {en},
	urldate = {2025-08-19},
	journal = {A\&A},
	author = {Jakobsen, P. and Ferruit, P. and Alves De Oliveira, C. and Arribas, S. and Bagnasco, G. and Barho, R. and Beck, T. L. and Birkmann, S. and Böker, T. and Bunker, A. J. and Charlot, S. and De Jong, P. and De Marchi, G. and Ehrenwinkler, R. and Falcolini, M. and Fels, R. and Franx, M. and Franz, D. and Funke, M. and Giardino, G. and Gnata, X. and Holota, W. and Honnen, K. and Jensen, P. L. and Jentsch, M. and Johnson, T. and Jollet, D. and Karl, H. and Kling, G. and Köhler, J. and Kolm, M.-G. and Kumari, N. and Lander, M. E. and Lemke, R. and López-Caniego, M. and Lützgendorf, N. and Maiolino, R. and Manjavacas, E. and Marston, A. and Maschmann, M. and Maurer, R. and Messerschmidt, B. and Moseley, S. H. and Mosner, P. and Mott, D. B. and Muzerolle, J. and Pirzkal, N. and Pittet, J.-F. and Plitzke, A. and Posselt, W. and Rapp, B. and Rauscher, B. J. and Rawle, T. and Rix, H.-W. and Rödel, A. and Rumler, P. and Sabbi, E. and Salvignol, J.-C. and Schmid, T. and Sirianni, M. and Smith, C. and Strada, P. and Te Plate, M. and Valenti, J. and Wettemann, T. and Wiehe, T. and Wiesmayer, M. and Willott, C. J. and Wright, R. and Zeidler, P. and Zincke, C.},
	month = may,
	year = {2022},
	pages = {A80},
}

@article{williams_detection_2009,
	title = {Detection of quiescent galaxies in a bicolor sequence from z=0-2},
	volume = {691},
	issn = {0004-637X, 1538-4357},
	url = {http://arxiv.org/abs/0806.0625},
	doi = {10.1088/0004-637X/691/2/1879},
	number = {2},
	urldate = {2025-08-07},
	journal = {ApJ},
	author = {Williams, Rik J. and Quadri, Ryan F. and Franx, Marijn and Dokkum, Pieter van and Labbe, Ivo},
	month = feb,
	year = {2009},
	pages = {1879--1895},
}

@article{genin_dawn_2025,
	title = {{DAWN} {JWST} {Archive}: {Morphology} from profile fitting of over 340 000 galaxies in major fields},
	volume = {699},
	issn = {0004-6361, 1432-0746},
	shorttitle = {{DAWN} {JWST} {Archive}},
	url = {http://arxiv.org/abs/2505.21622},
	doi = {10.1051/0004-6361/202555504},
	urldate = {2025-08-07},
	journal = {A\&A},
	author = {Genin, Aurélien and Shuntov, Marko and Brammer, Gabe and Allen, Natalie and Ito, Kei and Magdis, Georgios and Matharu, Jasleen and Oesch, Pascal A. and Toft, Sune and Valentino, Francesco},
	month = jul,
	year = {2025},
	pages = {A343},
}

@article{conroy_propagation_2010-1,
	title = {The {Propagation} of {Uncertainties} in {Stellar} {Population} {Synthesis} {Modeling}. {III}. {Model} {Calibration}, {Comparison}, and {Evaluation}},
	volume = {712},
	issn = {0004-637X},
	url = {https://ui.adsabs.harvard.edu/abs/2010ApJ...712..833C},
	doi = {10.1088/0004-637X/712/2/833},
	urldate = {2025-06-22},
	journal = {ApJ},
	publisher = {IOP},
	author = {Conroy, Charlie and Gunn, James E.},
	month = apr,
	year = {2010},
	pages = {833--857},
}

@article{conroy_propagation_2009,
	title = {The {Propagation} of {Uncertainties} in {Stellar} {Population} {Synthesis} {Modeling}. {I}. {The} {Relevance} of {Uncertain} {Aspects} of {Stellar} {Evolution} and the {Initial} {Mass} {Function} to the {Derived} {Physical} {Properties} of {Galaxies}},
	volume = {699},
	issn = {0004-637X},
	url = {https://ui.adsabs.harvard.edu/abs/2009ApJ...699..486C},
	doi = {10.1088/0004-637X/699/1/486},
	urldate = {2025-06-22},
	journal = {ApJ},
	publisher = {IOP},
	author = {Conroy, Charlie and Gunn, James E. and White, Martin},
	month = jul,
	year = {2009},
	pages = {486--506},
}

@article{rieke_smiles_2024,
	title = {{SMILES}: {A} {Prototype} {JWST} {Multiband} {Mid}-infrared {Survey}},
	volume = {975},
	issn = {0004-637X},
	shorttitle = {{SMILES}},
	url = {https://ui.adsabs.harvard.edu/abs/2024ApJ...975...83R},
	doi = {10.3847/1538-4357/ad6cd2},
	urldate = {2025-06-18},
	journal = {ApJ},
	publisher = {IOP},
	author = {Rieke, G. H. and Alberts, Stacey and Shivaei, Irene and Lyu, Jianwei and Willmer, Christopher N. A. and Pérez-González, Pablo and Williams, Christina C.},
	month = nov,
	year = {2024},
	pages = {83},
}

@article{prochaska_pypeitpypeit_2020,
	title = {pypeit/{PypeIt}: {Release} 1.0.0},
	shorttitle = {pypeit/{PypeIt}},
	url = {https://ui.adsabs.harvard.edu/abs/2020zndo...3743493P},
	doi = {10.5281/zenodo.3743493},
	urldate = {2025-06-17},
	journal = {Zenodo},
	publisher = {Zenodo},
	author = {Prochaska, J. Xavier and Hennawi, Joseph and Cooke, Ryan and Westfall, Kyle and Wang, Feige and {EmAstro} and {Tiffanyhsyu} and Wasserman, Asher and Villaume, Alexa and {Marijana777} and Schindler, JT and Young, David and Simha, Sunil and Wilde, Matt and Tejos, Nicolas and Isbell, Jacob and Flörs, Andreas and Sandford, Nathan and Vasović, Zlatan and Betts, Edward and Holden, Brad},
	month = apr,
	year = {2020},
}

@article{prochaska_pypeit_2020,
	title = {{PypeIt}: {The} {Python} {Spectroscopic} {Data} {Reduction} {Pipeline}},
	volume = {5},
	shorttitle = {{PypeIt}},
	url = {https://ui.adsabs.harvard.edu/abs/2020JOSS....5.2308P},
	doi = {10.21105/joss.02308},
	urldate = {2025-06-17},
	journal = {The Journal of Open Source Software},
	author = {Prochaska, J. and Hennawi, Joseph and Westfall, Kyle and Cooke, Ryan and Wang, Feige and Hsyu, Tiffany and Davies, Frederick and Farina, Emanuele and Pelliccia, Debora},
	month = dec,
	year = {2020},
	pages = {2308},
}

@article{inami_muse_2017,
	title = {The {MUSE} {Hubble} {Ultra} {Deep} {Field} {Survey} - {II}. {Spectroscopic} redshifts and comparisons to color selections of high-redshift galaxies},
	volume = {608},
	copyright = {© ESO, 2017},
	issn = {0004-6361, 1432-0746},
	url = {https://www.aanda.org/articles/aa/abs/2017/12/aa31195-17/aa31195-17.html},
	doi = {10.1051/0004-6361/201731195},
	language = {en},
	urldate = {2025-06-17},
	journal = {A\&A},
	publisher = {EDP Sciences},
	author = {Inami, H. and Bacon, R. and Brinchmann, J. and Richard, J. and Contini, T. and Conseil, S. and Hamer, S. and Akhlaghi, M. and Bouché, N. and Clément, B. and Desprez, G. and Drake, A. B. and Hashimoto, T. and Leclercq, F. and Maseda, M. and Michel-Dansac, L. and Paalvast, M. and Tresse, L. and Ventou, E. and Kollatschny, W. and Boogaard, L. A. and Finley, H. and Marino, R. A. and Schaye, J. and Wisotzki, L.},
	month = dec,
	year = {2017},
	pages = {A2},
}

@article{bacon_muse_2017,
	title = {The {MUSE} {Hubble} {Ultra} {Deep} {Field} {Survey} - {I}. {Survey} description, data reduction, and source detection},
	volume = {608},
	copyright = {© ESO, 2017},
	issn = {0004-6361, 1432-0746},
	url = {https://www.aanda.org/articles/aa/abs/2017/12/aa30833-17/aa30833-17.html},
	doi = {10.1051/0004-6361/201730833},
	language = {en},
	urldate = {2025-06-17},
	journal = {A\&A},
	publisher = {EDP Sciences},
	author = {Bacon, Roland and Conseil, Simon and Mary, David and Brinchmann, Jarle and Shepherd, Martin and Akhlaghi, Mohammad and Weilbacher, Peter M. and Piqueras, Laure and Wisotzki, Lutz and Lagattuta, David and Epinat, Benoit and Guerou, Adrien and Inami, Hanae and Cantalupo, Sebastiano and Courbot, Jean Baptiste and Contini, Thierry and Richard, Johan and Maseda, Michael and Bouwens, Rychard and Bouché, Nicolas and Kollatschny, Wolfram and Schaye, Joop and Marino, Raffaella Anna and Pello, Roser and Herenz, Christian and Guiderdoni, Bruno and Carollo, Marcella},
	month = dec,
	year = {2017},
	pages = {A1},
}

@article{steglich_electronic_2010,
	title = {{ELECTRONIC} {SPECTROSCOPY} {OF} {MEDIUM}-{SIZED} {POLYCYCLIC} {AROMATIC} {HYDROCARBONS}: {IMPLICATIONS} {FOR} {THE} {CARRIERS} {OF} {THE} 2175 Å {UV} {BUMP}},
	volume = {712},
	issn = {2041-8205},
	shorttitle = {{ELECTRONIC} {SPECTROSCOPY} {OF} {MEDIUM}-{SIZED} {POLYCYCLIC} {AROMATIC} {HYDROCARBONS}},
	url = {https://dx.doi.org/10.1088/2041-8205/712/1/L16},
	doi = {10.1088/2041-8205/712/1/L16},
	language = {en},
	number = {1},
	urldate = {2025-06-17},
	journal = {ApJL},
	publisher = {The American Astronomical Society},
	author = {Steglich, M. and Jäger, C. and Rouillé, G. and Huisken, F. and Mutschke, H. and Henning, Th.},
	month = feb,
	year = {2010},
	pages = {L16},
}

@article{papoular_polycrystalline_2009,
	title = {A polycrystalline graphite model for the 2175 Å interstellar extinction band},
	volume = {394},
	issn = {0035-8711},
	url = {https://doi.org/10.1111/j.1365-2966.2009.14484.x},
	doi = {10.1111/j.1365-2966.2009.14484.x},
	number = {4},
	urldate = {2025-06-17},
	journal = {MNRAS},
	author = {Papoular, Robert J. and Papoular, Renaud},
	month = apr,
	year = {2009},
	pages = {2175--2181},
}

@article{li_infrared_2001,
	title = {Infrared {Emission} from {Interstellar} {Dust}. {II}. {The} {Diffuse} {Interstellar} {Medium}},
	volume = {554},
	issn = {0004-637X},
	url = {https://iopscience.iop.org/article/10.1086/323147/meta},
	doi = {10.1086/323147},
	language = {en},
	number = {2},
	urldate = {2025-06-17},
	journal = {ApJ},
	publisher = {IOP Publishing},
	author = {Li, Aigen and Draine, B. T.},
	month = jun,
	year = {2001},
	pages = {778},
}

@article{bradley_astronomical_2005,
	title = {An {Astronomical} 2175 Å {Feature} in {Interplanetary} {Dust} {Particles}},
	volume = {307},
	url = {https://www.science.org/doi/10.1126/science.1106717},
	doi = {10.1126/science.1106717},
	number = {5707},
	urldate = {2025-06-17},
	journal = {Science},
	publisher = {American Association for the Advancement of Science},
	author = {Bradley, John and Dai, Zu Rong and Erni, Rolf and Browning, Nigel and Graham, Giles and Weber, Peter and Smith, Julie and Hutcheon, Ian and Ishii, Hope and Bajt, Sasa and Floss, Christine and Stadermann, Frank and Sandford, Scott},
	month = jan,
	year = {2005},
	pages = {244--247},
}

@article{joblin_contribution_1992,
	title = {Contribution of {Polycyclic} {Aromatic} {Hydrocarbon} {Molecules} to the {Interstellar} {Extinction} {Curve}},
	volume = {393},
	issn = {0004-637X},
	url = {https://ui.adsabs.harvard.edu/abs/1992ApJ...393L..79J},
	doi = {10.1086/186456},
	urldate = {2025-06-17},
	journal = {ApJ},
	publisher = {IOP},
	author = {Joblin, C. and Leger, A. and Martin, P.},
	month = jul,
	year = {1992},
	pages = {L79},
}

@article{stecher_graphite_1965,
	title = {On {Graphite} and {Interstellar} {Extinction}},
	volume = {142},
	issn = {0004-637X},
	url = {https://ui.adsabs.harvard.edu/abs/1965ApJ...142.1681S},
	doi = {10.1086/148461},
	urldate = {2025-06-17},
	journal = {ApJ},
	publisher = {IOP},
	author = {Stecher, T. P. and Donn, B.},
	month = nov,
	year = {1965},
	pages = {1681},
}

@article{gardner_james_2023,
	title = {The {James} {Webb} {Space} {Telescope} {Mission}},
	volume = {135},
	issn = {0004-6280},
	url = {https://ui.adsabs.harvard.edu/abs/2023PASP..135f8001G},
	doi = {10.1088/1538-3873/acd1b5},
	urldate = {2025-06-17},
	journal = {PASP},
	publisher = {IOP},
	author = {Gardner, Jonathan P. and Mather, John C. and Abbott, Randy and Abell, James S. and Abernathy, Mark and Abney, Faith E. and Abraham, John G. and Abraham, Roberto and Abul-Huda, Yasin M. and Acton, Scott and Adams, Cynthia K. and Adams, Evan and Adler, David S. and Adriaensen, Maarten and Aguilar, Jonathan Albert and Ahmed, Mansoor and Ahmed, Nasif S. and Ahmed, Tanjira and Albat, Rüdeger and Albert, Loïc and Alberts, Stacey and Aldridge, David and Allen, Mary Marsha and Allen, Shaune S. and Altenburg, Martin and Altunc, Serhat and Alvarez, Jose Lorenzo and Álvarez-Márquez, Javier and Alves de Oliveira, Catarina and Ambrose, Leslie L. and Anandakrishnan, Satya M. and Andersen, Gregory C. and Anderson, Harry James and Anderson, Jay and Anderson, Kristen and Anderson, Sara M. and Aprea, Julio and Archer, Benita J. and Arenberg, Jonathan W. and Argyriou, Ioannis and Arribas, Santiago and Artigau, Étienne and Arvai, Amanda Rose and Atcheson, Paul and Atkinson, Charles B. and Averbukh, Jesse and Aymergen, Cagatay and Bacinski, John J. and Baggett, Wayne E. and Bagnasco, Giorgio and Baker, Lynn L. and Balzano, Vicki Ann and Banks, Kimberly A. and Baran, David A. and Barker, Elizabeth A. and Barrett, Larry K. and Barringer, Bruce O. and Barto, Allison and Bast, William and Baudoz, Pierre and Baum, Stefi and Beatty, Thomas G. and Beaulieu, Mathilde and Bechtold, Kathryn and Beck, Tracy and Beddard, Megan M. and Beichman, Charles and Bellagama, Larry and Bely, Pierre and Berger, Timothy W. and Bergeron, Louis E. and Bernier, Antoine-Darveau and Bertch, Maria D. and Beskow, Charlotte and Betz, Laura E. and Biagetti, Carl P. and Birkmann, Stephan and Bjorklund, Kurt F. and Blackwood, James D. and Blazek, Ronald Paul and Blossfeld, Stephen and Bluth, Marcel and Boccaletti, Anthony and Boegner, Jr., Martin E. and Bohlin, Ralph C. and Boia, John Joseph and Böker, Torsten and Bonaventura, N. and Bond, Nicholas A. and Bosley, Kari Ann and Boucarut, Rene A. and Bouchet, Patrice and Bouwman, Jeroen and Bower, Gary and Bowers, Ariel S. and Bowers, Charles W. and Boyce, Leslye A. and Boyer, Christine T. and Boyer, Martha L. and Boyer, Michael and Boyer, Robert and Bradley, Larry D. and Brady, Gregory R. and Brandl, Bernhard R. and Brannen, Judith L. and Breda, David and Bremmer, Harold G. and Brennan, David and Bresnahan, Pamela A. and Bright, Stacey N. and Broiles, Brian J. and Bromenschenkel, Asa and Brooks, Brian H. and Brooks, Keira J. and Brown, Bob and Brown, Bruce and Brown, Thomas M. and Bruce, Barry W. and Bryson, Jonathan G. and Bujanda, Edwin D. and Bullock, Blake M. and Bunker, A. J. and Bureo, Rafael and Burt, Irving J. and Bush, James Aaron and Bushouse, Howard A. and Bussman, Marie C. and Cabaud, Olivier and Cale, Steven and Calhoon, Charles D. and Calvani, Humberto and Canipe, Alicia M. and Caputo, Francis M. and Cara, Mihai and Carey, Larkin and Case, Michael Eli and Cesari, Thaddeus and Cetorelli, Lee D. and Chance, Don R. and Chandler, Lynn and Chaney, Dave and Chapman, George N. and Charlot, S. and Chayer, Pierre and Cheezum, Jeffrey I. and Chen, Bin and Chen, Christine H. and Cherinka, Brian and Chichester, Sarah C. and Chilton, Zachary S. and Chittiraibalan, Dharini and Clampin, Mark and Clark, Charles R. and Clark, Kerry W. and Clark, Stephanie M. and Claybrooks, Edward E. and Cleveland, Keith A. and Cohen, Andrew L. and Cohen, Lester M. and Colón, Knicole D. and Coleman, Benee L. and Colina, Luis and Comber, Brian J. and Comeau, Thomas M. and Comer, Thomas and Conde Reis, Alain and Connolly, Dennis C. and Conroy, Kyle E. and Contos, Adam R. and Contreras, James and Cook, Neil J. and Cooper, James L. and Cooper, Rachel Aviva and Correia, Michael F. and Correnti, Matteo and Cossou, Christophe and Costanza, Brian F. and Coulais, Alain and Cox, Colin R. and Coyle, Ray T. and Cracraft, Misty M. and Crew, Keith A. and Curtis, Gary J. and Cusveller, Bianca and Da Costa Maciel, Cleyciane and Dailey, Christopher T. and Daugeron, Frédéric and Davidson, Greg S. and Davies, James E. and Davis, Katherine Anne and Davis, Michael S. and Day, Ratna and de Chambure, Daniel and de Jong, Pauline and De Marchi, Guido and Dean, Bruce H. and Decker, John E. and Delisa, Amy S. and Dell, Lawrence C. and Dellagatta, Gail},
	month = jun,
	year = {2023},
	pages = {068001},
}

@article{engelbracht_metallicity_2005,
	title = {Metallicity {Effects} on {Mid}-{Infrared} {Colors} and the 8 μm {PAH} {Emission} in {Galaxies}},
	volume = {628},
	issn = {0004-637X},
	url = {https://ui.adsabs.harvard.edu/abs/2005ApJ...628L..29E},
	doi = {10.1086/432613},
	urldate = {2025-06-17},
	journal = {ApJ},
	publisher = {IOP},
	author = {Engelbracht, C. W. and Gordon, K. D. and Rieke, G. H. and Werner, M. W. and Dale, D. A. and Latter, W. B.},
	month = jul,
	year = {2005},
	pages = {L29--L32},
}

@article{madden_ism_2006,
	title = {{ISM} properties in low-metallicity environments},
	volume = {446},
	issn = {0004-6361},
	url = {https://ui.adsabs.harvard.edu/abs/2006A&A...446..877M},
	doi = {10.1051/0004-6361:20053890},
	urldate = {2025-06-17},
	journal = {A\&A},
	author = {Madden, S. C. and Galliano, F. and Jones, A. P. and Sauvage, M.},
	month = feb,
	year = {2006},
	pages = {877--896},
}

@article{whitcomb_metallicity_2024,
	title = {The {Metallicity} {Dependence} of {PAH} {Emission} in {Galaxies}. {I}. {Insights} from {Deep} {Radial} {Spitzer} {Spectroscopy}},
	volume = {974},
	issn = {0004-637X},
	url = {https://ui.adsabs.harvard.edu/abs/2024ApJ...974...20W},
	doi = {10.3847/1538-4357/ad66c8},
	urldate = {2025-06-17},
	journal = {ApJ},
	publisher = {IOP},
	author = {Whitcomb, Cory M. and Smith, J. -D. T. and Sandstrom, Karin and Starkey, Carl A. and Donnelly, Grant P. and Draine, Bruce T. and Skillman, Evan D. and Dale, Daniel A. and Armus, Lee and Hensley, Brandon S. and Lai, Thomas S. -Y. and Kennicutt, Robert C.},
	month = oct,
	year = {2024},
	pages = {20},
}

@article{draine_interstellar_2003,
	title = {Interstellar {Dust} {Grains}},
	volume = {41},
	issn = {0066-4146},
	url = {https://ui.adsabs.harvard.edu/abs/2003ARA&A..41..241D},
	doi = {10.1146/annurev.astro.41.011802.094840},
	urldate = {2025-06-12},
	journal = {ARA\&A},
	author = {Draine, B. T.},
	month = jan,
	year = {2003},
	pages = {241--289},
}

@article{li_spitzers_2020,
	title = {Spitzer's perspective of polycyclic aromatic hydrocarbons in galaxies},
	volume = {4},
	issn = {2397-3366},
	url = {https://ui.adsabs.harvard.edu/abs/2020NatAs...4..339L},
	doi = {10.1038/s41550-020-1051-1},
	urldate = {2025-05-28},
	journal = {Nature Astronomy},
	author = {Li, Aigen},
	month = mar,
	year = {2020},
	pages = {339--351},
}

@article{maheson_big_2025,
	title = {Big, {Dusty} {Galaxies} in {Blue} {Jay}: {Insights} into the {Relationship} {Between} {Morphology} and {Dust} {Attenuation} at {Cosmic} {Noon}},
	shorttitle = {Big, {Dusty} {Galaxies} in {Blue} {Jay}},
	url = {http://arxiv.org/abs/2504.15346},
	doi = {10.48550/arXiv.2504.15346},
	language = {en},
	urldate = {2025-04-23},
	publisher = {arXiv},
	author = {Maheson, Gabriel and Tacchella, Sandro and Belli, Sirio and Park, Minjung and Danhaive, A. Lola and Bugiani, Letizia and Davies, Rebecca and Emami, Razieh and Khoram, Amir H. and Lam, Laurence and Leja, Joel and Mendel, Trevor and Nelson, Erica June},
	month = apr,
	year = {2025},
	note = {arXiv:2504.15346 [astro-ph]},
}

@article{shivaei_mosdef_2017,
	title = {The {MOSDEF} {Survey}: {Metallicity} {Dependence} of the {PAH} {Emission} at {High} {Redshift} and {Implications} for 24 micron-inferred {IR} {Luminosities} and {Star} {Formation} {Rates} at z{\textasciitilde}2},
	volume = {837},
	issn = {1538-4357},
	shorttitle = {The {MOSDEF} {Survey}},
	url = {http://arxiv.org/abs/1609.04814},
	doi = {10.3847/1538-4357/aa619c},
	language = {en},
	number = {2},
	urldate = {2025-02-21},
	journal = {ApJ},
	author = {Shivaei, Irene and Reddy, Naveen and Shapley, Alice and Siana, Brian and Kriek, Mariska and Mobasher, Bahram and Coil, Alison and Freeman, William and Sanders, Ryan and Price, Sedona and Azadi, Mojegan and Zick, Tom},
	month = mar,
	year = {2017},
	pages = {157},
}

@article{shivaei_new_2024,
	title = {A new census of dust and polycyclic aromatic hydrocarbons at z=0.7-2 with {JWST} {MIRI}},
	volume = {690},
	issn = {0004-6361, 1432-0746},
	url = {http://arxiv.org/abs/2402.07989},
	doi = {10.1051/0004-6361/202449579},
	language = {en},
	urldate = {2025-02-21},
	journal = {A\&A},
	author = {Shivaei, Irene and Alberts, Stacey and Florian, Michael and Rieke, George and Wuyts, Stijn and Bodansky, Sarah and Bunker, Andrew J. and Cameron, Alex J. and Curti, Mirko and D'Eugenio, Francesco and Dudzeviciute, Ugne and Kramarenko, Ivan and Ji, Zhiyuan and Johnson, Benjamin D. and Lyu, Jianwei and Matthee, Jorryt and Morrison, Jane and Naidu, Rohan and Reddy, Naveen and Robertson, Brant and Pérez-González, Pablo G. and Sun, Yang and Tacchella, Sandro and Whitaker, Katherine and Williams, Christina C. and Willmer, Christopher N. A. and Witstok, Joris and Xiao, Mengyuan and Zhu, Yongda},
	month = oct,
	year = {2024},
	pages = {A89},
}

@article{draine_dust_2007,
	title = {Dust {Masses}, {PAH} {Abundances}, and {Starlight} {Intensities} in the {SINGS} {Galaxy} {Sample}},
	volume = {663},
	issn = {0004-637X},
	url = {https://iopscience.iop.org/article/10.1086/518306/meta},
	doi = {10.1086/518306},
	language = {en},
	number = {2},
	urldate = {2025-04-04},
	journal = {ApJ},
	publisher = {IOP Publishing},
	author = {Draine, B. T. and Dale, D. A. and Bendo, G. and Gordon, K. D. and Smith, J. D. T. and Armus, L. and Engelbracht, C. W. and Helou, G. and R. C. Kennicutt, Jr and Li, A. and Roussel, H. and Walter, F. and Calzetti, D. and Moustakas, J. and Murphy, E. J. and Rieke, G. H. and Bot, C. and Hollenbach, D. J. and Sheth, K. and Teplitz, H. I.},
	month = jul,
	year = {2007},
	pages = {866},
}

@article{johnson_stellar_2021,
	title = {Stellar {Population} {Inference} with {Prospector}},
	volume = {254},
	issn = {0067-0049, 1538-4365},
	url = {http://arxiv.org/abs/2012.01426},
	doi = {10.3847/1538-4365/abef67},
	number = {2},
	urldate = {2025-04-04},
	journal = {ApJS},
	author = {Johnson, Benjamin D. and Leja, Joel and Conroy, Charlie and Speagle, Joshua S.},
	month = jun,
	year = {2021},
	pages = {22},
}

@article{leja_deriving_2017,
	title = {Deriving {Physical} {Properties} from {Broadband} {Photometry} with {Prospector}: {Description} of the {Model} and a {Demonstration} of its {Accuracy} {Using} 129 {Galaxies} in the {Local} {Universe}},
	volume = {837},
	issn = {0004-637X},
	shorttitle = {Deriving {Physical} {Properties} from {Broadband} {Photometry} with {Prospector}},
	url = {https://ui.adsabs.harvard.edu/abs/2017ApJ...837..170L},
	doi = {10.3847/1538-4357/aa5ffe},
	urldate = {2025-04-04},
	journal = {ApJ},
	publisher = {IOP},
	author = {Leja, Joel and Johnson, Benjamin D. and Conroy, Charlie and van Dokkum, Pieter G. and Byler, Nell},
	month = mar,
	year = {2017},
	pages = {170},
}

@article{galliano_interstellar_2018,
	title = {The {Interstellar} {Dust} {Properties} of {Nearby} {Galaxies}},
	volume = {56},
	issn = {0066-4146},
	url = {https://ui.adsabs.harvard.edu/abs/2018ARA&A..56..673G},
	doi = {10.1146/annurev-astro-081817-051900},
	urldate = {2025-03-28},
	journal = {ARA\&A},
	author = {Galliano, Frédéric and Galametz, Maud and Jones, Anthony P.},
	month = sep,
	year = {2018},
	pages = {673--713},
}

@article{tielens_interstellar_2008,
	title = {Interstellar polycyclic aromatic hydrocarbon molecules.},
	volume = {46},
	issn = {0066-4146},
	url = {https://ui.adsabs.harvard.edu/abs/2008ARA&A..46..289T},
	doi = {10.1146/annurev.astro.46.060407.145211},
	urldate = {2025-03-14},
	journal = {ARA\&A},
	author = {Tielens, A. G. G. M.},
	month = sep,
	year = {2008},
	pages = {289--337},
}

@article{shivaei_mosdef_2020,
	title = {The {MOSDEF} {Survey}: {The} {Variation} of the {Dust} {Attenuation} {Curve} with {Metallicity}},
	volume = {899},
	issn = {0004-637X},
	shorttitle = {The {MOSDEF} {Survey}},
	url = {https://ui.adsabs.harvard.edu/abs/2020ApJ...899..117S},
	doi = {10.3847/1538-4357/aba35e},
	urldate = {2025-02-21},
	journal = {ApJ},
	publisher = {IOP},
	author = {Shivaei, Irene and Reddy, Naveen and Rieke, George and Shapley, Alice and Kriek, Mariska and Battisti, Andrew and Mobasher, Bahram and Sanders, Ryan and Fetherolf, Tara and Azadi, Mojegan and Coil, Alison L. and Freeman, William R. and de Groot, Laura and Leung, Gene and Price, Sedona H. and Siana, Brian and Zick, Tom},
	month = aug,
	year = {2020},
	pages = {117},
}

@article{salim_dust_2020,
	title = {The {Dust} {Attenuation} {Law} in {Galaxies}},
	volume = {58},
	issn = {0066-4146, 1545-4282},
	url = {https://www.annualreviews.org/doi/10.1146/annurev-astro-032620-021933},
	doi = {10.1146/annurev-astro-032620-021933},
	language = {en},
	number = {1},
	urldate = {2025-02-21},
	journal = {ARA\&A},
	author = {Salim, Samir and Narayanan, Desika},
	month = aug,
	year = {2020},
	pages = {529--575},
}

@article{shivaei_uv_2022,
	title = {The {UV} {2175Å} {Attenuation} {Bump} and its {Correlation} with {PAH} {Emission} at z{\textasciitilde}2},
	volume = {514},
	issn = {0035-8711, 1365-2966},
	url = {http://arxiv.org/abs/2203.09153},
	doi = {10.1093/mnras/stac1313},
	language = {en},
	number = {2},
	urldate = {2025-02-21},
	journal = {MNRAS},
	author = {Shivaei, Irene and Boogaard, Leindert and Díaz-Santos, Tanio and Battisti, Andrew and Brinchmann, Jarle and Cunha, Elisabete da and Maseda, Michael and Matthee, Jorryt and Monreal-Ibero, Ana and Nanayakkara, Themiya and Popping, Gergö and Vidal-García, Alba and Weilbacher, Peter M.},
	month = jun,
	year = {2022},
	pages = {1886--1894},
}

@article{schneider_formation_2024,
	title = {The formation and cosmic evolution of dust in the early {Universe}: {I}. {Dust} sources},
	volume = {32},
	issn = {0935-4956},
	shorttitle = {The formation and cosmic evolution of dust in the early {Universe}},
	url = {https://ui.adsabs.harvard.edu/abs/2024A&ARv..32....2S/abstract},
	doi = {10.1007/s00159-024-00151-2},
	language = {en},
	number = {1},
	urldate = {2025-11-08},
	journal = {A\&ARv},
	author = {Schneider, Raffaella and Maiolino, Roberto},
	month = apr,
	year = {2024},
	pages = {2},
}

@article{witstok_carbonaceous_2023,
	title = {Carbonaceous dust grains seen in the first billion years of cosmic time},
	volume = {621},
	copyright = {2023 The Author(s)},
	issn = {1476-4687},
	url = {https://www.nature.com/articles/s41586-023-06413-w},
	doi = {10.1038/s41586-023-06413-w},
	language = {en},
	number = {7978},
	urldate = {2025-11-10},
	journal = {Nature},
	publisher = {Nature Publishing Group},
	author = {Witstok, Joris and Shivaei, Irene and Smit, Renske and Maiolino, Roberto and Carniani, Stefano and Curtis-Lake, Emma and Ferruit, Pierre and Arribas, Santiago and Bunker, Andrew J. and Cameron, Alex J. and Charlot, Stephane and Chevallard, Jacopo and Curti, Mirko and de Graaff, Anna and D’Eugenio, Francesco and Giardino, Giovanna and Looser, Tobias J. and Rawle, Tim and Rodríguez del Pino, Bruno and Willott, Chris and Alberts, Stacey and Baker, William M. and Boyett, Kristan and Egami, Eiichi and Eisenstein, Daniel J. and Endsley, Ryan and Hainline, Kevin N. and Ji, Zhiyuan and Johnson, Benjamin D. and Kumari, Nimisha and Lyu, Jianwei and Nelson, Erica and Perna, Michele and Rieke, Marcia and Robertson, Brant E. and Sandles, Lester and Saxena, Aayush and Scholtz, Jan and Sun, Fengwu and Tacchella, Sandro and Williams, Christina C. and Willmer, Christopher N. A.},
	month = sep,
	year = {2023},
	pages = {267--270},
}

@article{markov_dust_2023,
	title = {Dust attenuation law in {JWST} galaxies at z ∼ 7–8},
	volume = {679},
	copyright = {© The Authors 2023},
	issn = {0004-6361, 1432-0746},
	url = {https://www.aanda.org/articles/aa/abs/2023/11/aa46723-23/aa46723-23.html},
	doi = {10.1051/0004-6361/202346723},
	language = {en},
	urldate = {2025-11-10},
	journal = {A\&A},
	publisher = {EDP Sciences},
	author = {Markov, V. and Gallerani, S. and Pallottini, A. and Sommovigo, L. and Carniani, S. and Ferrara, A. and Parlanti, E. and Mascia, F. Di},
	month = nov,
	year = {2023},
	pages = {A12},
}

@article{markov_evolution_2025,
	title = {The evolution of dust attenuation in z ≈ 2-12 galaxies observed by {JWST}},
	volume = {9},
	issn = {2397-3366},
	url = {https://ui.adsabs.harvard.edu/abs/2025NatAs...9..458M/abstract},
	doi = {10.1038/s41550-024-02426-1},
	language = {en},
	urldate = {2025-11-10},
	journal = {Nature Astronomy},
	author = {Markov, Vladan and Gallerani, Simona and Ferrara, Andrea and Pallottini, Andrea and Parlanti, Eleonora and Mascia, Fabio Di and Sommovigo, Laura and Kohandel, Mahsa},
	month = mar,
	year = {2025},
	pages = {458--468},
}

@article{langan_alma_2024,
	title = {{ALMA} reveals a dust-obscured galaxy merger at cosmic noon},
	volume = {689},
	copyright = {https://creativecommons.org/licenses/by/4.0},
	issn = {0004-6361, 1432-0746},
	url = {https://www.aanda.org/10.1051/0004-6361/202449626},
	doi = {10.1051/0004-6361/202449626},
	language = {en},
	urldate = {2025-11-18},
	journal = {A\&A},
	author = {Langan, I. and Popping, G. and Ginolfi, M. and Gentile, F. and Valentino, F. and Kaasinen, M.},
	month = sep,
	year = {2024},
	pages = {A283},
}

@article{noll_gmass_2009,
	title = {{GMASS} ultradeep spectroscopy of galaxies at z {\textasciitilde} 2. {IV}. {The} variety of dust populations},
	volume = {499},
	issn = {0004-6361},
	url = {https://ui.adsabs.harvard.edu/abs/2009A%26A...499...69N/abstract},
	doi = {10.1051/0004-6361/200811526},
	language = {en},
	number = {1},
	urldate = {2025-12-02},
	journal = {A\&A},
	author = {Noll, S. and Pierini, D. and Cimatti, A. and Daddi, E. and Kurk, J. D. and Bolzonella, M. and Cassata, P. and Halliday, C. and Mignoli, M. and Pozzetti, L. and Renzini, A. and Berta, S. and Dickinson, M. and Franceschini, A. and Rodighiero, G. and Rosati, P. and Zamorani, G.},
	month = may,
	year = {2009},
	pages = {69--85},
}

@article{kashino_2175_2021,
	title = {The 2175 Å dust feature in star-forming galaxies at \$1.3{\textbackslash}le z{\textbackslash}le 1.8\$: the dependence on stellar mass and specific star formation rate},
	volume = {909},
	issn = {0004-637X, 1538-4357},
	shorttitle = {The 2175 Å dust feature in star-forming galaxies at \$1.3{\textbackslash}le z{\textbackslash}le 1.8\$},
	url = {http://arxiv.org/abs/2009.13582},
	doi = {10.3847/1538-4357/abdf62},
	number = {2},
	urldate = {2025-12-03},
	journal = {ApJ},
	author = {Kashino, Daichi and Lilly, Simon J. and Silverman, John D. and Renzini, Alvio and Daddi, Emanuele and Bardelli, Sandro and Cucciati, Olga and Kartaltepe, Jeyhan S. and Mainieri, Vincenzo and Pelló, Roser and Peng, Ying-jie and Sanders, David B. and Zucca, Elena},
	month = mar,
	year = {2021},
	pages = {213},
}

@article{cardelli_relationship_1989,
	title = {The relationship between infrared, optical, and ultraviolet extinction},
	volume = {345},
	issn = {0004-637X, 1538-4357},
	url = {http://adsabs.harvard.edu/doi/10.1086/167900},
	doi = {10.1086/167900},
	language = {en},
	urldate = {2025-12-03},
	journal = {ApJ},
	author = {Cardelli, Jason A. and Clayton, Geoffrey C. and Mathis, John S.},
	month = oct,
	year = {1989},
	pages = {245},
}

@article{prochaska_pypeit_2020-1,
	title = {{PypeIt}: {The} {Python} {Spectroscopic} {Data} {Reduction} {Pipeline}},
	volume = {5},
	issn = {2475-9066},
	shorttitle = {{PypeIt}},
	url = {https://joss.theoj.org/papers/10.21105/joss.02308},
	doi = {10.21105/joss.02308},
	language = {en},
	number = {56},
	urldate = {2025-12-03},
	journal = {Journal of Open Source Software},
	author = {Prochaska, J. Xavier and Hennawi, Joseph F. and Westfall, Kyle B. and Cooke, Ryan J. and Wang, Feige and Hsyu, Tiffany and Davies, Frederick B. and Farina, Emanuele Paolo and Pelliccia, Debora},
	month = dec,
	year = {2020},
	pages = {2308},
}

@inproceedings{mclean_mosfire_2012,
	title = {{MOSFIRE}, the multi-object spectrometer for infra-red exploration at the {Keck} {Observatory}},
	volume = {8446},
	url = {https://www.spiedigitallibrary.org/conference-proceedings-of-spie/8446/84460J/MOSFIRE-the-multi-object-spectrometer-for-infra-red-exploration-at/10.1117/12.924794.full},
	doi = {10.1117/12.924794},
	urldate = {2025-12-03},
	booktitle = {Ground-based and {Airborne} {Instrumentation} for {Astronomy} {IV}},
	publisher = {SPIE},
	author = {McLean, Ian S. and Steidel, Charles C. and Epps, Harland W. and Konidaris, Nicholas and Matthews, Keith Y. and Adkins, Sean and Aliado, Theodore and Brims, George and Canfield, John M. and Cromer, John L. and Fucik, Jason and Kulas, Kristin and Mace, Greg and Magnone, Ken and Rodriguez, Hector and Rudie, Gwen and Trainor, Ryan and Wang, Eric and Weber, Bob and Weiss, Jason},
	month = sep,
	year = {2012},
	pages = {149--163},
}

@article{valentino_atlas_2023,
	title = {An {Atlas} of {Color}-selected {Quiescent} {Galaxies} at \$z{\textgreater}3\$ in {Public} \${JWST}\$ {Fields}},
	volume = {947},
	issn = {0004-637X, 1538-4357},
	url = {http://arxiv.org/abs/2302.10936},
	doi = {10.3847/1538-4357/acbefa},
	number = {1},
	urldate = {2025-12-03},
	journal = {ApJ},
	author = {Valentino, Francesco and Brammer, Gabriel and Gould, Katriona M. L. and Kokorev, Vasily and Fujimoto, Seiji and Jespersen, Christian Kragh and Vijayan, Aswin P. and Weaver, John R. and Ito, Kei and Tanaka, Masayuki and Ilbert, Olivier and Magdis, Georgios E. and Whitaker, Katherine E. and Faisst, Andreas L. and Gallazzi, Anna and Gillman, Steven and Gimenez-Arteaga, Clara and Gomez-Guijarro, Carlos and Kubo, Mariko and Heintz, Kasper E. and Hirschmann, Michaela and Oesch, Pascal and Onodera, Masato and Rizzo, Francesca and Lee, Minju and Strait, Victoria and Toft, Sune},
	month = apr,
	year = {2023},
	pages = {20},
}

@article{narayanan_theory_2018,
	title = {A {Theory} for the {Variation} of {Dust} {Attenuation} {Laws} in {Galaxies}},
	volume = {869},
	issn = {0004-637X},
	url = {https://doi.org/10.3847/1538-4357/aaed25},
	doi = {10.3847/1538-4357/aaed25},
	language = {en},
	number = {1},
	urldate = {2025-12-04},
	journal = {ApJ},
	publisher = {The American Astronomical Society},
	author = {Narayanan, Desika and Conroy, Charlie and Davé, Romeel and Johnson, Benjamin D. and Popping, Gergö},
	month = dec,
	year = {2018},
	pages = {70},
}

@article{reddy_mosdef_2020,
	title = {The {MOSDEF} {Survey}: {The} {First} {Direct} {Measurements} of the {Nebular} {Dust} {Attenuation} {Curve} at {High} {Redshift}*},
	volume = {902},
	issn = {0004-637X},
	shorttitle = {The {MOSDEF} {Survey}},
	url = {https://doi.org/10.3847/1538-4357/abb674},
	doi = {10.3847/1538-4357/abb674},
	language = {en},
	number = {2},
	urldate = {2025-12-04},
	journal = {ApJ},
	publisher = {The American Astronomical Society},
	author = {Reddy, Naveen A. and Shapley, Alice E. and Kriek, Mariska and Steidel, Charles C. and Shivaei, Irene and Sanders, Ryan L. and Mobasher, Bahram and Coil, Alison L. and Siana, Brian and Freeman, William R. and Azadi, Mojegan and Fetherolf, Tara and Leung, Gene and Price, Sedona H. and Zick, Tom},
	month = oct,
	year = {2020},
	pages = {123},
}

@article{calzetti_dust_2000,
	title = {The {Dust} {Content} and {Opacity} of {Actively} {Star}-{Forming} {Galaxies}},
	volume = {533},
	issn = {0004-637X, 1538-4357},
	url = {http://arxiv.org/abs/astro-ph/9911459},
	doi = {10.1086/308692},
	number = {2},
	urldate = {2025-12-04},
	journal = {ApJ},
	author = {Calzetti, D. and Armus, L. and Bohlin, R. C. and Kinney, A. L. and Koornneef, J. and Storchi-Bergmann, T.},
	month = apr,
	year = {2000},
	pages = {682--695},
}

@book{carroll_introduction_1996,
	title = {An {Introduction} to {Modern} {Astrophysics}},
	author = {Carroll, Bradley W. and Ostlie, Dale A.},
	year = {1996},
}

@book{osterbrock_astrophysics_1989,
	title = {Astrophysics of gaseous nebulae and active galactic nuclei},
	author = {Osterbrock, Donald E.},
	year = {1989},
}

@article{markov_unveiling_2025,
	title = {Unveiling the trends between dust attenuation and galaxy properties at z ∼ 2‑12 with the {James} {Webb} {Space} {Telescope}},
	volume = {702},
	issn = {0004-6361},
	url = {https://ui.adsabs.harvard.edu/abs/2025A%26A...702A..33M/abstract},
	doi = {10.1051/0004-6361/202555182},
	language = {en},
	urldate = {2025-12-05},
	journal = {A\&A},
	author = {Markov, V. and Gallerani, S. and Pallottini, A. and Bradač, M. and Carniani, S. and Tripodi, R. and Noirot, G. and Di Mascia, F. and Parlanti, E. and Martis, N.},
	month = oct,
	year = {2025},
	pages = {A33},
}

@article{peroux_cosmic_2020,
	title = {The {Cosmic} {Baryon} and {Metal} {Cycles}},
	volume = {58},
	issn = {0066-4146},
	url = {https://ui.adsabs.harvard.edu/abs/2020ARA&A..58..363P/abstract},
	doi = {10.1146/annurev-astro-021820-120014},
	language = {en},
	urldate = {2025-12-08},
	journal = {ARA\&A},
	author = {Péroux, Céline and Howk, J. Christopher},
	month = aug,
	year = {2020},
	pages = {363--406},
}

@article{gall_production_2011,
	title = {Production of dust by massive stars at high redshift},
	volume = {19},
	issn = {0935-4956},
	url = {https://ui.adsabs.harvard.edu/abs/2011A&ARv..19...43G/abstract},
	doi = {10.1007/s00159-011-0043-7},
	language = {en},
	urldate = {2025-12-08},
	journal = {A\&ARv},
	author = {Gall, C. and Hjorth, J. and Andersen, A. C.},
	month = sep,
	year = {2011},
	pages = {43},
}

@article{micelotta_dust_2018,
	title = {Dust in {Supernovae} and {Supernova} {Remnants} {II}: {Processing} and {Survival}},
	volume = {214},
	issn = {0038-6308},
	shorttitle = {Dust in {Supernovae} and {Supernova} {Remnants} {II}},
	url = {https://ui.adsabs.harvard.edu/abs/2018SSRv..214...53M/abstract},
	doi = {10.1007/s11214-018-0484-7},
	language = {en},
	number = {2},
	urldate = {2025-12-08},
	journal = {Space Science Reviews},
	author = {Micelotta, E. R. and Matsuura, M. and Sarangi, A.},
	month = mar,
	year = {2018},
	pages = {53},
}

@article{shivaei_diversity_2025,
	title = {The {Diversity} and {Evolution} of {Dust} {Attenuation} {Curves} from {Redshift} z {\textasciitilde} 1 to 9},
	url = {https://ui.adsabs.harvard.edu/abs/2025arXiv250901795S/abstract},
	doi = {10.48550/arXiv.2509.01795},
	language = {en},
	urldate = {2025-12-08},
	publisher = {arXiv},
	author = {Shivaei, Irene and Naidu, Rohan P. and Rodríguez Montero, Francisco and Matsumoto, Kosei and Leja, Joel and Matthee, Jorryt and Johnson, Benjamin D. and Oesch, Pascal A. and Chevallard, Jacopo and Adamo, Angela and Bodansky, Sarah and Bunker, Andrew J. and Covelo Paz, Alba and Di Cesare, Claudia and Egami, Eiichi and Furtak, Lukas J. and Heintz, Kasper E. and Kramarenko, Ivan and Meyer, Romain A. and Reddy, Naveen A. and Rinaldi, Pierluigi and Tacchella, Sandro and Torralba, Alberto and Witstok, Joris and Wozniak, Michael A. and Xiao, Mengyuan},
	month = sep,
	year = {2025},
	note = {arXiv:2509.01795},
}

@article{chastenet_resolved_2024,
	title = {The {Resolved} {Behavior} of {Dust} {Mass}, {Polycyclic} {Aromatic} {Hydrocarbon} {Fraction}, and {Radiation} {Field} in ∼800 {Nearby} {Galaxies}},
	volume = {276},
	issn = {0067-0049},
	url = {https://doi.org/10.3847/1538-4365/ad8a5c},
	doi = {10.3847/1538-4365/ad8a5c},
	language = {en},
	number = {1},
	urldate = {2025-12-18},
	journal = {ApJS},
	publisher = {The American Astronomical Society},
	author = {Chastenet, Jérémy and Sandstrom, Karin and Leroy, Adam K. and Bot, Caroline and Chiang, I-Da and Chown, Ryan and Gordon, Karl D. and Koch, Eric W. and Roussel, Hélène and Sutter, Jessica and Williams, Thomas G.},
	month = dec,
	year = {2024},
	pages = {2},
}

@article{wild_empirical_2011,
	title = {Empirical determination of the shape of dust attenuation curves in star-forming galaxies},
	volume = {417},
	issn = {0035-8711},
	url = {https://doi.org/10.1111/j.1365-2966.2011.19367.x},
	doi = {10.1111/j.1365-2966.2011.19367.x},
	number = {3},
	urldate = {2025-12-19},
	journal = {MNRAS},
	author = {Wild, Vivienne and Charlot, Stéphane and Brinchmann, Jarle and Heckman, Timothy and Vince, Oliver and Pacifici, Camilla and Chevallard, Jacopo},
	month = nov,
	year = {2011},
	pages = {1760--1786},
}

@article{draine_graphite_1993,
	title = {On {Graphite} and the 2175 {Angstrom} {Extinction} {Profile}},
	volume = {414},
	issn = {0004-637X},
	url = {https://ui.adsabs.harvard.edu/abs/1993ApJ...414..632D/abstract},
	doi = {10.1086/173109},
	language = {en},
	urldate = {2025-12-19},
	journal = {ApJ},
	author = {Draine, B. T. and Malhotra, Sangeeta},
	month = sep,
	year = {1993},
	pages = {632},
}

@article{mennella_new_1998,
	title = {A {New} {Approach} to the {Puzzle} of the {Ultraviolet} {Interstellar} {Extinction} {Bump}},
	volume = {507},
	issn = {0004-637X},
	url = {https://ui.adsabs.harvard.edu/abs/1998ApJ...507L.177M/abstract},
	doi = {10.1086/311693},
	language = {en},
	number = {2},
	urldate = {2025-12-19},
	journal = {ApJ},
	author = {Mennella, V. and Colangeli, L. and Bussoletti, E. and Palumbo, P. and Rotundi, A.},
	month = nov,
	year = {1998},
	pages = {L177},
}

@article{wada_carbonaceous_1999,
	title = {Carbonaceous onion-like particles as a component of interstellar dust},
	volume = {345},
	issn = {0004-6361},
	url = {https://ui.adsabs.harvard.edu/abs/1999A%26A...345..259W/abstract},
	language = {en},
	urldate = {2025-12-19},
	journal = {A\&A},
	author = {Wada, Setsuko and Kaito, Chihiro and Kimura, Seiji and Ono, Hiroshi and Tokunaga, Alan T.},
	month = may,
	year = {1999},
	pages = {259},
}

@article{marchenko_search_2017,
	title = {Search for polycyclic aromatic hydrocarbons in the outflows from dust-producing {Wolf}-{Rayet} stars},
	volume = {468},
	issn = {0035-8711},
	url = {https://ui.adsabs.harvard.edu/abs/2017MNRAS.468.2416M/abstract},
	doi = {10.1093/mnras/stx563},
	language = {en},
	number = {2},
	urldate = {2026-01-09},
	journal = {MNRAS},
	author = {Marchenko, Sergey V. and Moffat, A. F. J.},
	month = jun,
	year = {2017},
	pages = {2416},
}

@article{rose_investigating_2025,
	title = {Investigating the {Electron}-capture {Supernova} {Candidate} {AT} 2019abn with {JWST} {Spectroscopy}},
	volume = {980},
	issn = {0004-637X},
	url = {https://ui.adsabs.harvard.edu/abs/2025ApJ...980L..14R/abstract},
	doi = {10.3847/2041-8213/adad61},
	language = {en},
	number = {1},
	urldate = {2026-01-09},
	journal = {ApJL},
	author = {Rose, Sam and Lau, Ryan M. and Jencson, Jacob E. and Kasliwal, Mansi M. and De, Kishalay and Ressler, Michael E. and Fox, Ori D. and Hankins, Matthew J.},
	month = feb,
	year = {2025},
	pages = {L14},
}

@article{calzetti_polycyclic_2011,
	title = {Polycyclic {Aromatic} {Hydrocarbons} as {Star} {Formation} {Rate} {Indicators}},
	volume = {46},
	issn = {1633-4760},
	url = {https://ui.adsabs.harvard.edu/abs/2011EAS....46..133C/abstract},
	doi = {10.1051/eas/1146014},
	language = {en},
	urldate = {2026-01-09},
	journal = {EAS Publications Series, Volume 46, 2011, pp.133-141},
	author = {Calzetti, D.},
	month = mar,
	year = {2011},
	pages = {133},
}

@article{galliano_stellar_2008,
	title = {Stellar {Evolutionary} {Effects} on the {Abundances} of {PAH} and {SN}-{Condensed} {Dust} in {Galaxies}},
	volume = {672},
	issn = {0004-637X, 1538-4357},
	url = {http://arxiv.org/abs/0708.0790},
	doi = {10.1086/523621},
	number = {1},
	urldate = {2026-01-09},
	journal = {ApJ},
	author = {Galliano, F. and Dwek, E. and Chanial, P.},
	month = jan,
	year = {2008},
	pages = {214--243},
}

@article{fitzpatrick_analysis_2007,
	title = {An {Analysis} of the {Shapes} of {Interstellar} {Extinction} {Curves}. {V}. {The} {IR}-through-{UV} {Curve} {Morphology}},
	volume = {663},
	issn = {0004-637X},
	url = {https://iopscience.iop.org/article/10.1086/518158},
	doi = {10.1086/518158},
	language = {en},
	number = {1},
	urldate = {2026-01-21},
	journal = {ApJ},
	publisher = {IOP Publishing},
	author = {Fitzpatrick, E. L. and Massa, D.},
	month = jul,
	year = {2007},
	pages = {320},
}

@article{tremonti_origin_2004,
	title = {The {Origin} of the {Mass}-{Metallicity} {Relation}: {Insights} from 53,000 {Star}-forming {Galaxies} in the {Sloan} {Digital} {Sky} {Survey}},
	volume = {613},
	issn = {0004-637X},
	shorttitle = {The {Origin} of the {Mass}-{Metallicity} {Relation}},
	url = {https://iopscience.iop.org/article/10.1086/423264},
	doi = {10.1086/423264},
	language = {en},
	number = {2},
	urldate = {2026-01-26},
	journal = {ApJ},
	publisher = {IOP Publishing},
	author = {Tremonti, Christy A. and Heckman, Timothy M. and Kauffmann, Guinevere and Brinchmann, Jarle and Charlot, Stéphane and White, Simon D. M. and Seibert, Mark and Peng, Eric W. and Schlegel, David J. and Uomoto, Alan and Fukugita, Masataka and Brinkmann, Jon},
	month = oct,
	year = {2004},
	pages = {898},
}

@article{johnson_jwst_2026,
	title = {{JWST} {Advanced} {Deep} {Extragalactic} {Survey} ({JADES}) {Data} {Release} 5: {NIRCam} {Imaging} in {GOODS}-{S} and {GOODS}-{N}},
	shorttitle = {{JWST} {Advanced} {Deep} {Extragalactic} {Survey} ({JADES}) {Data} {Release} 5},
	url = {http://arxiv.org/abs/2601.15954},
	doi = {10.48550/arXiv.2601.15954},
	urldate = {2026-01-27},
	publisher = {arXiv},
	author = {Johnson, Benjamin D. and Robertson, Brant E. and Eisenstein, Daniel J. and Tacchella, Sandro and Puskás, Dávid and Duan, Qiao and Wu, Zihao and Hainline, Kevin and Rieke, Marcia and Willott, Chris and Willmer, Christopher N. A. and Trussler, James A. A. and Alberts, Stacey and Arribas, Santiago and Baker, William M. and Bunker, Andrew J. and Cameron, Alex J. and Carniani, Stefano and Carreira, Courtney and Cargile, Phillip A. and Curtis-Lake, Emma and Egami, Eiichi and Hausen, Ryan and Helton, Jakob M. and Ji, Zhiyuan and Maiolino, Roberto and Pérez-González, Pablo G. and Rinaldi, Pierluigi and Sun, Fengwu and Sun, Yang and Villanueva, Natalia C. and Williams, Christina C. and Zhu, Yongda},
	month = jan,
	year = {2026},
	note = {arXiv:2601.15954},
}

@article{witt_multiple_1996,
	title = {Multiple {Scattering} in {Clumpy} {Media}. {I}. {Escape} of {Stellar} {Radiation} from a {Clumpy} {Scattering} {Environment}},
	volume = {463},
	issn = {0004-637X},
	url = {https://ui.adsabs.harvard.edu/abs/1996ApJ...463..681W/abstract},
	doi = {10.1086/177282},
	language = {en},
	urldate = {2026-01-30},
	journal = {ApJ},
	author = {Witt, Adolf N. and Gordon, Karl D.},
	month = jun,
	year = {1996},
	pages = {681},
}

@article{hunt_spitzer_2010,
	title = {The {Spitzer} {View} of {Low}-{Metallicity} {Star} {Formation}. {III}. {Fine}-{Structure} {Lines}, {Aromatic} {Features}, and {Molecules}},
	volume = {712},
	issn = {0004-637X},
	url = {https://ui.adsabs.harvard.edu/abs/2010ApJ...712..164H/abstract},
	doi = {10.1088/0004-637X/712/1/164},
	language = {en},
	number = {1},
	urldate = {2026-02-03},
	journal = {ApJ},
	author = {Hunt, Leslie K. and Thuan, Trinh X. and Izotov, Yuri I. and Sauvage, Marc},
	month = mar,
	year = {2010},
	pages = {164},
}

@article{hunter_matplotlib_2007,
	title = {Matplotlib: {A} {2D} {Graphics} {Environment}},
	volume = {9},
	issn = {1558-366X},
	shorttitle = {Matplotlib},
	url = {https://ieeexplore.ieee.org/document/4160265},
	doi = {10.1109/MCSE.2007.55},
	number = {3},
	urldate = {2026-02-10},
	journal = {Computing in Science \& Engineering},
	author = {Hunter, John D.},
	month = may,
	year = {2007},
	pages = {90--95},
}

@article{harris_array_2020,
	title = {Array programming with {NumPy}},
	volume = {585},
	copyright = {2020 The Author(s)},
	issn = {1476-4687},
	url = {https://www.nature.com/articles/s41586-020-2649-2},
	doi = {10.1038/s41586-020-2649-2},
	language = {en},
	number = {7825},
	urldate = {2026-02-10},
	journal = {Nature},
	publisher = {Nature Publishing Group},
	author = {Harris, Charles R. and Millman, K. Jarrod and van der Walt, Stéfan J. and Gommers, Ralf and Virtanen, Pauli and Cournapeau, David and Wieser, Eric and Taylor, Julian and Berg, Sebastian and Smith, Nathaniel J. and Kern, Robert and Picus, Matti and Hoyer, Stephan and van Kerkwijk, Marten H. and Brett, Matthew and Haldane, Allan and del Río, Jaime Fernández and Wiebe, Mark and Peterson, Pearu and Gérard-Marchant, Pierre and Sheppard, Kevin and Reddy, Tyler and Weckesser, Warren and Abbasi, Hameer and Gohlke, Christoph and Oliphant, Travis E.},
	month = sep,
	year = {2020},
	pages = {357--362},
}

@article{rousselot_night-sky_2000,
	title = {Night-sky spectral atlas of {OH} emission lines in the near-infrared},
	volume = {354},
	issn = {0004-6361},
	url = {https://ui.adsabs.harvard.edu/abs/2000A&A...354.1134R/abstract},
	language = {en},
	urldate = {2026-02-17},
	journal = {A\&A},
	author = {Rousselot, P. and Lidman, C. and Cuby, J.-G. and Moreels, G. and Monnet, G.},
	month = feb,
	year = {2000},
	pages = {1134--1150},
}

@article{horne_optimal_1986,
	title = {An optimal extraction algorithm for {CCD} spectroscopy.},
	volume = {98},
	issn = {0004-6280},
	url = {https://ui.adsabs.harvard.edu/abs/1986PASP...98..609H/abstract},
	doi = {10.1086/131801},
	language = {en},
	urldate = {2026-03-06},
	journal = {PASP},
	author = {Horne, K.},
	month = jun,
	year = {1986},
	pages = {609},
}

@book{kuckartz_statistik_2013,
	address = {Wiesbaden},
	title = {Statistik: {Eine} verständliche {Einführung}},
	copyright = {https://www.springernature.com/gp/researchers/text-and-data-mining},
	isbn = {978-3-531-19889-7 978-3-531-19890-3},
	shorttitle = {Statistik},
	url = {https://link.springer.com/10.1007/978-3-531-19890-3},
	doi = {10.1007/978-3-531-19890-3},
	language = {de},
	urldate = {2026-03-10},
	publisher = {VS Verlag für Sozialwissenschaften},
	author = {Kuckartz, Udo and Rädiker, Stefan and Ebert, Thomas and Schehl, Julia},
	year = {2013},
}

@article{lyu_active_2024,
	title = {Active {Galactic} {Nuclei} {Selection} and {Demographics}: {A} {New} {Age} with {JWST}/{MIRI}},
	volume = {966},
	issn = {0004-637X},
	shorttitle = {Active {Galactic} {Nuclei} {Selection} and {Demographics}},
	url = {https://doi.org/10.3847/1538-4357/ad3643},
	doi = {10.3847/1538-4357/ad3643},
	language = {en},
	number = {2},
	urldate = {2026-03-11},
	journal = {ApJ},
	publisher = {The American Astronomical Society},
	author = {Lyu, Jianwei and Alberts, Stacey and Rieke, George H. and Shivaei, Irene and Pérez-González, Pablo G. and Sun, Fengwu and Hainline, Kevin N. and Baum, Stefi and Bonaventura, Nina and Bunker, Andrew J. and Egami, Eiichi and Eisenstein, Daniel J. and Florian, Michael and Ji, Zhiyuan and Johnson, Benjamin D. and Morrison, Jane and Rieke, Marcia and Robertson, Brant and Rujopakarn, Wiphu and Tacchella, Sandro and Scholtz, Jan and Willmer, Christopher N. A.},
	month = may,
	year = {2024},
	pages = {229},
}

@article{dotter_mesa_2016,
	title = {{MESA} {Isochrones} and {Stellar} {Tracks} ({MIST}) 0: {Methods} for the {Construction} of {Stellar} {Isochrones}},
	volume = {222},
	issn = {0067-0049},
	shorttitle = {{MESA} {Isochrones} and {Stellar} {Tracks} ({MIST}) 0},
	url = {https://ui.adsabs.harvard.edu/abs/2016ApJS..222....8D/abstract},
	doi = {10.3847/0067-0049/222/1/8},
	language = {en},
	number = {1},
	urldate = {2026-03-13},
	journal = {The Astrophysical Journal Supplement Series},
	author = {Dotter, Aaron},
	month = jan,
	year = {2016},
	pages = {8},
}

@article{falcon-barroso_updated_2011,
	title = {An updated {MILES} stellar library and stellar population models},
	volume = {532},
	issn = {0004-6361},
	url = {https://ui.adsabs.harvard.edu/abs/2011A&A...532A..95F/abstract},
	doi = {10.1051/0004-6361/201116842},
	language = {en},
	urldate = {2026-03-13},
	journal = {A\&A},
	author = {Falcón-Barroso, J. and Sánchez-Blázquez, P. and Vazdekis, A. and Ricciardelli, E. and Cardiel, N. and Cenarro, A. J. and Gorgas, J. and Peletier, R. F.},
	month = aug,
	year = {2011},
	pages = {A95},
}

@article{rogers_chaos_2021,
	title = {{CHAOS}. {VI}. {Direct} {Abundances} in {NGC} 2403},
	volume = {915},
	issn = {0004-637X, 1538-4357},
	url = {https://iopscience.iop.org/article/10.3847/1538-4357/abf8b9},
	doi = {10.3847/1538-4357/abf8b9},
	number = {1},
	urldate = {2026-03-19},
	journal = {ApJ},
	author = {Rogers, Noah S. J. and Skillman, Evan D. and Pogge, Richard W. and Berg, Danielle A. and Moustakas, John and Croxall, Kevin V. and Sun, Jiayi},
	month = jul,
	year = {2021},
	pages = {21},
}

@article{xie_new_2019,
	title = {A {New} {Calibration} of {Star} {Formation} {Rate} in {Galaxies} {Based} on {Polycyclic} {Aromatic} {Hydrocarbon} {Emission}},
	volume = {884},
	issn = {0004-637X, 1538-4357},
	url = {https://iopscience.iop.org/article/10.3847/1538-4357/ab4200},
	doi = {10.3847/1538-4357/ab4200},
	number = {2},
	urldate = {2026-03-19},
	journal = {ApJ},
	author = {Xie, Yanxia and Ho, Luis C.},
	month = oct,
	year = {2019},
	pages = {136},
}

@article{chastenet_polycyclic_2019,
	title = {The {Polycyclic} {Aromatic} {Hydrocarbon} {Mass} {Fraction} on a 10 pc {Scale} in the {Magellanic} {Clouds}},
	volume = {876},
	issn = {0004-637X, 1538-4357},
	url = {https://iopscience.iop.org/article/10.3847/1538-4357/ab16cf},
	doi = {10.3847/1538-4357/ab16cf},
	number = {1},
	urldate = {2026-03-20},
	journal = {ApJ},
	author = {Chastenet, Jérémy and Sandstrom, Karin and Chiang, I-Da and Leroy, Adam K. and Utomo, Dyas and Bot, Caroline and Gordon, Karl D. and Draine, Bruce T. and Fukui, Yasuo and Onishi, Toshikazu and Tsuge, Kisetsu},
	month = may,
	year = {2019},
	pages = {62},
}

@article{rigopoulou_properties_2021,
	title = {The properties of polycyclic aromatic hydrocarbons in galaxies: constraints on {PAH} sizes, charge and radiation fields},
	volume = {504},
	copyright = {https://academic.oup.com/journals/pages/open\_access/funder\_policies/chorus/standard\_publication\_model},
	issn = {0035-8711, 1365-2966},
	shorttitle = {The properties of polycyclic aromatic hydrocarbons in galaxies},
	url = {https://academic.oup.com/mnras/article/504/4/5287/6223446},
	doi = {10.1093/mnras/stab959},
	language = {en},
	number = {4},
	urldate = {2026-03-20},
	journal = {MNRAS},
	author = {Rigopoulou, D and Barale, M and Clary, D C and Shan, X and Alonso-Herrero, A and García-Bernete, I and Hunt, L and Kerkeni, B and Pereira-Santaella, M and Roche, P F},
	month = may,
	year = {2021},
	pages = {5287--5300},
}

@article{salgado_orion_2016,
	title = {{THE} {ORION} {H} ii {REGION} {AND} {THE} {ORION} {BAR} {IN} {THE} {MID}-{INFRARED}},
	volume = {830},
	issn = {0004-637X, 1538-4357},
	url = {https://iopscience.iop.org/article/10.3847/0004-637X/830/2/118},
	doi = {10.3847/0004-637X/830/2/118},
	number = {2},
	urldate = {2026-03-20},
	journal = {ApJ},
	author = {Salgado, F. and Berné, O. and Adams, J. D. and Herter, T. L. and Keller, L. D. and Tielens, A. G. G. M.},
	month = oct,
	year = {2016},
	pages = {118},
}

@article{sutter_fraction_2024,
	title = {The {Fraction} of {Dust} {Mass} in the {Form} of {Polycyclic} {Aromatic} {Hydrocarbons} on 10–50 pc {Scales} in {Nearby} {Galaxies}},
	volume = {971},
	issn = {0004-637X, 1538-4357},
	url = {https://iopscience.iop.org/article/10.3847/1538-4357/ad54bd},
	doi = {10.3847/1538-4357/ad54bd},
	number = {2},
	urldate = {2026-03-20},
	journal = {ApJ},
	author = {Sutter, Jessica and Sandstrom, Karin and Chastenet, Jérémy and Leroy, Adam K. and Koch, Eric W. and Williams, Thomas G. and Chown, Ryan and Belfiore, Francesco and Bigiel, Frank and Boquien, Médéric and Cao, Yixian and Chevance, Mélanie and Dale, Daniel A. and Egorov, Oleg V. and Glover, Simon C. O. and Groves, Brent and Klessen, Ralf S. and Kreckel, Kathryn and Larson, Kirsten L. and Oakes, Elias K. and Pathak, Debosmita and Ramambason, Lise and Rosolowsky, Erik and Watkins, Elizabeth J.},
	month = aug,
	year = {2024},
	pages = {178},
}

@article{egorov_phangsjwst_2023,
	title = {{PHANGS}–{JWST} {First} {Results}: {Destruction} of the {PAH} {Molecules} in {H} ii {Regions} {Probed} by {JWST} and {MUSE}},
	volume = {944},
	issn = {2041-8205, 2041-8213},
	shorttitle = {{PHANGS}–{JWST} {First} {Results}},
	url = {https://iopscience.iop.org/article/10.3847/2041-8213/acac92},
	doi = {10.3847/2041-8213/acac92},
	number = {2},
	urldate = {2026-03-20},
	journal = {ApJL},
	author = {Egorov, Oleg V. and Kreckel, Kathryn and Sandstrom, Karin M. and Leroy, Adam K. and Glover, Simon C. O. and Groves, Brent and Kruijssen, J. M. Diederik and Barnes, Ashley. T. and Belfiore, Francesco and Bigiel, F. and Blanc, Guillermo A. and Boquien, Médéric and Cao, Yixian and Chastenet, Jérémy and Chevance, Mélanie and Congiu, Enrico and Dale, Daniel A. and Emsellem, Eric and Grasha, Kathryn and Klessen, Ralf S. and Larson, Kirsten L. and Liu, Daizhong and Murphy, Eric J. and Pan, Hsi-An and Pessa, Ismael and Pety, Jérôme and Rosolowsky, Erik and Scheuermann, Fabian and Schinnerer, Eva and Sutter, Jessica and Thilker, David A. and Watkins, Elizabeth J. and Williams, Thomas G.},
	month = feb,
	year = {2023},
	pages = {L16},
}

@article{boogaard_alma_2020,
	title = {The {ALMA} {Spectroscopic} {Survey} in the {Hubble} {Ultra} {Deep} {Field}: {CO} {Excitation} and {Atomic} {Carbon} in {Star}-forming {Galaxies} at z = 1–3},
	volume = {902},
	issn = {0004-637X, 1538-4357},
	shorttitle = {The {ALMA} {Spectroscopic} {Survey} in the {Hubble} {Ultra} {Deep} {Field}},
	url = {https://iopscience.iop.org/article/10.3847/1538-4357/abb82f},
	doi = {10.3847/1538-4357/abb82f},
	number = {2},
	urldate = {2026-03-22},
	journal = {ApJ},
	author = {Boogaard, Leindert A. and Werf, Paul Van Der and Weiss, Axel and Popping, Gergö and Decarli, Roberto and Walter, Fabian and Aravena, Manuel and Bouwens, Rychard and Riechers, Dominik and González-López, Jorge and Smail, Ian and Carilli, Chris and Kaasinen, Melanie and Daddi, Emanuele and Cox, Pierre and Díaz-Santos, Tanio and Inami, Hanae and Cortes, Paulo C. and Wagg, Jeff},
	month = oct,
	year = {2020},
	pages = {109},
}

@article{kriek_h_2011,
	title = {Hα {AND} 4000 Å {BREAK} {MEASUREMENTS} {FOR} ∼3500 \textit{{K}} -{SELECTED} {GALAXIES} {AT} 0.5 {\textless} \textit{z} {\textless} 2.0},
	volume = {743},
	issn = {0004-637X, 1538-4357},
	url = {https://iopscience.iop.org/article/10.1088/0004-637X/743/2/168},
	doi = {10.1088/0004-637X/743/2/168},
	number = {2},
	urldate = {2026-03-23},
	journal = {ApJ},
	author = {Kriek, Mariska and Van Dokkum, Pieter G. and Whitaker, Katherine E. and Labbé, Ivo and Franx, Marijn and Brammer, Gabriel B.},
	month = dec,
	year = {2011},
	pages = {168},
}

@article{van_dokkum_newfirm_2009,
	title = {The {NEWFIRM} {Medium}-{Band} {Survey}: {Filter} {Definitions} and {First} {Results}},
	volume = {121},
	issn = {0004-6280, 1538-3873},
	shorttitle = {The {NEWFIRM} {Medium}-{Band} {Survey}},
	url = {http://iopscience.iop.org/article/10.1086/597138},
	doi = {10.1086/597138},
	language = {en},
	number = {875},
	urldate = {2026-03-23},
	journal = {PASP},
	author = {Van Dokkum, Pieter G. and Labbé, Ivo and Marchesini, Danilo and Quadri, Ryan and Brammer, Gabriel and Whitaker, Katherine E. and Kriek, Mariska and Franx, Marijn and Rudnick, Gregory and Illingworth, Garth and Lee, Kyoung-Soo and Muzzin, Adam},
	month = jan,
	year = {2009},
	pages = {2--8},
}

@article{walter_alma_2016,
	title = {{ALMA} {SPECTROSCOPIC} {SURVEY} {IN} {THE} {HUBBLE} {ULTRA} {DEEP} {FIELD}: {SURVEY} {DESCRIPTION}},
	volume = {833},
	issn = {0004-637X, 1538-4357},
	shorttitle = {{ALMA} {SPECTROSCOPIC} {SURVEY} {IN} {THE} {HUBBLE} {ULTRA} {DEEP} {FIELD}},
	url = {https://iopscience.iop.org/article/10.3847/1538-4357/833/1/67},
	doi = {10.3847/1538-4357/833/1/67},
	number = {1},
	urldate = {2026-03-24},
	journal = {ApJ},
	author = {Walter, Fabian and Decarli, Roberto and Aravena, Manuel and Carilli, Chris and Bouwens, Rychard and Cunha, Elisabete Da and Daddi, Emanuele and Ivison, R. J. and Riechers, Dominik and Smail, Ian and Swinbank, Mark and Weiss, Axel and Anguita, Timo and Assef, Roberto and Bacon, Roland and Bauer, Franz and Bell, Eric F. and Bertoldi, Frank and Chapman, Scott and Colina, Luis and Cortes, Paulo C. and Cox, Pierre and Dickinson, Mark and Elbaz, David and Gónzalez-López, Jorge and Ibar, Edo and Inami, Hanae and Infante, Leopoldo and Hodge, Jacqueline and Karim, Alex and Fevre, Olivier Le and Magnelli, Benjamin and Neri, Roberto and Oesch, Pascal and Ota, Kazuaki and Popping, Gergö and Rix, Hans-Walter and Sargent, Mark and Sheth, Kartik and Wel, Arjen Van Der and Werf, Paul Van Der and Wagg, Jeff},
	month = dec,
	year = {2016},
	pages = {67},
}

@misc{newville_lmfit_2025,
	title = {{LMFIT}: {Non}-{Linear} {Least}-{Squares} {Minimization} and {Curve}-{Fitting} for {Python}},
	copyright = {Creative Commons Attribution 4.0 International},
	shorttitle = {{LMFIT}},
	url = {https://zenodo.org/doi/10.5281/zenodo.598352},
	doi = {10.5281/ZENODO.598352},
	urldate = {2026-03-25},
	publisher = {Zenodo},
	author = {Newville, Matthew and Otten, Renee and Nelson, Andrew and Stensitzki, Till and Ingargiola, Antonino and Allan, Daniel and Fox, Austin and Carter, Faustin and Rawlik, Michal},
	month = jul,
	year = {2025},
}

@article{marble_aromatic_2010,
	title = {{AN} {AROMATIC} {INVENTORY} {OF} {THE} {LOCAL} {VOLUME}},
	volume = {715},
	issn = {0004-637X, 1538-4357},
	url = {https://iopscience.iop.org/article/10.1088/0004-637X/715/1/506},
	doi = {10.1088/0004-637X/715/1/506},
	number = {1},
	urldate = {2026-03-30},
	journal = {ApJ},
	author = {Marble, A. R. and Engelbracht, C. W. and Van Zee, L. and Dale, D. A. and Smith, J. D. T. and Gordon, K. D. and Wu, Y. and Lee, J. C. and Kennicutt, R. C. and Skillman, E. D. and Johnson, L. C. and Block, M. and Calzetti, D. and Cohen, S. A. and Lee, H. and Schuster, M. D.},
	month = may,
	year = {2010},
	pages = {506--540},
}

@article{blasberger_observational_2017,
	title = {Observational {Evidence} {Linking} {Interstellar} {UV} {Absorption} to {PAH} {Molecules}},
	volume = {836},
	issn = {0004-637X, 1538-4357},
	url = {https://iopscience.iop.org/article/10.3847/1538-4357/aa5b8a},
	doi = {10.3847/1538-4357/aa5b8a},
	number = {2},
	urldate = {2026-03-30},
	journal = {ApJ},
	author = {Blasberger, Avi and Behar, Ehud and Perets, Hagai B. and Brosch, Noah and Tielens, Alexander G. G. M.},
	month = feb,
	year = {2017},
	pages = {173},
}

@article{gordon_expanded_2024,
	title = {Expanded {Sample} of {Small} {Magellanic} {Cloud} {Ultraviolet} {Dust} {Extinction} {Curves}: {Correlations} between the 2175 Å {Bump}, q$_{\textrm{{PAH}}}$ , {Ultraviolet} {Extinction} {Shape}, and {N}({H} i)/{A}({V})},
	volume = {970},
	issn = {0004-637X, 1538-4357},
	shorttitle = {Expanded {Sample} of {Small} {Magellanic} {Cloud} {Ultraviolet} {Dust} {Extinction} {Curves}},
	url = {https://iopscience.iop.org/article/10.3847/1538-4357/ad4be1},
	doi = {10.3847/1538-4357/ad4be1},
	number = {1},
	urldate = {2026-03-30},
	journal = {ApJ},
	author = {Gordon, Karl D. and Fitzpatrick, E. L. and Massa, Derck and Bohlin, Ralph and Chastenet, Jérémy and Murray, Claire E. and Clayton, Geoffrey C. and Lennon, Daniel J. and Misselt, Karl A. and Sandstrom, Karin},
	month = jul,
	year = {2024},
	pages = {51},
}

@article{weingartner_dust_2001,
	title = {Dust {Grain}–{Size} {Distributions} and {Extinction} in the {Milky} {Way}, {Large} {Magellanic} {Cloud}, and {Small} {Magellanic} {Cloud}},
	volume = {548},
	issn = {0004-637X, 1538-4357},
	url = {https://iopscience.iop.org/article/10.1086/318651},
	doi = {10.1086/318651},
	language = {en},
	number = {1},
	urldate = {2026-03-31},
	journal = {ApJ},
	author = {Weingartner, Joseph C. and Draine, B. T.},
	month = feb,
	year = {2001},
	pages = {296--309},
}

@article{jones_evolution_2013,
	title = {The evolution of amorphous hydrocarbons in the {ISM}: dust modelling from a new vantage point},
	volume = {558},
	issn = {0004-6361, 1432-0746},
	shorttitle = {The evolution of amorphous hydrocarbons in the {ISM}},
	url = {http://www.aanda.org/10.1051/0004-6361/201321686},
	doi = {10.1051/0004-6361/201321686},
	urldate = {2026-03-31},
	journal = {A\&A},
	author = {Jones, A. P. and Fanciullo, L. and Köhler, M. and Verstraete, L. and Guillet, V. and Bocchio, M. and Ysard, N.},
	month = oct,
	year = {2013},
	pages = {A62},
}

@article{hensley_astrodustpah_2023,
	title = {The {Astrodust}+{PAH} {Model}: {A} {Unified} {Description} of the {Extinction}, {Emission}, and {Polarization} from {Dust} in the {Diffuse} {Interstellar} {Medium}},
	volume = {948},
	issn = {0004-637X, 1538-4357},
	shorttitle = {The {Astrodust}+{PAH} {Model}},
	url = {https://iopscience.iop.org/article/10.3847/1538-4357/acc4c2},
	doi = {10.3847/1538-4357/acc4c2},
	number = {1},
	urldate = {2026-03-31},
	journal = {ApJ},
	author = {Hensley, Brandon S. and Draine, B. T.},
	month = may,
	year = {2023},
	pages = {55},
}

@article{witt_multiple_2000,
	title = {Multiple {Scattering} in {Clumpy} {Media}. {II}. {Galactic} {Environments}},
	volume = {528},
	issn = {0004-637X, 1538-4357},
	url = {https://iopscience.iop.org/article/10.1086/308197},
	doi = {10.1086/308197},
	language = {en},
	number = {2},
	urldate = {2026-03-31},
	journal = {ApJ},
	author = {Witt, Adolf N. and Gordon, Karl D.},
	month = jan,
	year = {2000},
	pages = {799--816},
}

@article{massa_relations_2022,
	title = {Relations between {Mid}-infrared {Dust} {Emission} and {UV} {Extinction}},
	volume = {925},
	issn = {0004-637X, 1538-4357},
	url = {https://iopscience.iop.org/article/10.3847/1538-4357/ac3825},
	doi = {10.3847/1538-4357/ac3825},
	number = {1},
	urldate = {2026-04-07},
	journal = {ApJ},
	author = {Massa, Derck and Gordon, Karl D. and Fitzpatrick, E. L.},
	month = jan,
	year = {2022},
	pages = {19},
}

@article{hirashita_remodelling_2019,
	title = {Remodelling the evolution of grain size distribution in galaxies},
	volume = {482},
	copyright = {https://academic.oup.com/journals/pages/open\_access/funder\_policies/chorus/standard\_publication\_model},
	issn = {0035-8711, 1365-2966},
	url = {https://academic.oup.com/mnras/article/482/2/2555/5142317},
	doi = {10.1093/mnras/sty2838},
	language = {en},
	number = {2},
	urldate = {2026-04-13},
	journal = {MNRAS},
	author = {Hirashita, Hiroyuki and Aoyama, Shohei},
	month = jan,
	year = {2019},
	pages = {2555--2572},
}

@article{relano_dust_2022,
	title = {Dust grain size evolution in local galaxies: a comparison between observations and simulations},
	volume = {515},
	copyright = {https://academic.oup.com/journals/pages/open\_access/funder\_policies/chorus/standard\_publication\_model},
	issn = {0035-8711, 1365-2966},
	shorttitle = {Dust grain size evolution in local galaxies},
	url = {https://academic.oup.com/mnras/article/515/4/5306/6652124},
	doi = {10.1093/mnras/stac2108},
	language = {en},
	number = {4},
	urldate = {2026-04-13},
	journal = {MNRAS},
	author = {Relaño, M and De Looze, I and Saintonge, A and Hou, K-C and Romano, L E C and Nagamine, K and Hirashita, H and Aoyama, S and Lamperti, I and Lisenfeld, U and Smith, M W L and Chastenet, J and Xiao, T and Gao, Y and Sargent, M and van der Giessen, S A},
	month = aug,
	year = {2022},
	pages = {5306--5334},
}

@article{hou_dust_2019,
	title = {Dust scaling relations in a cosmological simulation},
	volume = {485},
	copyright = {https://academic.oup.com/journals/pages/open\_access/funder\_policies/chorus/standard\_publication\_model},
	issn = {0035-8711, 1365-2966},
	url = {https://academic.oup.com/mnras/article/485/2/1727/5308843},
	doi = {10.1093/mnras/stz121},
	language = {en},
	number = {2},
	urldate = {2026-04-13},
	journal = {MNRAS},
	author = {Hou, Kuan-Chou and Aoyama, Shohei and Hirashita, Hiroyuki and Nagamine, Kentaro and Shimizu, Ikkoh},
	month = may,
	year = {2019},
	pages = {1727--1744},
}

@article{meurer_dust_1999,
	title = {Dust {Absorption} and the {Ultraviolet} {Luminosity} {Density} at \textit{z} ≈ 3 as {Calibrated} by {Local} {Starburst} {Galaxies}},
	volume = {521},
	issn = {0004-637X, 1538-4357},
	url = {https://iopscience.iop.org/article/10.1086/307523},
	doi = {10.1086/307523},
	language = {en},
	number = {1},
	urldate = {2026-04-16},
	journal = {ApJ},
	author = {Meurer, Gerhardt R. and Heckman, Timothy M. and Calzetti, Daniela},
	month = aug,
	year = {1999},
	pages = {64--80},
}

@article{hollenbach_molecule_1979,
	title = {Molecule formation and infrared emission in fast interstellar shocks. {I} {Physical} processes},
	volume = {41},
	issn = {0067-0049, 1538-4365},
	url = {http://adsabs.harvard.edu/doi/10.1086/190631},
	doi = {10.1086/190631},
	language = {en},
	urldate = {2026-04-16},
	journal = {ApJS},
	author = {Hollenbach, D. and McKee, C. F.},
	month = nov,
	year = {1979},
	pages = {555},
}

@article{planck_collaboration_planck_2016,
	title = {\textit{{Planck}} 2015 results: {XIII}. {Cosmological} parameters},
	volume = {594},
	issn = {0004-6361, 1432-0746},
	shorttitle = {\textit{{Planck}} 2015 results},
	url = {http://www.aanda.org/10.1051/0004-6361/201525830},
	doi = {10.1051/0004-6361/201525830},
	urldate = {2026-04-17},
	journal = {A\&A},
	author = {{Planck Collaboration} and Ade, P. A. R. and Aghanim, N. and Arnaud, M. and Ashdown, M. and Aumont, J. and Baccigalupi, C. and Banday, A. J. and Barreiro, R. B. and Bartlett, J. G. and Bartolo, N. and Battaner, E. and Battye, R. and Benabed, K. and Benoît, A. and Benoit-Lévy, A. and Bernard, J.-P. and Bersanelli, M. and Bielewicz, P. and Bock, J. J. and Bonaldi, A. and Bonavera, L. and Bond, J. R. and Borrill, J. and Bouchet, F. R. and Boulanger, F. and Bucher, M. and Burigana, C. and Butler, R. C. and Calabrese, E. and Cardoso, J.-F. and Catalano, A. and Challinor, A. and Chamballu, A. and Chary, R.-R. and Chiang, H. C. and Chluba, J. and Christensen, P. R. and Church, S. and Clements, D. L. and Colombi, S. and Colombo, L. P. L. and Combet, C. and Coulais, A. and Crill, B. P. and Curto, A. and Cuttaia, F. and Danese, L. and Davies, R. D. and Davis, R. J. and De Bernardis, P. and De Rosa, A. and De Zotti, G. and Delabrouille, J. and Désert, F.-X. and Di Valentino, E. and Dickinson, C. and Diego, J. M. and Dolag, K. and Dole, H. and Donzelli, S. and Doré, O. and Douspis, M. and Ducout, A. and Dunkley, J. and Dupac, X. and Efstathiou, G. and Elsner, F. and Enßlin, T. A. and Eriksen, H. K. and Farhang, M. and Fergusson, J. and Finelli, F. and Forni, O. and Frailis, M. and Fraisse, A. A. and Franceschi, E. and Frejsel, A. and Galeotta, S. and Galli, S. and Ganga, K. and Gauthier, C. and Gerbino, M. and Ghosh, T. and Giard, M. and Giraud-Héraud, Y. and Giusarma, E. and Gjerløw, E. and González-Nuevo, J. and Górski, K. M. and Gratton, S. and Gregorio, A. and Gruppuso, A. and Gudmundsson, J. E. and Hamann, J. and Hansen, F. K. and Hanson, D. and Harrison, D. L. and Helou, G. and Henrot-Versillé, S. and Hernández-Monteagudo, C. and Herranz, D. and Hildebrandt, S. R. and Hivon, E. and Hobson, M. and Holmes, W. A. and Hornstrup, A. and Hovest, W. and Huang, Z. and Huffenberger, K. M. and Hurier, G. and Jaffe, A. H. and Jaffe, T. R. and Jones, W. C. and Juvela, M. and Keihänen, E. and Keskitalo, R. and Kisner, T. S. and Kneissl, R. and Knoche, J. and Knox, L. and Kunz, M. and Kurki-Suonio, H. and Lagache, G. and Lähteenmäki, A. and Lamarre, J.-M. and Lasenby, A. and Lattanzi, M. and Lawrence, C. R. and Leahy, J. P. and Leonardi, R. and Lesgourgues, J. and Levrier, F. and Lewis, A. and Liguori, M. and Lilje, P. B. and Linden-Vørnle, M. and López-Caniego, M. and Lubin, P. M. and Macías-Pérez, J. F. and Maggio, G. and Maino, D. and Mandolesi, N. and Mangilli, A. and Marchini, A. and Maris, M. and Martin, P. G. and Martinelli, M. and Martínez-González, E. and Masi, S. and Matarrese, S. and McGehee, P. and Meinhold, P. R. and Melchiorri, A. and Melin, J.-B. and Mendes, L. and Mennella, A. and Migliaccio, M. and Millea, M. and Mitra, S. and Miville-Deschênes, M.-A. and Moneti, A. and Montier, L. and Morgante, G. and Mortlock, D. and Moss, A. and Munshi, D. and Murphy, J. A. and Naselsky, P. and Nati, F. and Natoli, P. and Netterfield, C. B. and Nørgaard-Nielsen, H. U. and Noviello, F. and Novikov, D. and Novikov, I. and Oxborrow, C. A. and Paci, F. and Pagano, L. and Pajot, F. and Paladini, R. and Paoletti, D. and Partridge, B. and Pasian, F. and Patanchon, G. and Pearson, T. J. and Perdereau, O. and Perotto, L. and Perrotta, F. and Pettorino, V. and Piacentini, F. and Piat, M. and Pierpaoli, E. and Pietrobon, D. and Plaszczynski, S. and Pointecouteau, E. and Polenta, G. and Popa, L. and Pratt, G. W. and Prézeau, G. and Prunet, S. and Puget, J.-L. and Rachen, J. P. and Reach, W. T. and Rebolo, R. and Reinecke, M. and Remazeilles, M. and Renault, C. and Renzi, A. and Ristorcelli, I. and Rocha, G. and Rosset, C. and Rossetti, M. and Roudier, G. and Rouillé d’Orfeuil, B. and Rowan-Robinson, M. and Rubiño-Martín, J. A. and Rusholme, B. and Said, N. and Salvatelli, V. and Salvati, L. and Sandri, M. and Santos, D. and Savelainen, M. and Savini, G. and Scott, D. and Seiffert, M. D. and Serra, P. and Shellard, E. P. S. and Spencer, L. D. and Spinelli, M. and Stolyarov, V. and Stompor, R. and Sudiwala, R. and Sunyaev, R. and Sutton, D. and Suur-Uski, A.-S. and Sygnet, J.-F. and Tauber, J. A. and Terenzi, L. and Toffolatti, L. and Tomasi, M. and Tristram, M. and Trombetti, T. and Tucci, M. and Tuovinen, J. and Türler, M. and Umana, G. and Valenziano, L. and Valiviita, J. and Van Tent, F. and Vielva, P. and Villa, F. and Wade, L. A. and Wandelt, B. D. and Wehus, I. K. and White, M. and White, S. D. M. and Wilkinson, A. and Yvon, D. and Zacchei, A. and Zonca, A.},
	month = oct,
	year = {2016},
	pages = {A13},
}

@article{whitaker_hubble_2019,
	title = {The {Hubble} {Legacy} {Field} {GOODS}-{S} {Photometric} {Catalog}},
	volume = {244},
	issn = {0067-0049, 1538-4365},
	url = {https://iopscience.iop.org/article/10.3847/1538-4365/ab3853},
	doi = {10.3847/1538-4365/ab3853},
	number = {1},
	urldate = {2026-04-22},
	journal = {ApJS},
	author = {Whitaker, Katherine E. and Ashas, Mohammad and Illingworth, Garth and Magee, Daniel and Leja, Joel and Oesch, Pascal and Van Dokkum, Pieter and Mowla, Lamiya and Bouwens, Rychard and Franx, Marijn and Holden, Bradford and Labbé, Ivo and Rafelski, Marc and Teplitz, Harry and Gonzalez, Valentino},
	month = sep,
	year = {2019},
	pages = {16},
}

@article{gruppioni_alpine-alma_2020,
	title = {The {ALPINE}-{ALMA} [{CII}] survey: {The} nature, luminosity function, and star formation history of dusty galaxies up to \textit{z} ≃ 6},
	volume = {643},
	copyright = {https://www.edpsciences.org/en/authors/copyright-and-licensing},
	issn = {0004-6361, 1432-0746},
	shorttitle = {The {ALPINE}-{ALMA} [{CII}] survey},
	url = {https://www.aanda.org/10.1051/0004-6361/202038487},
	doi = {10.1051/0004-6361/202038487},
	urldate = {2026-04-26},
	journal = {A\&A},
	author = {Gruppioni, C. and Béthermin, M. and Loiacono, F. and Le Fèvre, O. and Capak, P. and Cassata, P. and Faisst, A. L. and Schaerer, D. and Silverman, J. and Yan, L. and Bardelli, S. and Boquien, M. and Carraro, R. and Cimatti, A. and Dessauges-Zavadsky, M. and Ginolfi, M. and Fujimoto, S. and Hathi, N. P. and Jones, G. C. and Khusanova, Y. and Koekemoer, A. M. and Lagache, G. and Lemaux, B. C. and Oesch, P. A. and Pozzi, F. and Riechers, D. A. and Rodighiero, G. and Romano, M. and Talia, M. and Vallini, L. and Vergani, D. and Zamorani, G. and Zucca, E.},
	month = nov,
	year = {2020},
	pages = {A8},
}

@article{le_fevre_alpine-alma_2020,
	title = {The {ALPINE}-{ALMA} [{CII}] survey: {Survey} strategy, observations, and sample properties of 118 star-forming galaxies at 4 {\textless} \textit{z} {\textless} 6},
	volume = {643},
	copyright = {https://creativecommons.org/licenses/by/4.0},
	issn = {0004-6361, 1432-0746},
	shorttitle = {The {ALPINE}-{ALMA} [{CII}] survey},
	url = {https://www.aanda.org/10.1051/0004-6361/201936965},
	doi = {10.1051/0004-6361/201936965},
	urldate = {2026-04-26},
	journal = {A\&A},
	author = {Le Fèvre, O. and Béthermin, M. and Faisst, A. and Jones, G. C. and Capak, P. and Cassata, P. and Silverman, J. D. and Schaerer, D. and Yan, L. and Amorin, R. and Bardelli, S. and Boquien, M. and Cimatti, A. and Dessauges-Zavadsky, M. and Giavalisco, M. and Hathi, N. P. and Fudamoto, Y. and Fujimoto, S. and Ginolfi, M. and Gruppioni, C. and Hemmati, S. and Ibar, E. and Koekemoer, A. and Khusanova, Y. and Lagache, G. and Lemaux, B. C. and Loiacono, F. and Maiolino, R. and Mancini, C. and Narayanan, D. and Morselli, L. and Méndez-Hernàndez, Hugo and Oesch, P. A. and Pozzi, F. and Romano, M. and Riechers, D. and Scoville, N. and Talia, M. and Tasca, L. A. M. and Thomas, R. and Toft, S. and Vallini, L. and Vergani, D. and Walter, F. and Zamorani, G. and Zucca, E.},
	month = nov,
	year = {2020},
	pages = {A1},
}

@article{bethermin_alpine-alma_2020,
	title = {The {ALPINE}-{ALMA} [{CII}] survey: {Data} processing, catalogs, and statistical source properties},
	volume = {643},
	copyright = {https://creativecommons.org/licenses/by/4.0},
	issn = {0004-6361, 1432-0746},
	shorttitle = {The {ALPINE}-{ALMA} [{CII}] survey},
	url = {https://www.aanda.org/10.1051/0004-6361/202037649},
	doi = {10.1051/0004-6361/202037649},
	urldate = {2026-04-26},
	journal = {A\&A},
	author = {Béthermin, M. and Fudamoto, Y. and Ginolfi, M. and Loiacono, F. and Khusanova, Y. and Capak, P. L. and Cassata, P. and Faisst, A. and Le Fèvre, O. and Schaerer, D. and Silverman, J. D. and Yan, L. and Amorin, R. and Bardelli, S. and Boquien, M. and Cimatti, A. and Davidzon, I. and Dessauges-Zavadsky, M. and Fujimoto, S. and Gruppioni, C. and Hathi, N. P. and Ibar, E. and Jones, G. C. and Koekemoer, A. M. and Lagache, G. and Lemaux, B. C. and Moreau, C. and Oesch, P. A. and Pozzi, F. and Riechers, D. A. and Talia, M. and Toft, S. and Vallini, L. and Vergani, D. and Zamorani, G. and Zucca, E.},
	month = nov,
	year = {2020},
	pages = {A2},
}

@article{faisst_alpinealma_2020,
	title = {The {ALPINE}–{ALMA} [{C} ii] {Survey}: {Multiwavelength} {Ancillary} {Data} and {Basic} {Physical} {Measurements}},
	volume = {247},
	issn = {0067-0049, 1538-4365},
	shorttitle = {The {ALPINE}–{ALMA} [{C} ii] {Survey}},
	url = {https://iopscience.iop.org/article/10.3847/1538-4365/ab7ccd},
	doi = {10.3847/1538-4365/ab7ccd},
	number = {2},
	urldate = {2026-04-26},
	journal = {ApJS},
	author = {Faisst, A. L. and Schaerer, D. and Lemaux, B. C. and Oesch, P. A. and Fudamoto, Y. and Cassata, P. and Béthermin, M. and Capak, P. L. and Le Fèvre, O. and Silverman, J. D. and Yan, L. and Ginolfi, M. and Koekemoer, A. M. and Morselli, L. and Amorín, R. and Bardelli, S. and Boquien, M. and Brammer, G. and Cimatti, A. and Dessauges-Zavadsky, M. and Fujimoto, S. and Gruppioni, C. and Hathi, N. P. and Hemmati, S. and Ibar, E. and Jones, G. C. and Khusanova, Y. and Loiacono, F. and Pozzi, F. and Talia, M. and Tasca, L. A. M. and Riechers, D. A. and Rodighiero, G. and Romano, M. and Scoville, N. and Toft, S. and Vallini, L. and Vergani, D. and Zamorani, G. and Zucca, E.},
	month = apr,
	year = {2020},
	pages = {61},
}

@article{wright_mid-infrared_2023,
	title = {The {Mid}-infrared {Instrument} for {JWST} and {Its} {In}-flight {Performance}},
	volume = {135},
	issn = {0004-6280, 1538-3873},
	url = {https://iopscience.iop.org/article/10.1088/1538-3873/acbe66},
	doi = {10.1088/1538-3873/acbe66},
	number = {1046},
	urldate = {2026-04-26},
	journal = {PASP},
	author = {Wright, Gillian S. and Rieke, George H. and Glasse, Alistair and Ressler, Michael and García Marín, Macarena and Aguilar, Jonathan and Alberts, Stacey and Álvarez-Márquez, Javier and Argyriou, Ioannis and Banks, Kimberly and Baudoz, Pierre and Boccaletti, Anthony and Bouchet, Patrice and Bouwman, Jeroen and Brandl, Bernard R. and Breda, David and Bright, Stacey and Cale, Steven and Colina, Luis and Cossou, Christophe and Coulais, Alain and Cracraft, Misty and De Meester, Wim and Dicken, Daniel and Engesser, Michael and Etxaluze, Mireya and Fox, Ori D. and Friedman, Scott and Fu, Henry and Gasman, Danny and Gáspár, András and Gastaud, René and Geers, Vincent and Glauser, Adrian Michael and Gordon, Karl D. and Greene, Thomas and Greve, Thomas R. and Grundy, Timothy and Güdel, Manuel and Guillard, Pierre and Haderlein, Peter and Hashimoto, Ryan and Henning, Thomas and Hines, Dean and Holler, Bryan and Detre, Örs Hunor and Jahromi, Amir and James, Bryan and Jones, Olivia C. and Justtanont, Kay and Kavanagh, Patrick and Kendrew, Sarah and Klaassen, Pamela and Krause, Oliver and Labiano, Alvaro and Lagage, Pierre-Olivier and Lambros, Scott and Larson, Kirsten and Law, David and Lee, David and Libralato, Mattia and Alverez, Jose Lorenzo and Meixner, Margaret and Morrison, Jane and Mueller, Migo and Murray, Katherine and Mycroft, Matthew and Myers, Richard and Nayak, Omnarayani and Naylor, Bret and Nickson, Bryony and Noriega-Crespo, Alberto and Östlin, Göran and O’Sullivan, Brian and Ottens, Richard and Patapis, Polychronis and Penanen, Konstantin and Pietraszkiewicz, Martin and Ray, Tom and Regan, Michael and Roteliuk, Anthony and Royer, Pierre and Samara-Ratna, Piyal and Samuelson, Bridget and Sargent, Beth A. and Scheithauer, Silvia and Schneider, Analyn and Schreiber, Jürgen and Shaughnessy, Bryan and Sheehan, Evan and Shivaei, Irene and Sloan, G. C. and Tamas, Laszlo and Teague, Kelly and Temim, Tea and Tikkanen, Tuomo and Tustain, Samuel and Van Dishoeck, Ewine F. and Vandenbussche, Bart and Weilert, Mark and Whitehouse, Paul and Wolff, Schuyler},
	month = apr,
	year = {2023},
	pages = {048003},
}

@article{stecher_interstellar_1965,
	title = {Interstellar {Ectinction} in the {Ultraviolet}.},
	volume = {142},
	issn = {0004-637X, 1538-4357},
	url = {http://adsabs.harvard.edu/doi/10.1086/148462},
	doi = {10.1086/148462},
	language = {en},
	urldate = {2026-04-27},
	journal = {ApJ},
	author = {Stecher, Theodore P.},
	month = nov,
	year = {1965},
	pages = {1683},
}

@article{otaki_effective_2026,
	title = {Effective supernova dust yields from rotating and nonrotating stellar progenitors},
	volume = {708},
	copyright = {https://creativecommons.org/licenses/by/4.0},
	issn = {0004-6361, 1432-0746},
	url = {https://www.aanda.org/10.1051/0004-6361/202555906},
	doi = {10.1051/0004-6361/202555906},
	urldate = {2026-04-27},
	journal = {A\&A},
	author = {Otaki, Koki and Schneider, Raffaella and Graziani, Luca and Bonella, Alessandro and Marassi, Stefania and Limongi, Marco and Bianchi, Simone},
	month = apr,
	year = {2026},
	pages = {A136},
}

@article{alberts_smiles_2024,
	title = {{SMILES} {Initial} {Data} {Release}: {Unveiling} the {Obscured} {Universe} with {MIRI} {Multiband} {Imaging}},
	volume = {976},
	issn = {0004-637X, 1538-4357},
	shorttitle = {{SMILES} {Initial} {Data} {Release}},
	url = {https://iopscience.iop.org/article/10.3847/1538-4357/ad7396},
	doi = {10.3847/1538-4357/ad7396},
	number = {2},
	urldate = {2026-04-27},
	journal = {ApJ},
	author = {Alberts, Stacey and Lyu, Jianwei and Shivaei, Irene and Rieke, George H. and Pérez-González, Pablo G. and Bonaventura, Nina and Zhu, Yongda and Helton, Jakob M. and Ji, Zhiyuan and Morrison, Jane and Robertson, Brant E. and Stone, Meredith A. and Sun, Yang and Williams, Christina C. and Willmer, Christopher N. A.},
	month = dec,
	year = {2024},
	pages = {224},
}

@article{alberts_calibrating_2026,
	title = {Calibrating {Photometric} {Mid}-{Infrared} {Star} {Formation} {Rates} for {JWST}},
	url = {http://arxiv.org/abs/2604.01326},
	doi = {10.48550/arXiv.2604.01326},
	urldate = {2026-04-27},
	publisher = {arXiv},
	author = {Alberts, Stacey and Rieke, George H. and Shivaei, Irene and Ji, Zhiyuan and Oesch, Pascal and Brammer, Gabriel and Helton, Jakob M. and Lyu, Jianwei and Nelson, Erica J. and Reddy, Naveen and Rinaldi, Pierluigi and Sun, Yang and Whitaker, Katherine E. and Williams, Christina C. and Willmer, Christopher N. A. and Wuyts, Stijn},
	month = apr,
	year = {2026},
	note = {arXiv:2604.01326 [astro-ph]},
}

@article{allamandola_polycyclic_1985,
	title = {Polycyclic aromatic hydrocarbons and the unidentified infrared emission bands - {Auto} exhaust along the {Milky} {Way}},
	volume = {290},
	issn = {0004-637X, 1538-4357},
	url = {http://adsabs.harvard.edu/doi/10.1086/184435},
	doi = {10.1086/184435},
	language = {en},
	urldate = {2026-04-27},
	journal = {ApJ},
	author = {Allamandola, L. J. and Tielens, A. G. G. M. and Barker, J. R.},
	month = mar,
	year = {1985},
	pages = {L25},
}

@incollection{arriscado_metallicity-pah_2025,
	title = {The metallicity-{PAH} relationship from the local universe to cosmic noon},
	volume = {2},
	url = {https://ui.adsabs.harvard.edu/abs/2025prim.book...32A},
	urldate = {2026-04-27},
	booktitle = {{PRIMA} {General} {Observer} {Science} {Book} {Volume} 2},
	author = {Arriscado, Leonor and Shivaei, Irene and Boogaard, Leindert and Díaz-Santos, Tanio and McKinney, Jed},
	month = dec,
	year = {2025},
	pages = {32--37},
}

@article{asano_dust_2013,
	title = {Dust formation history of galaxies: {A} critical role of metallicity dust mass growth by accreting materials in the interstellar medium},
	volume = {65},
	issn = {1343-8832, 1880-5981},
	shorttitle = {Dust formation history of galaxies},
	url = {http://link.springer.com/10.5047/eps.2012.04.014},
	doi = {10.5047/eps.2012.04.014},
	language = {en},
	number = {3},
	urldate = {2026-04-27},
	journal = {Earth Planet Sp},
	author = {Asano, Ryosuke S. and Takeuchi, Tsutomu T. and Hirashita, Hiroyuki and Inoue, Akio K.},
	month = mar,
	year = {2013},
	pages = {213--222},
}

@article{asplund_chemical_2009,
	title = {The {Chemical} {Composition} of the {Sun}},
	volume = {47},
	issn = {0066-4146, 1545-4282},
	url = {https://www.annualreviews.org/doi/10.1146/annurev.astro.46.060407.145222},
	doi = {10.1146/annurev.astro.46.060407.145222},
	language = {en},
	number = {1},
	urldate = {2026-04-27},
	journal = {ARA\&A},
	author = {Asplund, Martin and Grevesse, Nicolas and Sauval, A. Jacques and Scott, Pat},
	month = sep,
	year = {2009},
	pages = {481--522},
}

@article{the_astropy_collaboration_astropy_2018,
	title = {The {Astropy} {Project}: {Building} an {Open}-science {Project} and {Status} of the v2.0 {Core} {Package}$^{\textrm{*}}$},
	volume = {156},
	issn = {0004-6256, 1538-3881},
	shorttitle = {The {Astropy} {Project}},
	url = {https://iopscience.iop.org/article/10.3847/1538-3881/aabc4f},
	doi = {10.3847/1538-3881/aabc4f},
	number = {3},
	urldate = {2026-04-27},
	journal = {AJ},
	author = {{The Astropy Collaboration} and Price-Whelan, A. M. and Sipőcz, B. M. and Günther, H. M. and Lim, P. L. and Crawford, S. M. and Conseil, S. and Shupe, D. L. and Craig, M. W. and Dencheva, N. and Ginsburg, A. and VanderPlas, J. T. and Bradley, L. D. and Pérez-Suárez, D. and De Val-Borro, M. and {(Primary Paper Contributors)} and Aldcroft, T. L. and Cruz, K. L. and Robitaille, T. P. and Tollerud, E. J. and {(Astropy Coordination Committee)} and Ardelean, C. and Babej, T. and Bach, Y. P. and Bachetti, M. and Bakanov, A. V. and Bamford, S. P. and Barentsen, G. and Barmby, P. and Baumbach, A. and Berry, K. L. and Biscani, F. and Boquien, M. and Bostroem, K. A. and Bouma, L. G. and Brammer, G. B. and Bray, E. M. and Breytenbach, H. and Buddelmeijer, H. and Burke, D. J. and Calderone, G. and Rodríguez, J. L. Cano and Cara, M. and Cardoso, J. V. M. and Cheedella, S. and Copin, Y. and Corrales, L. and Crichton, D. and D’Avella, D. and Deil, C. and Depagne, É. and Dietrich, J. P. and Donath, A. and Droettboom, M. and Earl, N. and Erben, T. and Fabbro, S. and Ferreira, L. A. and Finethy, T. and Fox, R. T. and Garrison, L. H. and Gibbons, S. L. J. and Goldstein, D. A. and Gommers, R. and Greco, J. P. and Greenfield, P. and Groener, A. M. and Grollier, F. and Hagen, A. and Hirst, P. and Homeier, D. and Horton, A. J. and Hosseinzadeh, G. and Hu, L. and Hunkeler, J. S. and Ivezić, Ž. and Jain, A. and Jenness, T. and Kanarek, G. and Kendrew, S. and Kern, N. S. and Kerzendorf, W. E. and Khvalko, A. and King, J. and Kirkby, D. and Kulkarni, A. M. and Kumar, A. and Lee, A. and Lenz, D. and Littlefair, S. P. and Ma, Z. and Macleod, D. M. and Mastropietro, M. and McCully, C. and Montagnac, S. and Morris, B. M. and Mueller, M. and Mumford, S. J. and Muna, D. and Murphy, N. A. and Nelson, S. and Nguyen, G. H. and Ninan, J. P. and Nöthe, M. and Ogaz, S. and Oh, S. and Parejko, J. K. and Parley, N. and Pascual, S. and Patil, R. and Patil, A. A. and Plunkett, A. L. and Prochaska, J. X. and Rastogi, T. and Janga, V. Reddy and Sabater, J. and Sakurikar, P. and Seifert, M. and Sherbert, L. E. and Sherwood-Taylor, H. and Shih, A. Y. and Sick, J. and Silbiger, M. T. and Singanamalla, S. and Singer, L. P. and Sladen, P. H. and Sooley, K. A. and Sornarajah, S. and Streicher, O. and Teuben, P. and Thomas, S. W. and Tremblay, G. R. and Turner, J. E. H. and Terrón, V. and Kerkwijk, M. H. Van and De La Vega, A. and Watkins, L. L. and Weaver, B. A. and Whitmore, J. B. and Woillez, J. and Zabalza, V. and {(Astropy Contributors)}},
	month = sep,
	year = {2018},
	pages = {123},
}

@article{the_astropy_collaboration_astropy_2022,
	title = {The {Astropy} {Project}: {Sustaining} and {Growing} a {Community}-oriented {Open}-source {Project} and the {Latest} {Major} {Release} (v5.0) of the {Core} {Package}*},
	volume = {935},
	issn = {0004-637X, 1538-4357},
	shorttitle = {The {Astropy} {Project}},
	url = {https://iopscience.iop.org/article/10.3847/1538-4357/ac7c74},
	doi = {10.3847/1538-4357/ac7c74},
	number = {2},
	urldate = {2026-04-27},
	journal = {ApJ},
	author = {{The Astropy Collaboration} and Price-Whelan, Adrian M. and Lim, Pey Lian and Earl, Nicholas and Starkman, Nathaniel and Bradley, Larry and Shupe, David L. and Patil, Aarya A. and Corrales, Lia and Brasseur, C. E. and Nöthe, Maximilian and Donath, Axel and Tollerud, Erik and Morris, Brett M. and Ginsburg, Adam and Vaher, Eero and Weaver, Benjamin A. and Tocknell, James and Jamieson, William and Van Kerkwijk, Marten H. and Robitaille, Thomas P. and Merry, Bruce and Bachetti, Matteo and Günther, H. Moritz and {Paper Authors} and Aldcroft, Thomas L. and Alvarado-Montes, Jaime A. and Archibald, Anne M. and Bódi, Attila and Bapat, Shreyas and Barentsen, Geert and Bazán, Juanjo and Biswas, Manish and Boquien, Médéric and Burke, D. J. and Cara, Daria and Cara, Mihai and Conroy, Kyle E and Conseil, Simon and Craig, Matthew W. and Cross, Robert M. and Cruz, Kelle L. and D’Eugenio, Francesco and Dencheva, Nadia and Devillepoix, Hadrien A. R. and Dietrich, Jörg P. and Eigenbrot, Arthur Davis and Erben, Thomas and Ferreira, Leonardo and Foreman-Mackey, Daniel and Fox, Ryan and Freij, Nabil and Garg, Suyog and Geda, Robel and Glattly, Lauren and Gondhalekar, Yash and Gordon, Karl D. and Grant, David and Greenfield, Perry and Groener, Austen M. and Guest, Steve and Gurovich, Sebastian and Handberg, Rasmus and Hart, Akeem and Hatfield-Dodds, Zac and Homeier, Derek and Hosseinzadeh, Griffin and Jenness, Tim and Jones, Craig K. and Joseph, Prajwel and Kalmbach, J. Bryce and Karamehmetoglu, Emir and Kałuszyński, Mikołaj and Kelley, Michael S. P. and Kern, Nicholas and Kerzendorf, Wolfgang E. and Koch, Eric W. and Kulumani, Shankar and Lee, Antony and Ly, Chun and Ma, Zhiyuan and MacBride, Conor and Maljaars, Jakob M. and Muna, Demitri and Murphy, N. A. and Norman, Henrik and O’Steen, Richard and Oman, Kyle A. and Pacifici, Camilla and Pascual, Sergio and Pascual-Granado, J. and Patil, Rohit R. and Perren, Gabriel I and Pickering, Timothy E. and Rastogi, Tanuj and Roulston, Benjamin R. and Ryan, Daniel F and Rykoff, Eli S. and Sabater, Jose and Sakurikar, Parikshit and Salgado, Jesús and Sanghi, Aniket and Saunders, Nicholas and Savchenko, Volodymyr and Schwardt, Ludwig and Seifert-Eckert, Michael and Shih, Albert Y. and Jain, Anany Shrey and Shukla, Gyanendra and Sick, Jonathan and Simpson, Chris and Singanamalla, Sudheesh and Singer, Leo P. and Singhal, Jaladh and Sinha, Manodeep and Sipőcz, Brigitta M. and Spitler, Lee R. and Stansby, David and Streicher, Ole and Šumak, Jani and Swinbank, John D. and Taranu, Dan S. and Tewary, Nikita and Tremblay, Grant R. and Val-Borro, Miguel De and Van Kooten, Samuel J. and Vasović, Zlatan and Verma, Shresth and De Miranda Cardoso, José Vinícius and Williams, Peter K. G. and Wilson, Tom J. and Winkel, Benjamin and Wood-Vasey, W. M. and Xue, Rui and Yoachim, Peter and Zhang, Chen and Zonca, Andrea and {Astropy Project Contributors}},
	month = aug,
	year = {2022},
	pages = {167},
}

@article{the_astropy_collaboration_astropy_2013,
	title = {Astropy: {A} community {Python} package for astronomy},
	volume = {558},
	issn = {0004-6361, 1432-0746},
	shorttitle = {Astropy},
	url = {http://www.aanda.org/10.1051/0004-6361/201322068},
	doi = {10.1051/0004-6361/201322068},
	urldate = {2026-04-27},
	journal = {A\&A},
	author = {{The Astropy Collaboration} and Robitaille, Thomas P. and Tollerud, Erik J. and Greenfield, Perry and Droettboom, Michael and Bray, Erik and Aldcroft, Tom and Davis, Matt and Ginsburg, Adam and Price-Whelan, Adrian M. and Kerzendorf, Wolfgang E. and Conley, Alexander and Crighton, Neil and Barbary, Kyle and Muna, Demitri and Ferguson, Henry and Grollier, Frédéric and Parikh, Madhura M. and Nair, Prasanth H. and Günther, Hans M. and Deil, Christoph and Woillez, Julien and Conseil, Simon and Kramer, Roban and Turner, James E. H. and Singer, Leo and Fox, Ryan and Weaver, Benjamin A. and Zabalza, Victor and Edwards, Zachary I. and Azalee Bostroem, K. and Burke, D. J. and Casey, Andrew R. and Crawford, Steven M. and Dencheva, Nadia and Ely, Justin and Jenness, Tim and Labrie, Kathleen and Lim, Pey Lian and Pierfederici, Francesco and Pontzen, Andrew and Ptak, Andy and Refsdal, Brian and Servillat, Mathieu and Streicher, Ole},
	month = oct,
	year = {2013},
	pages = {A33},
}

@inproceedings{bacon_muse_2010,
	address = {San Diego, California, USA},
	title = {The {MUSE} second-generation {VLT} instrument},
	volume = {7735},
	url = {http://proceedings.spiedigitallibrary.org/proceeding.aspx?doi=10.1117/12.856027},
	doi = {10.1117/12.856027},
	urldate = {2026-04-27},
	booktitle = {Proceedings of the {SPIE}},
	author = {Bacon, R. and Accardo, M. and Adjali, L. and Anwand, H. and Bauer, S. and Biswas, I. and Blaizot, J. and Boudon, D. and Brau-Nogue, S. and Brinchmann, J. and Caillier, P. and Capoani, L. and Carollo, C. M. and Contini, T. and Couderc, P. and Daguisé, E. and Deiries, S. and Delabre, B. and Dreizler, S. and Dubois, J. and Dupieux, M. and Dupuy, C. and Emsellem, E. and Fechner, T. and Fleischmann, A. and François, M. and Gallou, G. and Gharsa, T. and Glindemann, A. and Gojak, D. and Guiderdoni, B. and Hansali, G. and Hahn, T. and Jarno, A. and Kelz, A. and Koehler, C. and Kosmalski, J. and Laurent, F. and Le Floch, M. and Lilly, S. J. and Lizon, J.-L. and Loupias, M. and Manescau, A. and Monstein, C. and Nicklas, H. and Olaya, J.-C. and Pares, L. and Pasquini, L. and Pécontal-Rousset, A. and Pelló, R. and Petit, C. and Popow, E. and Reiss, R. and Remillieux, A. and Renault, E. and Roth, M. and Rupprecht, G. and Serre, D. and Schaye, J. and Soucail, G. and Steinmetz, M. and Streicher, O. and Stuik, R. and Valentin, , H and Vernet, J. and Weilbacher, P. and Wisotzki, L. and Yerle, N.},
	editor = {McLean, Ian S. and Ramsay, Suzanne K. and Takami, Hideki},
	month = jul,
	year = {2010},
	pages = {773508},
}

@article{battisti_characterizing_2016,
	title = {{CHARACTERIZING} {DUST} {ATTENUATION} {IN} {LOCAL} {STAR}-{FORMING} {GALAXIES}: {UV} {AND} {OPTICAL} {REDDENING}},
	volume = {818},
	issn = {0004-637X, 1538-4357},
	shorttitle = {{CHARACTERIZING} {DUST} {ATTENUATION} {IN} {LOCAL} {STAR}-{FORMING} {GALAXIES}},
	url = {https://iopscience.iop.org/article/10.3847/0004-637X/818/1/13},
	doi = {10.3847/0004-637X/818/1/13},
	number = {1},
	urldate = {2026-04-27},
	journal = {ApJ},
	author = {Battisti, A. J. and Calzetti, D. and Chary, R.-R.},
	month = feb,
	year = {2016},
	pages = {13},
}

@article{battisti_strength_2020,
	title = {The {Strength} of the 2175 Å {Feature} in the {Attenuation} {Curves} of {Galaxies} at 0.1 {\textless} z ≲ 3},
	volume = {888},
	issn = {0004-637X, 1538-4357},
	url = {https://iopscience.iop.org/article/10.3847/1538-4357/ab5fdd},
	doi = {10.3847/1538-4357/ab5fdd},
	number = {2},
	urldate = {2026-04-27},
	journal = {ApJ},
	author = {Battisti, A. J. and Cunha, E. Da and Shivaei, I. and Calzetti, D. and {(COSMOS collaboration)}},
	month = jan,
	year = {2020},
	pages = {108},
}

@article{battisti_constraining_2025,
	title = {Constraining the link between the {2175Å} dust absorption feature and {PAHs} in {Nearby} {Star}-{Forming} {Galaxies} using {Swift}/{UVOT} and {JWST}/{MIRI}},
	volume = {42},
	issn = {1323-3580, 1448-6083},
	url = {https://www.cambridge.org/core/product/identifier/S1323358024001292/type/journal_article},
	doi = {10.1017/pasa.2024.129},
	language = {en},
	urldate = {2026-04-27},
	journal = {Publ. Astron. Soc. Aust.},
	author = {Battisti, A. and Shivaei, I. and Park, H.-J. and Decleir, M. and Calzetti, D. and Mathew, J. and Wisnioski, E. and Da Cunha, E.},
	year = {2025},
	pages = {e022},
}

@inproceedings{bertin_sourcextractor_2020,
	title = {The {SourceXtractor}++ {Software}},
	volume = {527},
	url = {https://ui.adsabs.harvard.edu/abs/2020ASPC..527..461B},
	urldate = {2026-04-27},
	booktitle = {{ASP} {Conference} {Series}},
	author = {Bertin, E. and Schefer, M. and Apostolakos, N. and Álvarez-Ayllón, A. and Dubath, P. and Kümmel, M.},
	month = jan,
	year = {2020},
	pages = {461},
}

@article{calzetti_calibration_2007,
	title = {The {Calibration} of {Mid}‐{Infrared} {Star} {Formation} {Rate} {Indicators}},
	volume = {666},
	issn = {0004-637X, 1538-4357},
	url = {https://iopscience.iop.org/article/10.1086/520082},
	doi = {10.1086/520082},
	language = {en},
	number = {2},
	urldate = {2026-04-27},
	journal = {ApJ},
	author = {Calzetti, D. and Kennicutt, R. C. and Engelbracht, C. W. and Leitherer, C. and Draine, B. T. and Kewley, L. and Moustakas, J. and Sosey, M. and Dale, D. A. and Gordon, K. D. and Helou, G. X. and Hollenbach, D. J. and Armus, L. and Bendo, G. and Bot, C. and Buckalew, B. and Jarrett, T. and Li, A. and Meyer, M. and Murphy, E. J. and Prescott, M. and Regan, M. W. and Rieke, G. H. and Roussel, H. and Sheth, K. and Smith, J. D. T. and Thornley, M. D. and Walter, F.},
	month = sep,
	year = {2007},
	pages = {870--895},
}

@article{chabrier_galactic_2003,
	title = {Galactic {Stellar} and {Substellar} {Initial} {Mass} {Function}},
	volume = {115},
	issn = {0004-6280, 1538-3873},
	url = {http://iopscience.iop.org/article/10.1086/376392},
	doi = {10.1086/376392},
	language = {en},
	number = {809},
	urldate = {2026-04-27},
	journal = {PUBL ASTRON SOC PAC},
	author = {Chabrier, Gilles},
	month = jul,
	year = {2003},
	pages = {763--795},
}

@article{conroy_modeling_2013,
	title = {Modeling the {Panchromatic} {Spectral} {Energy} {Distributions} of {Galaxies}},
	volume = {51},
	issn = {0066-4146, 1545-4282},
	url = {https://www.annualreviews.org/doi/10.1146/annurev-astro-082812-141017},
	doi = {10.1146/annurev-astro-082812-141017},
	language = {en},
	number = {1},
	urldate = {2026-04-27},
	journal = {ARA\&A},
	author = {Conroy, Charlie},
	month = aug,
	year = {2013},
	pages = {393--455},
}

@article{curtis-lake_jades_2025,
	title = {{JADES} {Data} {Release} 4 {Paper} {I}: {Sample} {Selection}, {Observing} {Strategy} and {Redshifts} of the complete spectroscopic sample},
	url = {http://arxiv.org/abs/2510.01033},
	doi = {10.48550/arXiv.2510.01033},
	urldate = {2026-04-27},
	publisher = {arXiv},
	author = {Curtis-Lake, Emma and Cameron, Alex J. and Bunker, Andrew J. and Scholtz, Jan and Carniani, Stefano and Parlanti, Eleonora and D'Eugenio, Francesco and Jakobsen, Peter and Willmer, Christopher N. A. and Arribas, Santiago and Baker, William M. and Charlot, Stéphane and Chevallard, Jacopo and Circosta, Chiara and Curti, Mirko and Eisenstein, Daniel J. and Hainline, Kevin and Ji, Zhiyuan and Johnson, Benjamin D. and Jones, Gareth C. and Maiolino, Roberto and Maseda, Michael V. and Pérez-González, Pablo G. and Rawle, Tim and Rieke, Marcia and Rinaldi, Pierluigi and Robertson, Brant and Pino, Bruno Rodrígez Del and Saxena, Aayush and Shivaei, Irene and Smit, Renske and Tacchella, Sandro and Übler, Hannah and Venturi, Giacomo and Williams, Christina C. and Willott, Chris and Duan, Qiao},
	month = oct,
	year = {2025},
	note = {arXiv:2510.01033},
}

@article{draine_optical_1984,
	title = {Optical {Properties} of {Interstellar} {Graphite} and {Silicate} {Grains}},
	volume = {285},
	issn = {0004-637X},
	url = {https://ui.adsabs.harvard.edu/abs/1984ApJ...285...89D},
	doi = {10.1086/162480},
	urldate = {2026-04-27},
	journal = {ApJ},
	publisher = {IOP},
	author = {Draine, B. T. and Lee, H. M.},
	month = oct,
	year = {1984},
	pages = {89},
}

@book{draine_physics_2011,
	title = {Physics of the {Interstellar} and {Intergalactic} {Medium}},
	url = {https://ui.adsabs.harvard.edu/abs/2011piim.book.....D},
	urldate = {2026-04-27},
	author = {Draine, Bruce T.},
	month = jan,
	year = {2011},
}

@article{endo_detection_2022,
	title = {Detection of a {Broad} 8 μm {UIR} {Feature} in the {Mid}-infrared {Spectrum} of {WR} 125 {Observed} with {Subaru}/{COMICS}},
	volume = {930},
	issn = {0004-637X, 1538-4357},
	url = {https://iopscience.iop.org/article/10.3847/1538-4357/ac63bd},
	doi = {10.3847/1538-4357/ac63bd},
	number = {2},
	urldate = {2026-04-27},
	journal = {ApJ},
	author = {Endo, Izumi and Lau, Ryan M. and Sakon, Itsuki and Onaka, Takashi and Williams, Peredur M. and Shenavrin, Victor I.},
	month = may,
	year = {2022},
	pages = {116},
}

@article{glenn_prima_2025,
	title = {{PRIMA} mission concept},
	volume = {11},
	url = {https://ui.adsabs.harvard.edu/abs/2025JATIS..11c1628G},
	doi = {10.1117/1.JATIS.11.3.031628},
	urldate = {2026-04-27},
	journal = {Journal of Astronomical Telescopes, Instruments, and Systems},
	author = {Glenn, Jason and Meixner, Margaret and Bradford, Charles M. and Pontoppidan, Klaus and Pope, Alexandra and Kataria, Tiffany and Rocca, Jennifer and Luthman, Elizabeth and Armus, Lee and Baselmans, Jochem and Battersby, Cara and Bollato, Alberto and Burgarella, Denis and Chen, Weibo and Ciesla, Laure and Day, Peter and Di Giorgio, Anna and Dipirro, Michael and Dowell, Charles Darren and Echternach, Pierre and Essinger-Hileman, Thomas and Foote, Marc and Gruppioni, Carlotta and Hensley, Brandon and Henning, Thomas and Jellema, Willem and Johnson, Matthew and Kogut, Alan and Krause, Oliver and McGuire, James and Mills, Elisabeth and Moullet, Arielle and Rodgers, Michael and Sauvage, Marc and Smith, John D. and Somerville, Rachel and Staguhn, Johannes and Stevenson, Thomas and Tucker, Carole and Unwin, Stephen and Ziemer, John and Cannella, Matthew and Dissly, Richard},
	month = jul,
	year = {2025},
	pages = {031628},
}

@article{gregg_calibration_2026,
	title = {The {Calibration} of {Short}-wavelength {Polycyclic} {Aromatic} {Hydrocarbon} {Emission} as {Star} {Formation} {Rate} {Indicators} with {JWST}},
	volume = {997},
	issn = {0004-637X, 1538-4357},
	url = {https://iopscience.iop.org/article/10.3847/1538-4357/ae1ca8},
	doi = {10.3847/1538-4357/ae1ca8},
	number = {1},
	urldate = {2026-04-27},
	journal = {ApJ},
	author = {Gregg, Benjamin and Calzetti, Daniela and Adamo, Angela and Pedrini, Alex and Linden, Sean T. and Bajaj, Varun and Ryon, Jenna E. and Bik, Arjan and Bortolini, Giacomo and Correnti, Matteo and Draine, Bruce T. and Elmegreen, Bruce G. and Faustino Vieira, Helena and Gallagher, John S. and Grasha, Kathryn and Johnson, Kelsey E. and Lai, Thomas S.-Y. and Messa, Matteo and Östlin, Göran and Smith, Linda J. and Tosi, Monica},
	month = jan,
	year = {2026},
	pages = {20},
}

@article{illingworth_hubble_2017,
	title = {The {Hubble} {Legacy} {Fields} ({HLF}-{GOODS}-{S}) v1.5 {Data} {Products}: {Combining} 2442 {Orbits} of {GOODS}-{S}/{CDF}-{S} {Region} {ACS} and {WFC3}/{IR} {Images}},
	shorttitle = {The {Hubble} {Legacy} {Fields} ({HLF}-{GOODS}-{S}) v1.5 {Data} {Products}},
	url = {http://arxiv.org/abs/1606.00841},
	doi = {10.48550/arXiv.1606.00841},
	urldate = {2026-04-27},
	publisher = {arXiv},
	author = {Illingworth, Garth and Magee, Daniel and Bouwens, Rychard and Oesch, Pascal and Labbe, Ivo and Dokkum, Pieter van and Whitaker, Katherine and Holden, Bradford and Franx, Marijn and Gonzalez, Valentino},
	month = may,
	year = {2017},
	note = {arXiv:1606.00841},
}

@inproceedings{kummel_working_2020,
	title = {Working {With} the {SourceXtractor}++ {Software}},
	volume = {527},
	url = {https://ui.adsabs.harvard.edu/abs/2020ASPC..527...29K},
	urldate = {2026-04-27},
	booktitle = {{ASP} {Conference} {Series}},
	author = {Kümmel, M. and Bertin, E. and Schefer, M. and Apostolakos, N. and Álvarez-Ayllón, A. and Dubath, P.},
	month = jan,
	year = {2020},
	pages = {29},
}

@article{lee_ecogal_2025,
	title = {{ECOGAL} {I}. {Project} design and the first catalogue},
	url = {http://arxiv.org/abs/2511.20751},
	doi = {10.48550/arXiv.2511.20751},
	urldate = {2026-04-27},
	publisher = {arXiv},
	author = {Lee, Minju M. and Magdis, Georgios and Brammer, Gabriel and Liu, Daizhong and Magnelli, Benjamin and Gillman, Steven and Gullberg, Bitten and Ito, Kei and Sillassen, Nikolaj and Valentino, Francesco and Gottumukkala, Rashmi and Hegde, Chandana and Greve, Thomas R.},
	month = nov,
	year = {2025},
	note = {arXiv:2511.20751 [astro-ph]},
}

@article{lin_polycyclic_2025,
	title = {Polycyclic aromatic hydrocarbon and the ultraviolet extinction bump at the cosmic dawn},
	volume = {694},
	copyright = {https://creativecommons.org/licenses/by/4.0},
	issn = {0004-6361, 1432-0746},
	url = {https://www.aanda.org/10.1051/0004-6361/202452372},
	doi = {10.1051/0004-6361/202452372},
	urldate = {2026-04-27},
	journal = {A\&A},
	author = {Lin, Qi and Yang, Xuejuan and Li, Aigen and Witstok, Joris},
	month = feb,
	year = {2025},
	pages = {A84},
}

@inproceedings{mckinney_data_2010,
	address = {Austin, Texas},
	title = {Data {Structures} for {Statistical} {Computing} in {Python}},
	url = {https://doi.curvenote.com/10.25080/Majora-92bf1922-00a},
	doi = {10.25080/Majora-92bf1922-00a},
	urldate = {2026-04-27},
	booktitle = {Python in {Science} {Conference}},
	author = {McKinney, Wes},
	year = {2010},
	pages = {56--61},
}

@inproceedings{mclean_design_2010,
	title = {Design and development of {MOSFIRE}: the multi-object spectrometer for infrared exploration at the {Keck} {Observatory}},
	volume = {7735},
	shorttitle = {Design and development of {MOSFIRE}},
	url = {https://ui.adsabs.harvard.edu/abs/2010SPIE.7735E..1EM},
	doi = {10.1117/12.856715},
	urldate = {2026-04-27},
	booktitle = {Ground-based and {Airborne} {Instrumentation} for {Astronomy} {III}},
	author = {McLean, Ian S. and Steidel, Charles C. and Epps, Harland and Matthews, Keith and Adkins, Sean and Konidaris, Nicholas and Weber, Bob and Aliado, Ted and Brims, George and Canfield, John and Cromer, John and Fucik, Jason and Kulas, Kristin and Mace, Greg and Magnone, Ken and Rodriguez, Hector and Wang, Eric and Weiss, Jason},
	month = jul,
	year = {2010},
	pages = {77351E},
}

@article{ormerod_detection_2025,
	title = {Detection of the 2175 Å {UV} bump at \textit{z} \&gt; 7: evidence for rapid dust evolution in a merging reionization-era galaxy},
	volume = {542},
	copyright = {https://creativecommons.org/licenses/by/4.0/},
	issn = {0035-8711, 1365-2966},
	shorttitle = {Detection of the 2175 Å {UV} bump at \textit{z} \&gt; 7},
	url = {https://academic.oup.com/mnras/article/542/2/1136/8213528},
	doi = {10.1093/mnras/staf1228},
	language = {en},
	number = {2},
	urldate = {2026-04-27},
	journal = {MNRAS},
	author = {Ormerod, Katherine and Witstok, Joris and Smit, Renske and de Graaff, Anna and Helton, Jakob M and Maseda, Michael V and Shivaei, Irene and Bunker, Andrew J and Carniani, Stefano and D’Eugenio, Francesco and Bhatawdekar, Rachana and Chevallard, Jacopo and Franx, Marijn and Kumari, Nimisha and Maiolino, Roberto and Rinaldi, Pierluigi and Robertson, Brant and Tacchella, Sandro},
	month = aug,
	year = {2025},
	pages = {1136--1154},
}

@article{padovani_extremely_2023,
	title = {The {Extremely} {Large} {Telescope}},
	volume = {64},
	issn = {0010-7514, 1366-5812},
	url = {https://www.tandfonline.com/doi/full/10.1080/00107514.2023.2266921},
	doi = {10.1080/00107514.2023.2266921},
	language = {en},
	number = {1},
	urldate = {2026-04-27},
	journal = {Contemporary Physics},
	author = {Padovani, Paolo and Cirasuolo, Michele},
	month = jan,
	year = {2023},
	pages = {47--64},
}

@misc{the_pandas_development_team_pandas-devpandas_2026,
	title = {pandas-dev/pandas: {Pandas}},
	copyright = {BSD 3-Clause "New" or "Revised" License},
	shorttitle = {pandas-dev/pandas},
	url = {https://zenodo.org/doi/10.5281/zenodo.19340003},
	doi = {10.5281/ZENODO.19340003},
	urldate = {2026-04-27},
	publisher = {Zenodo},
	author = {The pandas development team},
	month = mar,
	year = {2026},
}

@article{remy-ruyer_linking_2015,
	title = {Linking dust emission to fundamental properties in galaxies: the low-metallicity picture},
	volume = {582},
	issn = {0004-6361, 1432-0746},
	shorttitle = {Linking dust emission to fundamental properties in galaxies},
	url = {http://www.aanda.org/10.1051/0004-6361/201526067},
	doi = {10.1051/0004-6361/201526067},
	urldate = {2026-04-27},
	journal = {A\&A},
	author = {Rémy-Ruyer, A. and Madden, S. C. and Galliano, F. and Lebouteiller, V. and Baes, M. and Bendo, G. J. and Boselli, A. and Ciesla, L. and Cormier, D. and Cooray, A. and Cortese, L. and De Looze, I. and Doublier-Pritchard, V. and Galametz, M. and Jones, A. P. and Karczewski, O. Ł. and Lu, N. and Spinoglio, L.},
	month = oct,
	year = {2015},
	pages = {A121},
}

@article{rinaldi_filling_2026,
	title = {Filling the {Gap} in {Cluster} {Evolution}: {JWST}'s {Glimpse} into a {Young}, {Star}-{Forming} {Cluster} at {Cosmic} {Noon}},
	shorttitle = {Filling the {Gap} in {Cluster} {Evolution}},
	url = {http://arxiv.org/abs/2602.24162},
	doi = {10.48550/arXiv.2602.24162},
	urldate = {2026-04-27},
	publisher = {arXiv},
	author = {Rinaldi, Pierluigi and Alberts, Stacey and Willmer, Christopher N. A. and Carreira, Courtney and Williams, Christina C. and Noirot, Gaël and Gilbert, Carys J. E. and Bunker, Andrew J. and Baker, William M. and Barchiesi, Luigi and Ji, Zhiyuan and Lyu, Jianwei and Tacchella, Sandro and Wu, Zihao and Zhu, Yongda},
	month = feb,
	year = {2026},
	note = {arXiv:2602.24162},
}

@article{robertson_jwst_2026,
	title = {{JWST} {Advanced} {Deep} {Extragalactic} {Survey} ({JADES}) {Data} {Release} 5: {Photometric} {Catalog}},
	shorttitle = {{JWST} {Advanced} {Deep} {Extragalactic} {Survey} ({JADES}) {Data} {Release} 5},
	url = {https://ui.adsabs.harvard.edu/abs/2026arXiv260115956R},
	doi = {10.48550/arXiv.2601.15956},
	urldate = {2026-04-27},
	publisher = {arXiv},
	author = {Robertson, Brant E. and Johnson, Benjamin D. and Tacchella, Sandro and Eisenstein, Daniel J. and Hainline, Kevin and Alberts, Stacey and Arribas, Santiago and Baker, William M. and Bunker, Andrew J. and Cameron, Alex J. and Carniani, Stefano and Carreira, Courtney and Chevallard, Jacopo and Circosta, Chiara and Curtis-Lake, Emma and Danhaive, A. Lola and Duan, Qiao and Egami, Eiichi and Hausen, Ryan and Helton, Jakob M. and Ji, Zhiyuan and Maiolino, Roberto and Pérez-González, Pablo G. and Puskás, Dávid and Rieke, Marcia and Rinaldi, Pierluigi and Sun, Fengwu and Sun, Yang and Übler, Hannah and Trussler, James A. A. and Villanueva, Natalia C. and Whitler, Lily and Williams, Christina C. and Willmer, Christopher N. A. and Willott, Chris and Wu, Zihao and Zhu, Yongda},
	month = jan,
	year = {2026},
	note = {ADS Bibcode: 2026arXiv260115956R},
}

@article{sanders_co_2023,
	title = {{CO} {Emission}, {Molecular} {Gas}, and {Metallicity} in {Main}-sequence {Star}-forming {Galaxies} at z ∼ 2.3*},
	volume = {942},
	issn = {0004-637X, 1538-4357},
	url = {https://iopscience.iop.org/article/10.3847/1538-4357/aca46f},
	doi = {10.3847/1538-4357/aca46f},
	number = {1},
	urldate = {2026-04-27},
	journal = {ApJ},
	author = {Sanders, Ryan L. and Shapley, Alice E. and Jones, Tucker and Shivaei, Irene and Popping, Gergö and Reddy, Naveen A. and Davé, Romeel and Price, Sedona H. and Mobasher, Bahram and Kriek, Mariska and Coil, Alison L. and Siana, Brian},
	month = jan,
	year = {2023},
	pages = {24},
}

@article{scholtz_jades_2025,
	title = {{JADES} {Data} {Release} 4 -- {Paper} {II}: {Data} reduction, analysis and emission-line fluxes of the complete spectroscopic sample},
	shorttitle = {{JADES} {Data} {Release} 4 -- {Paper} {II}},
	url = {http://arxiv.org/abs/2510.01034},
	doi = {10.48550/arXiv.2510.01034},
	urldate = {2026-04-27},
	publisher = {arXiv},
	author = {Scholtz, J. and Carniani, S. and Parlanti, E. and D'Eugenio, F. and Curtis-Lake, E. and Jakobsen, P. and Bunker, A. J. and Cameron, A. J. and Arribas, S. and Baker, W. M. and Charlot, S. and Chevellard, J. and Circosta, C. and Curti, M. and Duan, Q. and Eisenstein, D. J. and Hainline, K. and Ji, Z. and Johnson, B. D. and Jones, G. C. and Kumari, N. and Maiolino, R. and Maseda, M. V. and Perna, M. and Pérez-González, P. G. and Rawle, T. and Rieke, M. and Rinaldi, P. and Robertson, B. and Saxena, A. and Shivaei, I. and Silcock, M. S. and Sun, Y. and Pino, B. Rodríguez Del and Tacchella, S. and Übler, H. and Venturi, G. and Williams, C. C. and Willmer, C. N. A. and Willott, C. and Witstok, J.},
	month = oct,
	year = {2025},
	note = {arXiv:2510.01034},
}

@article{shim_metallicitypah_2023,
	title = {Metallicity–{PAH} {Relation} of {MIR}-selected {Star}-forming {Galaxies} in {AKARI} {North} {Ecliptic} {Pole}-wide {Survey}},
	volume = {165},
	issn = {0004-6256, 1538-3881},
	url = {https://iopscience.iop.org/article/10.3847/1538-3881/aca09c},
	doi = {10.3847/1538-3881/aca09c},
	number = {2},
	urldate = {2026-04-27},
	journal = {AJ},
	author = {Shim, Hyunjin and Hwang, Ho Seong and Jeong, Woong-Seob and Toba, Yoshiki and Kim, Minjin and Kim, Dohyeong and Song, Hyunmi and Hashimoto, Tetsuya and Nakagawa, Takago and Nanni, Ambra and Pearson, William J. and Takagi, Toshinobu},
	month = feb,
	year = {2023},
	pages = {31},
}

@article{zavala_evolution_2021,
	title = {The {Evolution} of the {IR} {Luminosity} {Function} and {Dust}-obscured {Star} {Formation} over the {Past} 13 {Billion} {Years}},
	volume = {909},
	issn = {0004-637X, 1538-4357},
	url = {https://iopscience.iop.org/article/10.3847/1538-4357/abdb27},
	doi = {10.3847/1538-4357/abdb27},
	number = {2},
	urldate = {2026-04-27},
	journal = {ApJ},
	author = {Zavala, J. A. and Casey, C. M. and Manning, S. M. and Aravena, M. and Bethermin, M. and Caputi, K. I. and Clements, D. L. and Cunha, E. Da and Drew, P. and Finkelstein, S. L. and Fujimoto, S. and Hayward, C. and Hodge, J. and Kartaltepe, J. S. and Knudsen, K. and Koekemoer, A. M. and Long, A. S. and Magdis, G. E. and Man, A. W. S. and Popping, G. and Sanders, D. and Scoville, N. and Sheth, K. and Staguhn, J. and Toft, S. and Treister, E. and Vieira, J. D. and Yun, M. S.},
	month = mar,
	year = {2021},
	pages = {165},
}


\begin{appendix} 

\section{MOSFIRE data reduction and emission line fitting}\label{app:mosfire}

    The data was collected over 12 nights using two slit mask configurations (hereafter mask1 and mask2), each targeting a different set of sources. Each mask-night combination constitutes a distinct observational setup and was reduced independently. Tab.~\ref{tab01} summarizes the observing times for each setup. A standard star was observed each night in both bands, and sensitivity functions were derived for each band–mask–night combination. 
    
    The data was reduced using the Python-based \texttt{PypeIt} package (v1.7.4; \citealt{prochaska_pypeit_2020, prochaska_pypeitpypeit_2020, prochaska_pypeit_2020-1}), following the standard pipeline with modifications to the object detection and subtraction procedures. Specifically, object traces were determined from the position of a bright star in the slit mask rather than from the dither offsets recorded in the headers, since small unrecorded positional drifts occur in MOSFIRE data throughout the night. In addition, the SNR threshold for object detection in individual frames was lowered to 4, to enable the detection of our faint sources. Wavelength calibration was performed using either OH sky lines from the science frames or Ne/Ar arc lamp exposures, depending on the observational setup. For the K band, we consistently used arc lamp exposures. For the H band, we adopted the method that yielded the best results, as the quality of the night and the exposure times also affect the visibility of the OH lines \citep{rousselot_night-sky_2000}. The methods were assessed using the arc-frame diagnostic plots produced by \texttt{PypeIt}; the adopted solution was the one that provided a larger number of detected lines, a lower RMS ($<0.1$\,px)\footnote{Corresponding to a wavelength uncertainty of $\lesssim0.16$\,\AA \ in H band and $\lesssim0.22$\,\AA \ in K band, given dispersions of $\sim1.6$ and $\sim2.2$\,\AA/px, respectively.}, and satisfactory visual inspection (straight spectral traces).

    Individual science exposures were retained or discarded based on the seeing, defined as the mean spatial FWHM (in arcseconds) of all optimally extracted objects. We adopted thresholds of 0.8$''$, 1$''$, 1$''$, and 1$''$ for mask1–$H$, mask2–$H$, mask1–$K$, and mask2–$K$, respectively. Retained exposures were 2D-coadded within each setup and flux-calibrated using the derived sensitivity functions. 
    The resulting flux-calibrated spectra from each setup were then combined to produce a single spectrum per source in each band. From these, we optimally extracted 1D spectra at the centroid of the spatial profile using the technique described in \cite{horne_optimal_1986}. No slit-loss correction was applied to the spectra, which does not affect our results since we use only emission line ratios of lines that are close in wavelength. An example is shown in Fig.~\ref{fig01}.

\subsection{Emission line fitting}
\label{app:mosfire:lines}

Emission lines in the MOSFIRE spectra are fitted using the \texttt{Astropy} package in Python, specifically the \texttt{astropy.modeling} module. Each emission line (e.g., H$\beta$) is modeled with a one-dimensional Gaussian profile using the \texttt{TRFLSQFitter} routine. The H$\alpha$+[N\,\textsc{ii}] complex is fitted simultaneously, while all other lines are fitted independently. The amplitude of [O\,\textsc{iii}]$\lambda4960$ is fixed to one third of the [O\,\textsc{iii}]$\lambda5008$ amplitude, following atomic physics constraints \citep{osterbrock_astrophysics_1989}.

The underlying continuum is modeled locally as a linear component in $F_\lambda$, which provides an adequate approximation over the narrow wavelength range of each emission line. This component is included only when its contribution exceeds an SNR (per spectral resolution element) threshold of 1. Flux uncertainties are estimated via a Monte Carlo approach. We generate 500 realizations of each spectrum by perturbing the fluxes with Gaussian noise drawn from the \texttt{PypeIt}-provided uncertainties. The final line flux is taken as the mean of the resulting distribution, and the uncertainty as its standard deviation.

\begin{table}
    \centering
    \scriptsize 
    \begin{tabular}{cccrccr}
        \multirow{2}{*}{band} & \multicolumn{3}{c}{mask1} & \multicolumn{3}{c}{mask2} \\ \cline{2-7}
         & setup & night & $t_{\rm obs}$ [h:min] & setup & night & $t_{\rm obs}$ [h:min] \\ \hline \hline
        \multirow{3}{*}{H} & H1-A & 12/14/22 & 2:42 & H2-A & 12/14/22 & 0:40 \\ 
         & H1-B & 11/10/24 & 1:24 & H2-B & 01/13/22 & 1:36 \\ 
         & H1-C & 12/30/24 & 2:45 & H2-C & 12/29/24 & 4:49 \\ \hline
        \multirow{6}{*}{K} & K1-A & 12/15/22 & 1:13 & K2-A & 12/28/24 & 4:07 \\ 
         & K1-B & 01/13/23 & 2:00 & K2-B & 12/30/24 & 2:00 \\ 
         & K1-C & 11/10/24 & 2:13 & ~ & ~ & ~ \\ 
         & K1-D & 12/24/24 & 3:38 & ~ & ~ & ~ \\ 
         & K1-E & 12/25/24 & 0:53 & ~ & ~ & ~ \\ 
         & K1-F & 12/27/24 & 4:33 & ~ & ~ & ~ \\ 
    \end{tabular}
    \caption{Observing times for each MOSFIRE setup. Total exposure times are 7\,h for mask1–$H$ (H1 in the table), 7\,h for mask2–$H$ (H2), 14.5\,h for mask1–$K$ (K1), and 6\,h for mask2–$K$ (K2). Mask 1 covers 16 sources from the MUSE parent sample. Mask 2 covers 10.}
    \label{tab01}
\end{table}

\begin{figure}
   \resizebox{\hsize}{!}
            {\includegraphics[]{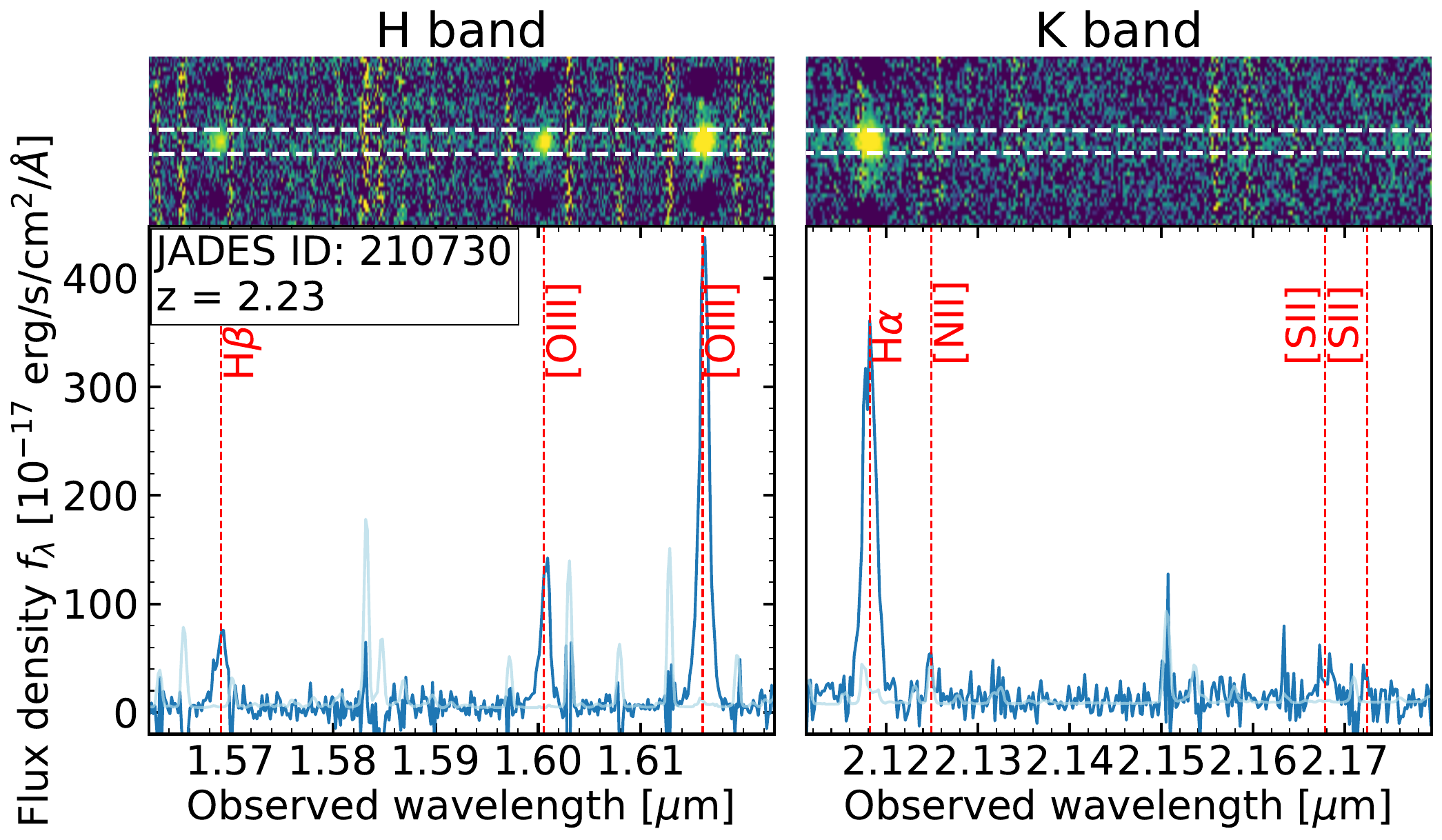}}
      \caption{Example MOSFIRE spectrum for one galaxy in the sample, in the $H$ (left) and $K$ (right) bands. The top panels show the 2D spectra, with the spatial extraction region used to obtain the 1D spectra indicated by white dashed lines. The bottom panels show the extracted 1D spectra (blue) with the associated uncertainties (light blue). The emission lines used for metallicity diagnostics are marked with red dashed lines. The source’s ID in the JADES NIRCam catalog (Sect.~\ref{sect02:phot}) and its spectroscopic redshift from the MUSE HUDF Survey are also indicated.}
      \label{fig01}
\end{figure}

\section{ALMA data and $q_{\rm PAH}$ estimates}\label{app:alma}

To assess the impact of far-IR constraints on the inferred PAH mass fractions, we compare $q_{\rm PAH}$ values derived with and without ALMA data for the subset of galaxies with available observations. Figure~\ref{fig:alma_qpah} shows that the inferred $q_{\rm PAH}$ values are generally consistent within uncertainties, with no systematic offset relative to the 1:1 relation. This indicates that the inclusion of ALMA data does not significantly change the inferred PAH fractions.

However, ALMA data provide direct constraints on the long-wavelength dust emission, which  better constrains the dust SED. This is illustrated in Fig.~\ref{fig:alma_sed}, where we show an example SED fit for the same galaxy with and without ALMA data. While both fits recover similar $q_{\rm PAH}$ values, the inclusion of ALMA measurements reduces degeneracies in the dust component and leads to a more physically constrained solution.

\begin{figure}
   \resizebox{\hsize}{!}
            {\includegraphics[]{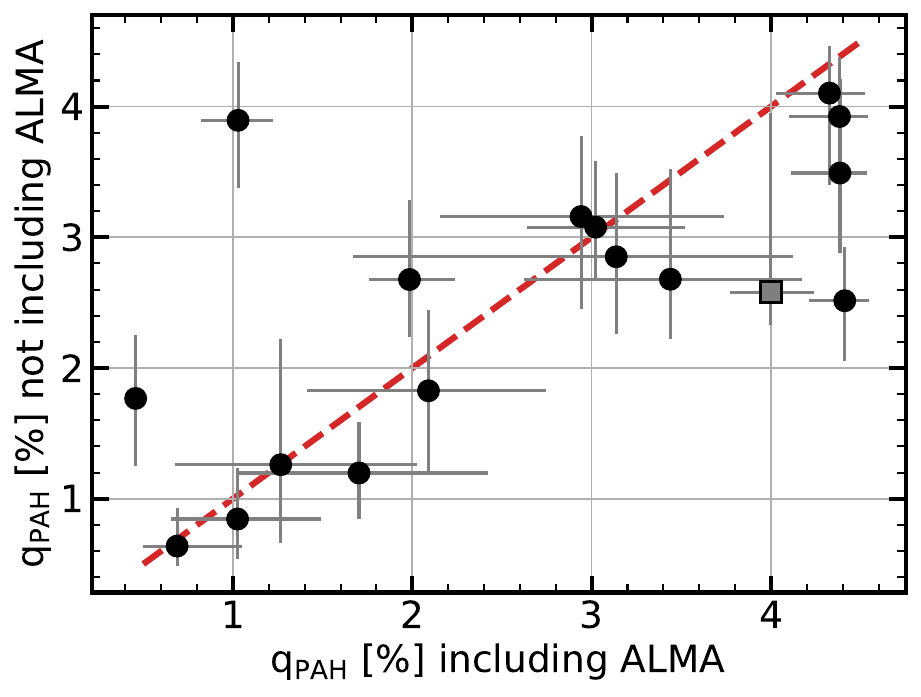}}
      \caption{Comparison between $q_{\rm PAH}$ obtained from including ALMA data in the SED fitting ($x$ axis) and excluding it ($y$ axis), for the 17 galaxies for which we have ALMA and in some cases \textit{Herschel} detections. The red dashed line indicates a 1:1 relation. The gray square marks one galaxy that is not included in our final sample for lack of MIRI data covering the PAH region (see selection steps in Sect.~\ref{sect02:sample}).}
      \label{fig:alma_qpah}
\end{figure}

\begin{figure}
   \resizebox{\hsize}{!}
            {\includegraphics[]{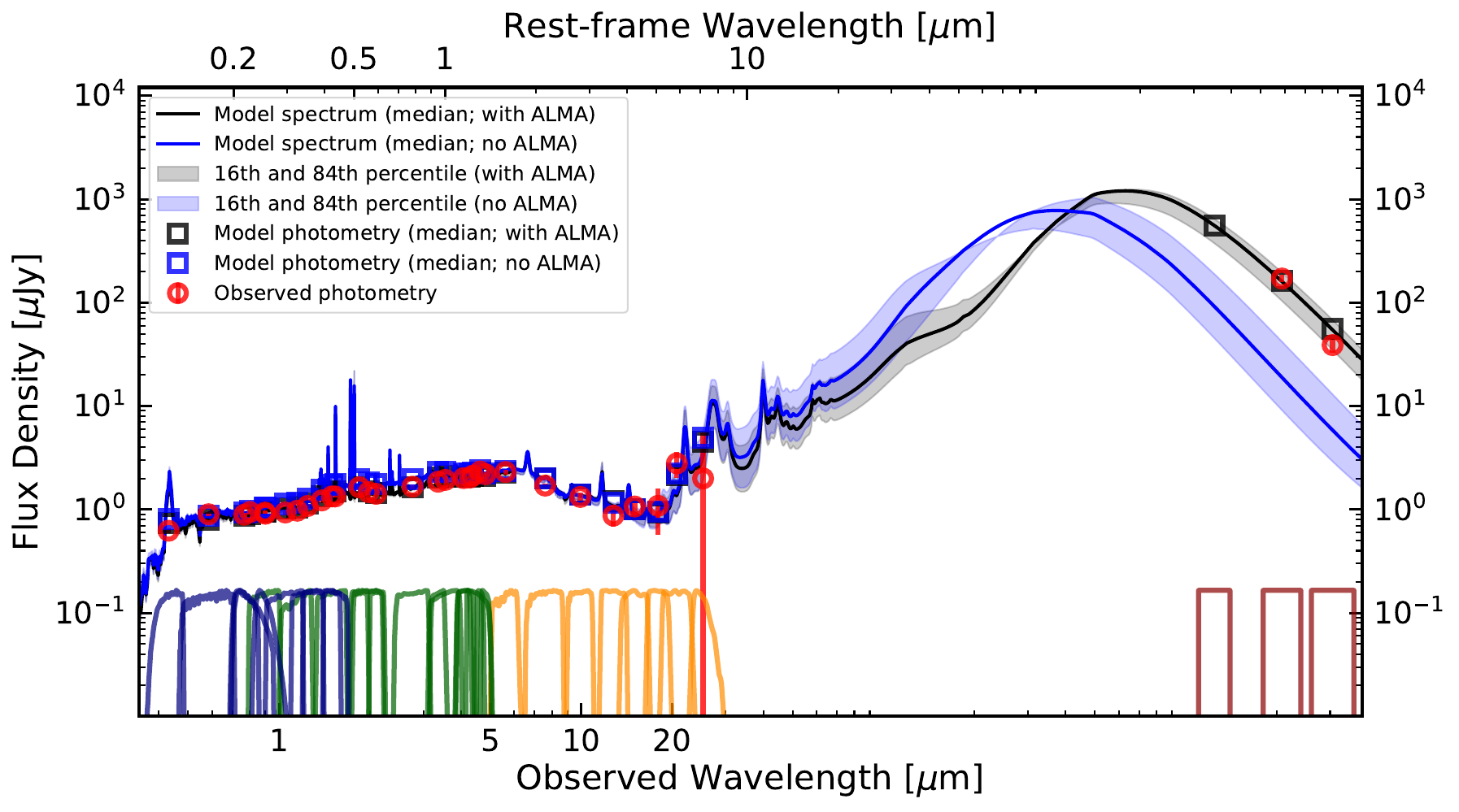}}
      \caption{Example \texttt{Prospector} SED fits including ALMA data (black) and excluding it (blue) for the galaxy with JADES ID 209116 ($z=2.54$). Filter transmission curves for \textsl{HST}, \textsl{JWST}/NIRCam, \textsl{JWST}/MIRI, and ALMA are shown at the bottom in blue, green, orange, and red, respectively.}
      \label{fig:alma_sed}
\end{figure}

\section{Details on SED fitting with \texttt{Prospector}}\label{app:prosp}

\begin{table}
\centering
\begin{tabular}{lc}
\hline
Parameter & Prior \\
\hline
$\tau_V$ & Tophat $[0,\,4]$ \\
$\delta$ & Tophat $[-0.6,\,0.3]$ \\
$q_{\rm PAH}$ [\%] & Tophat $[0.4,\,4.6]$ \\
$U_{\rm min}$ & Tophat $[0.1,\,15]$ \\
$\gamma$ & Free (prior not specified) \\
$\tau$ (SFH) [Gyr] & Log-uniform $[0.1,\,10]$ \\
$t$ (stellar age) [Gyr] & Tophat $[0.001,\,t_{\rm univ}(z)]$ \\
$\log_{10}(M_*/M_\odot)$ & Log-uniform $[10^6,\,10^{12}]$ \\
$Z/Z_\odot$ & Tophat $[-2.5,\,0.19]$ \\
$f_{\rm AGN}$ & Free (prior not specified) \\
$\tau_{\rm AGN}$ & Free (prior not specified) \\
IGM scaling factor & Clipped normal $\mu=1$, $\sigma=0.3$, $[0,\,2]$ \\
\hline
\end{tabular}
\caption{Summary of the main free parameters used in the \texttt{Prospector} SED fitting and their adopted priors. $t_{\rm univ}(z)$ is the cosmological age corresponding to the galaxy redshift.}
\label{tab:prospector}
\end{table}

\begin{figure}[t]
    \centering
    \begin{subfigure}[ht]{.85\columnwidth}
        \centering
        \includegraphics[width=\textwidth]{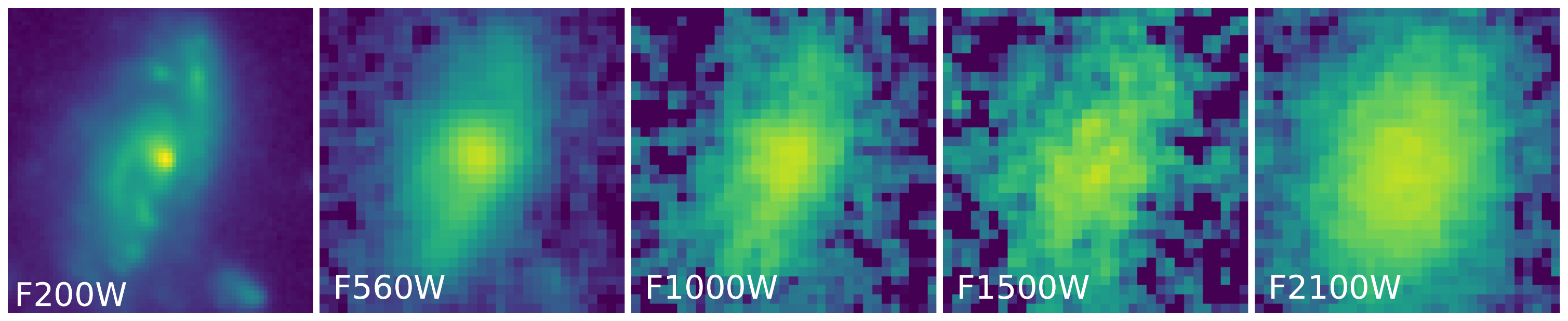}  
        \vspace{-.4cm}
        \label{fig05a}
    \end{subfigure}
    \begin{subfigure}[ht]{\columnwidth}
        \hspace{-.2cm}
        \centering
        \includegraphics[width=\textwidth]{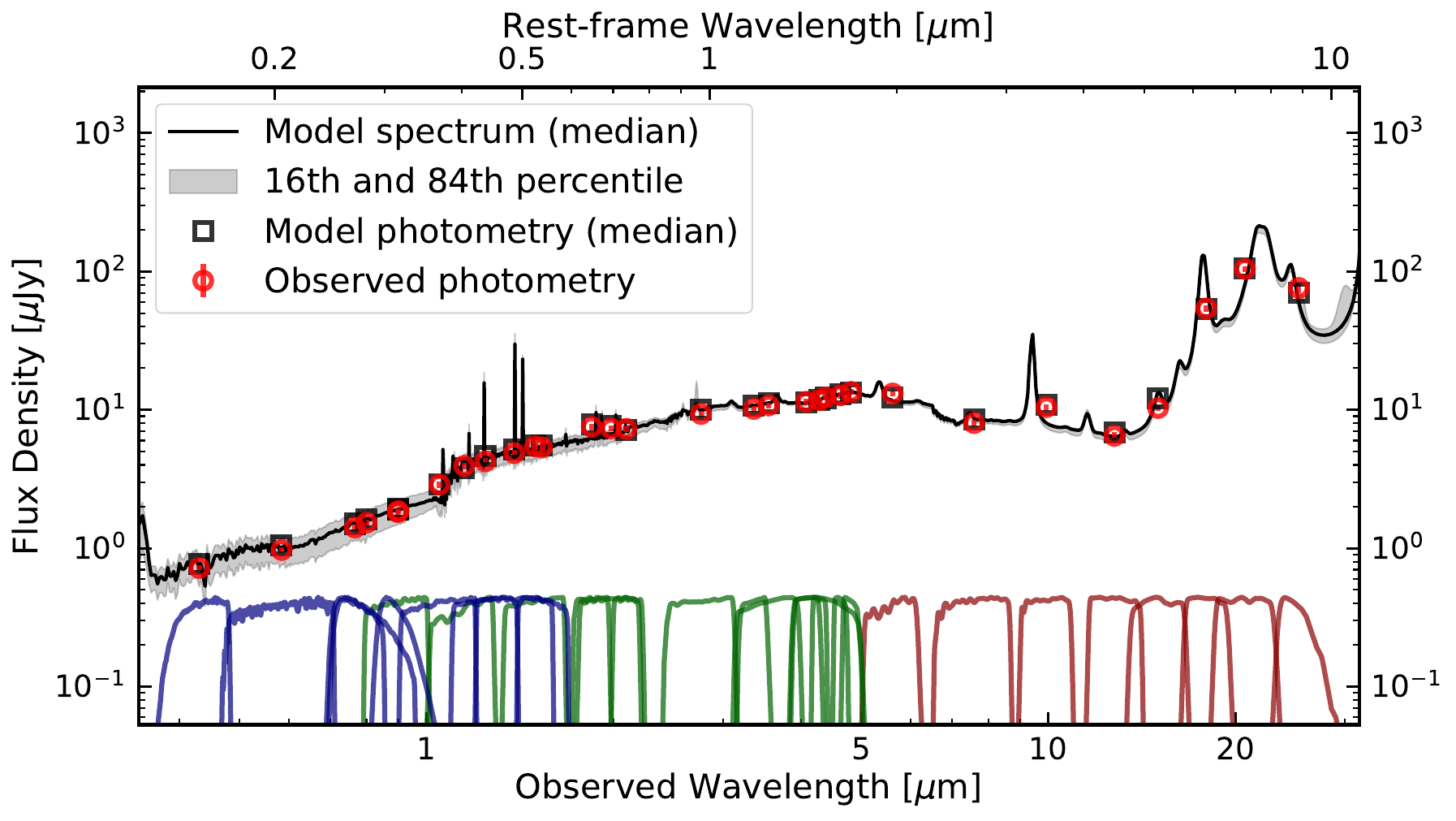}
        \label{fig05b}
    \end{subfigure}
\caption{\textsl{Top:} 2"x2" cutouts from JADES F200W mosaic and SMILES F560W, F1000W, F1500W, and F2100W mosaics, for galaxy with JADES ID~208134 at $z=1.85$. This galaxy lies at the center of a recently studied young galaxy cluster \citep{rinaldi_filling_2026}. \textsl{Bottom:} Example SED from \texttt{Prospector} for the same galaxy. Filter transmission curves for \textsl{HST}, \textsl{JWST}/NIRCam, and \textsl{JWST}/MIRI are shown at the bottom in blue, green, and red, respectively. At this redshift, 1"$=8.66$\,kpc.}
\label{fig05}
\end{figure}

An example SED fit is shown in Fig.~\ref{fig05}. The attenuation curve follows the prescription of \citet{noll_gmass_2009} and \citet{kriek_dust_2013} (\texttt{dust\_type} $=4$ in \texttt{Prospector}), in which the slope of the curve ($\delta$) is linked to the strength of the UV bump ($E_b$):

\begin{equation}
    A(\lambda)=\frac{A_V}{4.05}\Bigg(k'(\lambda)+\frac{E_b(\lambda \ \Delta\lambda)^2}{(\lambda^2-\lambda_0^2)^2+(\lambda \ \Delta\lambda)^2}\Bigg)\Bigg(\frac{\lambda}{\lambda_V}\Bigg)^\delta,
    \label{eq:02}
\end{equation}
where $A_V$ is the attenuation at $\lambda_V=5500$\,\AA\,, $k'(\lambda)$ is the reddening curve from \cite{calzetti_dust_2000}, and $\lambda_0=2175$\,\AA\, is the central wavelength of the UV bump. The optical depth at 5500\,\AA\ follows a tophat prior between 0 and 4, and the power-law index of the attenuation curve (or slope of the attenuation curve, $\delta$) follows a tophat prior between $-0.6$ and 0.3. We note that we do not use the bump amplitude from the SED fits, because we are only fitting photometric data to the SED, which is not sufficient to constrain the wide UV bump attenuation feature, and because the \texttt{Prospector}-derived $E_b$ is tied to $\delta$, as mentioned above; instead, the bump amplitude in this work is derived from rest-frame UV spectra as described in Sect.~\ref{sect02:spec}.

Dust emission is modeled following \citet{draine_infrared_2007}, which employs a silicate–graphite–PAH grain mixture. The PAH mass to total dust mass fraction ($q_{\rm PAH}$) follows a tophat prior between 0.4 and 4.6\%, while the minimum intensity of the interstellar radiation field ($U_{\rm min}$) follows a tophat prior between 0.1 and 15, and the fraction of dust heated by starlight with $U>U_{\rm min}$ ($\gamma$) is a free parameter. The prior on $q_{\rm PAH}$ is limited by the range in $q_{\rm PAH}$ given by the \cite{draine_infrared_2007} model. The priors on $U_{\rm min}$ and $\gamma$ are motivated by the analysis performed by \cite{leja_deriving_2017}. 

The $e$-folding timescale $\tau$ follows a log-uniform prior between 0.1 and 10\,Gyr, and the stellar age $t$ follows a tophat prior between 0.001\,Gyr and the cosmological age corresponding to the galaxy redshift. We assume a \citet{chabrier_galactic_2003} initial mass function (IMF), with the formed stellar mass following a log-uniform prior between $10^6$ and $10^{12}\,M_\odot$. In this paper, we adopt the surviving stellar mass, which considers mass losses due to stellar evolution, and is obtained by multiplying the formed stellar mass by the ratio between these two quantities provided by \texttt{Prospector} (parameter \texttt{mfrac}). Metallicity follows a tophat prior between $-2.5$ and $0.19\,Z_\odot$.

For the AGN dust emission model, the AGN fraction ($f_{\rm AGN}$) and the optical depth of an individual dust clump at 5500\r{A} ($\tau_{\rm AGN}$) are free parameters. The IGM opacity scaling factor is drawn from a clipped normal prior centered on 1, with a dispersion of 0.3 and bounded between 0 and 2. sSFRs in this work are derived from the \texttt{Prospector}-inferred star formation rates and stellar masses. All parameters and priors used in \texttt{Prospector} are summarized in Tab.~\ref{tab:prospector}.

A caveat of our SED fitting analysis is that the available photometric filters do not provide a clean constraint on the dust continuum baseline (e.g., at $\lambda_{\rm rest}\sim 20\,\mu$m). As a result, the inferred PAH mass fractions, $q_{\rm PAH}$, depend to some extent on the adopted model priors, particularly those governing dust attenuation and dust temperature, both of which influence the mid-infrared continuum level. This limitation is mitigated for the subset of 17 galaxies with ALMA/\textsl{Hershel} coverage that more directly constrains the dust continuum. For these sources, we verify that the derived $q_{\rm PAH}$ values are broadly consistent, within uncertainties, with those obtained without the far-IR data (Fig.~\ref{fig:alma_qpah}).


\section{Normalized $E_{\rm bump}$}\label{app:02}

As described in Sect.~\ref{sect03:uvbump}, bump measurements normalized to $A_V$ and $E(B-V)_{\rm neb}$ are presented in this appendix, where we examine how normalization affects the trends identified with $E_{\rm bump}$. This quantity, as defined in Sect.~\ref{sect03:uvbump}, represents the \textit{apparent} bump amplitude and contains a dependence on reddening ($E(B-V)$). As a result, it is affected by the dust column density (or reddening), such that regions with higher dust content will exhibit larger apparent bump amplitudes. Because the UV attenuation bump is measured relative to the stellar continuum, its amplitude is expected to correlate with the stellar continuum reddening, $E(B-V)_{\rm stars}$. Using the relation between optical depth and dust attenuation, $\tau_\lambda = 0.921\,A(\lambda)$ \citep{carroll_introduction_1996}, we convert the \texttt{Prospector}-derived $\tau_V$ into $A_V$ and use it to normalize the bump. However, this new quantity ($E_{\rm bump}/A_V$) is inherently dependent on the SED-inferred stellar attenuation curve. We therefore also test a second normalization in which $E_{\rm bump}$ is scaled by the dust reddening $E(B-V)$. Measuring precise $E(B-V)_{\rm stars}$ is challenging; however, nebular reddening $E(B-V)_{\rm neb}$, derived from strong emission lines in the ionized gas, can be measured from at least two hydrogen recombination lines and has been shown to scale with stellar reddening in both the local universe and at cosmic noon (e.g., \citealt{calzetti_dust_1994, battisti_characterizing_2016, shivaei_mosdef_2020, reddy_mosdef_2020}). We compute $E(B-V)_{\rm neb}$ from the Balmer decrement following \citet{calzetti_dust_1994, calzetti_dust_2000}:
\begin{equation*}
    E(B-V)_{\rm neb}=\frac{1}{0.4(\kappa(\text{H}\beta)-\kappa(\text{H}\alpha))}\log_{10}\frac{(F(\text{H}\alpha)/F(\text{H}\beta))_{\rm obs}}{(F(\text{H}\alpha)/F(\text{H}\beta))_{\rm int}},
\end{equation*}
where $(F(\mathrm{H}\alpha)/F(\mathrm{H}\beta))_{\rm int} = 2.86$ is the intrinsic unreddened ratio, assuming Case B recombination with $T_e = 10^4$\,K and $n_e = 100\,\mathrm{cm}^{-3}$ \citep{osterbrock_astrophysics_1989}. We adopt $\kappa(\mathrm{H}\beta) - \kappa(\mathrm{H}\alpha)$ from the \citet{cardelli_relationship_1989} Milky Way extinction curve (with $R_V = 3.1$), obtaining a value of $\sim 0.347$.

In Fig.~\ref{app_fig04}, we plot the UV bump amplitude normalized by the total dust attenuation at 5500\,\AA\ ($E_{\rm bump}/A_V$) against $q_{\rm PAH}$, color-coded by stellar mass (left panel) and oxygen abundance (right panel). We find a correlation between $E_{\rm bump}/A_V$ and $q_{\rm PAH}$. In the left panel, we identify six outliers in the final sample ($\sim 13$\,\% of the sources), five with $E_{\rm bump}/A_V>0.4$ and $q_{\rm PAH}<2$\,\%, and one with $E_{\rm bump}/A_V>2.0$ and $q_{\rm PAH}>3$\,\%. These span a wide range of stellar masses. We note that a third of these outliers are AGN, although we do not find any significant correlation between $E_{\rm bump}/A_V$ and the AGN bolometric fraction $f_{\rm AGN}$. We note that AGN in this sample are at the higher end of the redshift distribution ($z\gtrsim2.4$; see Fig.~\ref{fig11}) and this is most likely due to high-redshift galaxies having more uncertainties associated to them.

We find a moderate correlation between $E_{\rm bump}/A_V$ and stellar mass ($\rho = 0.40$, $p = 4.8\cdot 10^{-3}$), but no evidence for a correlation between $E_{\rm bump}/A_V$ and oxygen abundance in the metallicity subsample, due to a very high $p-$value ($\rho = 0.33$, $p = 0.16$).

We also explored the relation between $E_{\rm bump}/A_V$, $q_{\rm PAH}$, and several galaxy properties, including optical depth, redshift, and rest-frame $V\!-\!J$ color. The clearest trend is that galaxies with higher optical depths tend to exhibit higher PAH abundances ($q_{\rm PAH}\gtrsim2$\,\%), whereas systems with low PAH abundances are preferentially found at low optical depths. Rest-frame $V\!-\!J$ color shows the strongest correlation with both $E_{\rm bump}/A_V$ and $q_{\rm PAH}$, consistent with its connection to dust attenuation and reddening. Together, these trends indicate that the galaxies displaying high $E_{\rm bump}/A_V$ but low $q_{\rm PAH}$ are characterized by very low dust content, despite exhibiting unusually strong intrinsic UV bumps. We also note that several of these outliers lie at the high-redshift end of the sample, although the sample size is too small to establish any significant redshift dependence.

\begin{figure*}
   \resizebox{\hsize}{!}
            {\includegraphics[]{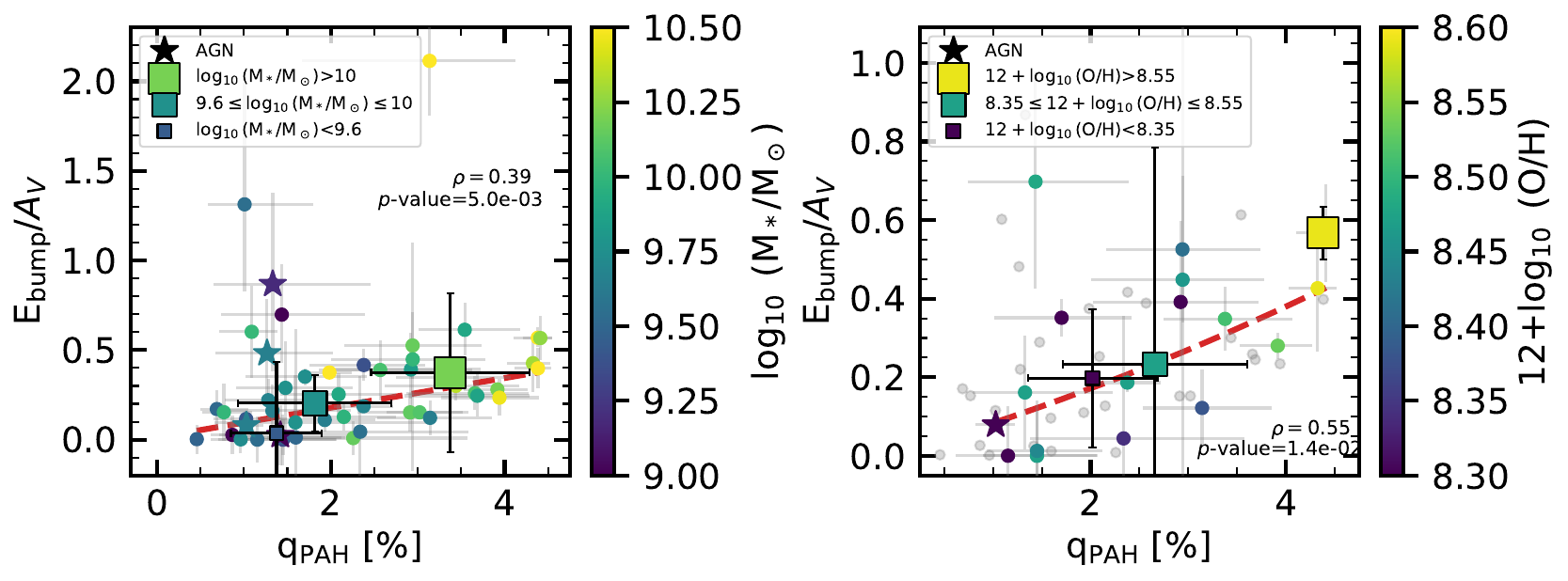}}
      \caption{Intrinsic 2175\,\AA\ UV attenuation bump amplitude ($E_{\rm bump}/A_V$) as a function of the PAH-to-total dust mass fraction ($q_{\rm PAH}$). Points are color-coded by stellar mass (left; 49 galaxies) and oxygen abundance (right; 19 galaxies). In the right panel, gray points indicate galaxies without metallicity measurements and are shown for reference only; they are not included in the correlation reported in this panel. The data is divided into three bins: in the left panel, $\log_{10}(M/M_\odot)<9.6$ (dark blue; 14 galaxies), $9.6 \leq \log_{10}(M/M_\odot) \leq 10$ (light blue; 18 galaxies), and $\log_{10}(M/M_\odot)>10$ (green; 17 galaxies); in the right panel, $Z<8.35$ (purple; 4 galaxies), $8.35 \leq Z \leq 8.55$ (green; 12 galaxies), and $Z>8.55$ (yellow; 3 galaxies). The median and standard deviation in each bin are shown as squares and reported in Tab.~\ref{tab:app02:01}. We fit the data points with a quadric model, shown as red dashed lines; the coefficients are reported in Tab.~\ref{tab:app02:02}. For both samples, we report the Spearman rank correlation coefficient and the corresponding $p-$value between $E_{\rm bump}/A_V$ and $q_{\rm PAH}$.}
      \label{app_fig04}
\end{figure*}

\begin{table*}
\centering
\begin{tabular}{lccccc}
\hline
Bin & Number of sources & $E_{\rm bump}/A_V$ & $q_{\rm PAH}$ [\%] & $\log_{10}(M_*/M_\odot)$ & $12+\log_{10}(\rm O/H)$ \\
\hline
Low stellar mass &
14 &
$0.04 \pm 0.40$ &
$1.38 \pm 0.52$ &
$9.43 \pm 0.35$ &
-- \\
Intermediate stellar mass &
18 &
$0.20 \pm 0.16$ &
$1.82 \pm 0.88$ &
$9.75 \pm 0.13$ &
-- \\
High stellar mass &
17 &
$0.37 \pm 0.44$ &
$3.38 \pm 0.91$ &
$10.19 \pm 0.20$ &
-- \\
\hline\hline
Low metallicity &
4 &
$0.20 \pm 0.18$ &
$2.02 \pm 0.66$ &
-- &
$8.28 \pm 0.03$ \\
Intermediate metallicity &
12 &
$0.23 \pm 0.55$ &
$2.66 \pm 0.94$ &
-- &
$8.47 \pm 0.04$ \\
High metallicity &
3 &
$0.57 \pm 0.07$ &
$4.38 \pm 0.04$ &
-- &
$8.59 \pm 0.01$ \\
\hline
\end{tabular}
\caption{Median values of $E_{\rm bump}/A_V$, $q_{\rm PAH}$, stellar mass, and oxygen abundance in bins of stellar mass and metallicity (Fig.~\ref{app_fig04}).}
\label{tab:app02:01}
\end{table*}

\begin{table*}
\centering
\begin{tabular}{lcccc}
\hline
 & a & b & c \\
\hline
Final sample & $(-1.57\pm161.31)\times 10^{-4}$ & $(8.36\pm7.74)\times 10^{-2}$ & $(1.18\pm7.61)\times 10^{-2}$ \\
Metallicity subsample & $(5.50\pm25.29)\times 10^{-3}$ & $(7.12\pm12.53)\times 10^{-2}$ & $(6.82\pm118.82)\times 10^{-3}$ \\
\hline
\end{tabular}
\caption{Best-fit coefficients describing the relation between $E_{\rm bump}/A_V$ and $q_{\rm PAH}$ using for the final sample (left panel of Fig.~\ref{app_fig04}) and metallicity subsample (right panel of Fig.~\ref{app_fig04}). Fits are of the form $y = ax^2 + bx + c$, where $y = E_{\rm bump}/A_V$ and $x = q_{\rm PAH}$. All fits are performed using the \texttt{lmfit} Python package, using the least-squares algorithm.}
\label{tab:app02:02}
\end{table*}

\begin{figure*}
   \resizebox{\hsize}{!}
            {\includegraphics[]{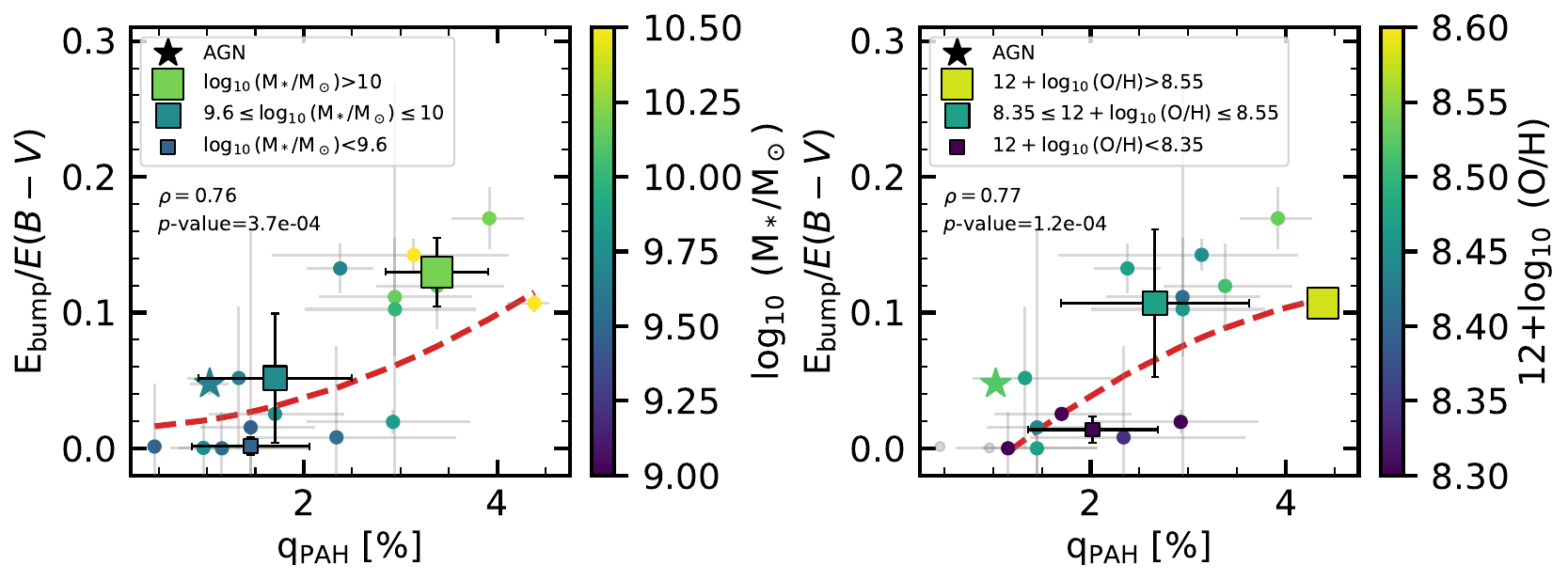}}
      \caption{Same as Fig.~\ref{app_fig04}, but with the intrinsic 2175\,\AA\ UV attenuation bump amplitude normalized by dust reddening $E_{\rm bump}/E(B-V)$. Points are color-coded by stellar mass (left; 17 galaxies) and oxygen abundance (right; 15 galaxies). Median values and fitting are reported in Tabs.~\ref{tab:app02:03} and \ref{tab:app02:04}.}
      \label{app_fig08}
\end{figure*}

Additionally, we perform a comparison with the reddening-normalized UV bump amplitude, $E_{\rm bump}/E(B-V)$, and we present the results in Fig.~\ref{app_fig08}. In the left panel, we show the 17 galaxies for which $E(B-V)$ measurements from the Balmer decrement are available (Sect.~\ref{sect03:uvbump}) and examine the relation between $q_{\rm PAH}$ and $E_{\rm bump}/E(B-V)$, color-coded by stellar mass. We identify a strong correlation between the intrinsic UV bump strength and $q_{\rm PAH}$ ($\rho=0.76$, $p=3.7\cdot 10^{-4}$). This is consistent with the results presented in Sect.~\ref{sect05:01}; we also recover strong correlations between stellar mass and both $q_{\rm PAH}$ and $E_{\rm bump}/E(B-V)$ ($\rho=0.80$ with $p=1.2\cdot 10^{-4}$ and $\rho=0.79$ with $p=1.5\cdot 10^{-4}$, respectively).

In the right panel of Fig.~\ref{app_fig08}, we present the 15 galaxies with both metallicity and $E(B-V)$ measurements. The correlation between $q_{\rm PAH}$ and the intrinsic bump amplitude remains consistent with that observed in the left panel ($\rho=0.77$, $p=1.2\cdot 10^{-4}$). The correlations with stellar mass are also consistent with those found in the left panel. For oxygen abundance, we don't find a correlation with $q_{\rm PAH}$ due to a high $p-$value ($\rho=0.36$, $p= 0.18$). We identify a correlation between oxygen abundance and the intrinsic UV bump amplitude ($\rho=0.52$), although with a rather large $p-$value of $\sim 0.05$.

Overall, the results obtained using the apparent UV bump definition ($E_{\rm bump}$) and the reddening-normalized UV bump amplitude $E_{\rm bump}/E(B-V)$ are consistent. In contrast, adopting a UV bump definition normalized by the $V$-band attenuation $A_V$ ($E_{\rm bump}/A_V$) yields qualitatively different results: it significantly reduces the scatter observed at high $q_{\rm PAH}$ and produces systematically weaker correlations between the UV bump amplitude and other galaxy properties. This suggests that dust-related effects may be responsible for the scatter observed at $q_{\rm PAH}>3$~\% in Fig.~\ref{fig07}.

These differences arising from the choice of UV bump parametrization may help explain the diversity and, in some cases, apparent discrepancies, reported in the literature regarding the relationship between UV bump strength and stellar mass. While some studies find no correlation (e.g., \citealt{battisti_strength_2020, maheson_big_2025}), others report a positive correlation (e.g., \citealt{shivaei_uv_2022}).

\begin{table*}
\centering
\begin{tabular}{lccccc}
\hline
Bin & Number of sources & $E_{\rm bump}/E(B-V)$ & $q_{\rm PAH}$ [\%] & $\log_{10}(M_*/M_\odot)$ & $12+\log_{10}(\rm O/H)$ \\
\hline
Low stellar mass &
5 &
$0.00 \pm 0.01$ &
$1.45 \pm 0.60$ &
$9.48 \pm 0.07$ &
-- \\
Intermediate stellar mass &
7 &
$0.05 \pm 0.05$ &
$1.70 \pm 0.79$ &
$9.72 \pm 0.11$ &
-- \\
High stellar mass &
5 &
$0.13 \pm 0.03$ &
$3.38 \pm 0.53$ &
$10.19 \pm 0.32$ &
-- \\
\hline\hline
Low metallicity &
4 &
$0.01 \pm 0.01$ &
$2.02 \pm 0.66$ &
-- &
$8.28 \pm 0.03$ \\
Intermediate metallicity &
10 &
$0.11 \pm 0.05$ &
$2.66 \pm 0.96$ &
-- &
$8.47 \pm 0.04$ \\
High metallicity &
1 &
$0.11 \pm 0.00$ &
$4.38 \pm 0.00$ &
-- &
$8.58 \pm 0.00$ \\
\hline
\end{tabular}
\caption{Median values of $E_{\rm bump}/E(B-V)$, $q_{\rm PAH}$, stellar mass, and oxygen abundance in bins of stellar mass and metallicity (Fig.~\ref{app_fig08}).}
\label{tab:app02:03}
\end{table*}

\begin{table*}
\centering
\begin{tabular}{lcccc}
\hline
 & a & b & c \\
\hline
Mass bins & $(4.95\pm7.74)\times 10^{-3}$ & $(1.19\pm43.87)\times 10^{-3}$ & $(1.47\pm5.21)\times 10^{-2}$ \\
Metallicity bins & $(5.97\pm11.60)\times 10^{-3}$ & $(6.81\pm6.86)\times 10^{-2}$ & $(-7.35\pm8.63)\times 10^{-2}$ \\
\hline
\end{tabular}
\caption{Best-fit coefficients describing the relation between $E_{\rm bump}/E(B-V)$ and $q_{\rm PAH}$ using for the final sample (left panel of Fig.~\ref{app_fig08}) and metallicity subsample (right panel of Fig.~\ref{app_fig08}). Fits are of the form $y = ax^2 + bx + c$, where $y = E_{\rm bump}/E(B-V)$ and $x = q_{\rm PAH}$. All fits are performed using the \texttt{lmfit} Python package, using the least-squares algorithm.}
\label{tab:app02:04}
\end{table*}

\newpage

\paragraph{}

\newpage

\section{Permutation tests}\label{app:permutation}

To assess the robustness of the main Spearman correlations against the small sample sizes, particularly for the metallicity subsample, we construct empirical null distributions using permutation tests. For each relation, the values of one variable are randomly permuted while the values of the other variable are kept fixed, and Spearman's $\rho$ is recomputed for each of $10^5$ permutations. The two-sided permutation $p-$value is defined as the fraction of permutations satisfying $|\rho_{\rm perm}|\geq|\rho_{\rm obs}|$.

Figure~\ref{fig:permutations} shows the resulting null distributions for the $E_{\rm bump}-q_{\rm PAH}$ relation in the final sample and metallicity subsample, and for the $E_{\rm bump}-$metallicity and $q_{\rm PAH}-$metallicity relations in the metallicity subsample. In all cases, the null distributions are centred close to zero, as expected under the hypothesis of no association. The observed correlations lie in the tails of the null distributions for the $E_{\rm bump}-q_{\rm PAH}$ and $E_{\rm bump}-$metallicity relations, whereas the $q_{\rm PAH}-$metallicity correlation is less strongly inconsistent with the null hypothesis.

\begin{figure*}
   \resizebox{\hsize}{!}
            {\includegraphics[]{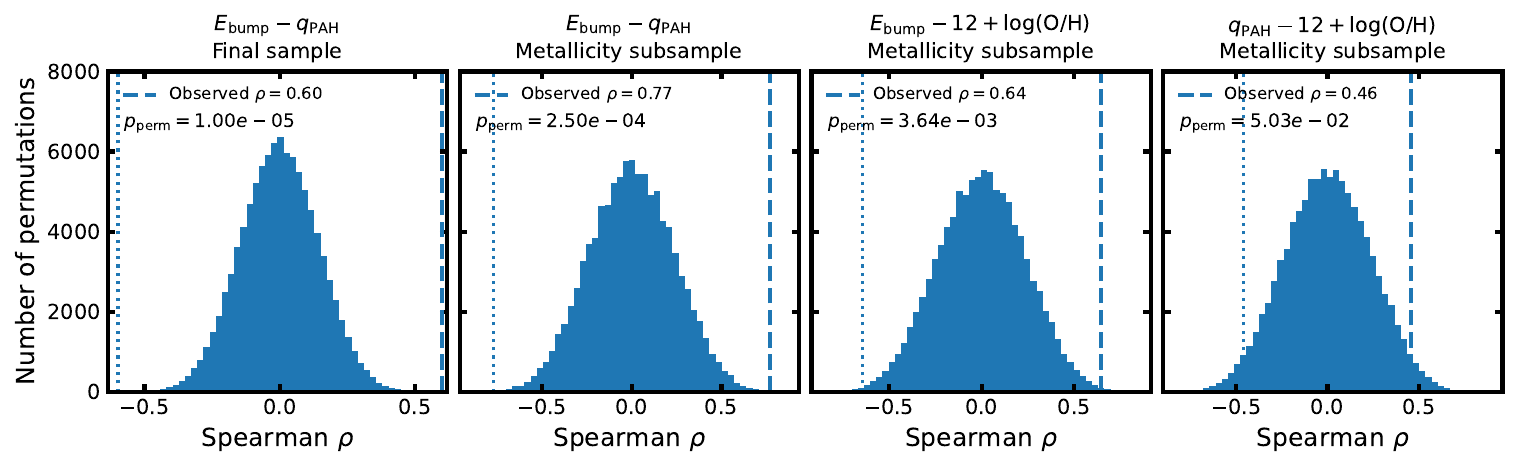}}
      \caption{Permutation tests performed for the final sample (49 galaxies) and metallicity subsample (19 galaxies), testing the robustness of the Spearman rank correlations between $E_{\rm bump}$, $q_{\rm PAH}$ and metallicity.}
      \label{fig:permutations}
\end{figure*}

\newpage

\paragraph{}

\newpage

\section{Correlation diagrams and figures}\label{app:01}

In Fig.~\ref{fig06}, we present the pairwise Spearman rank correlation coefficients ($\rho$) between $q_{\rm PAH}$ (left)/$E_{\rm bump}$ (right) and all properties with $\rho\geq 0.3$, following the procedures described in Sect.~\ref{sect04}. In Figs.~\ref{app_fig01} and ~\ref{app_fig02}, we show correlation matrices (heatmaps) of the pairwise Spearman rank correlation coefficients and corresponding $p-$values for all parameters.

\begin{figure*}[ht!]
   \resizebox{\hsize}{!}
            {\includegraphics[]{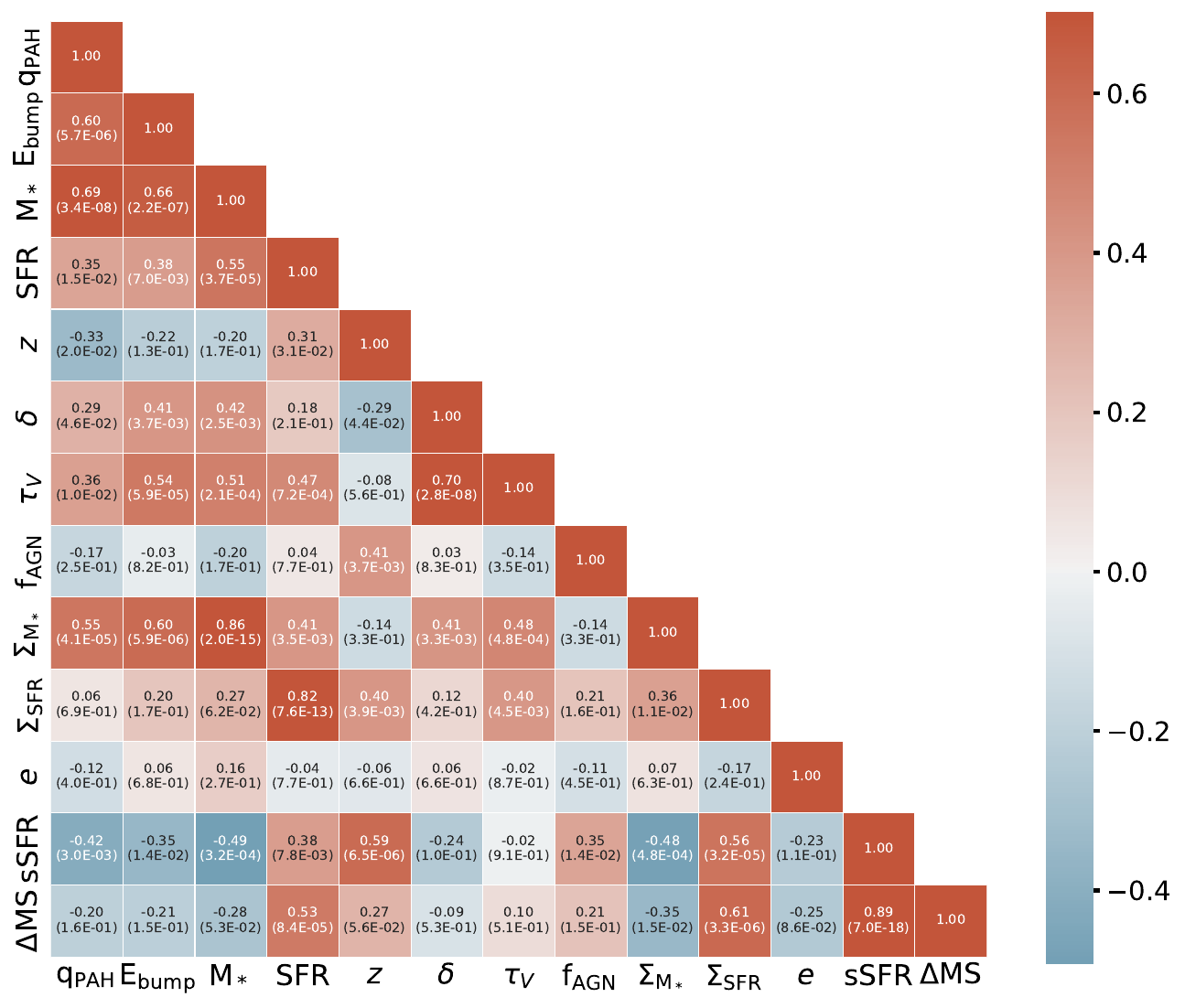}}
      \caption{Pairwise correlation matrix (heatmap) of various galaxy parameters for the final sample (49 galaxies; Sect.~\ref{sect02:sample}): PAH mass to total dust mass fraction ($q_{\rm PAH}$), UV bump amplitude ($E_{\rm bump}$), stellar mass ($M_*$), star formation rate (SFR), redshift ($z$), slope of the attenuation curve ($\delta$), optical depth at 5500\,\AA\ ($\tau_V$), AGN bolometric fraction ($f_{\rm AGN}$), stellar surface mass density ($\Sigma_{M_\ast}$), surface SFR density ($\Sigma_{\rm SFR}$), ellipticity ($e$), specific star formation rate (sSFR), and offset from the star-forming main sequence ($\Delta$MS). The color scale indicates the Spearman correlation coefficient ($\rho$), which quantifies the strength and sign of each correlation. The corresponding $p-$values are shown in parentheses below each coefficient.}
      \label{app_fig01}
\end{figure*}

\begin{figure*}
   \resizebox{\hsize}{!}
            {\includegraphics[]{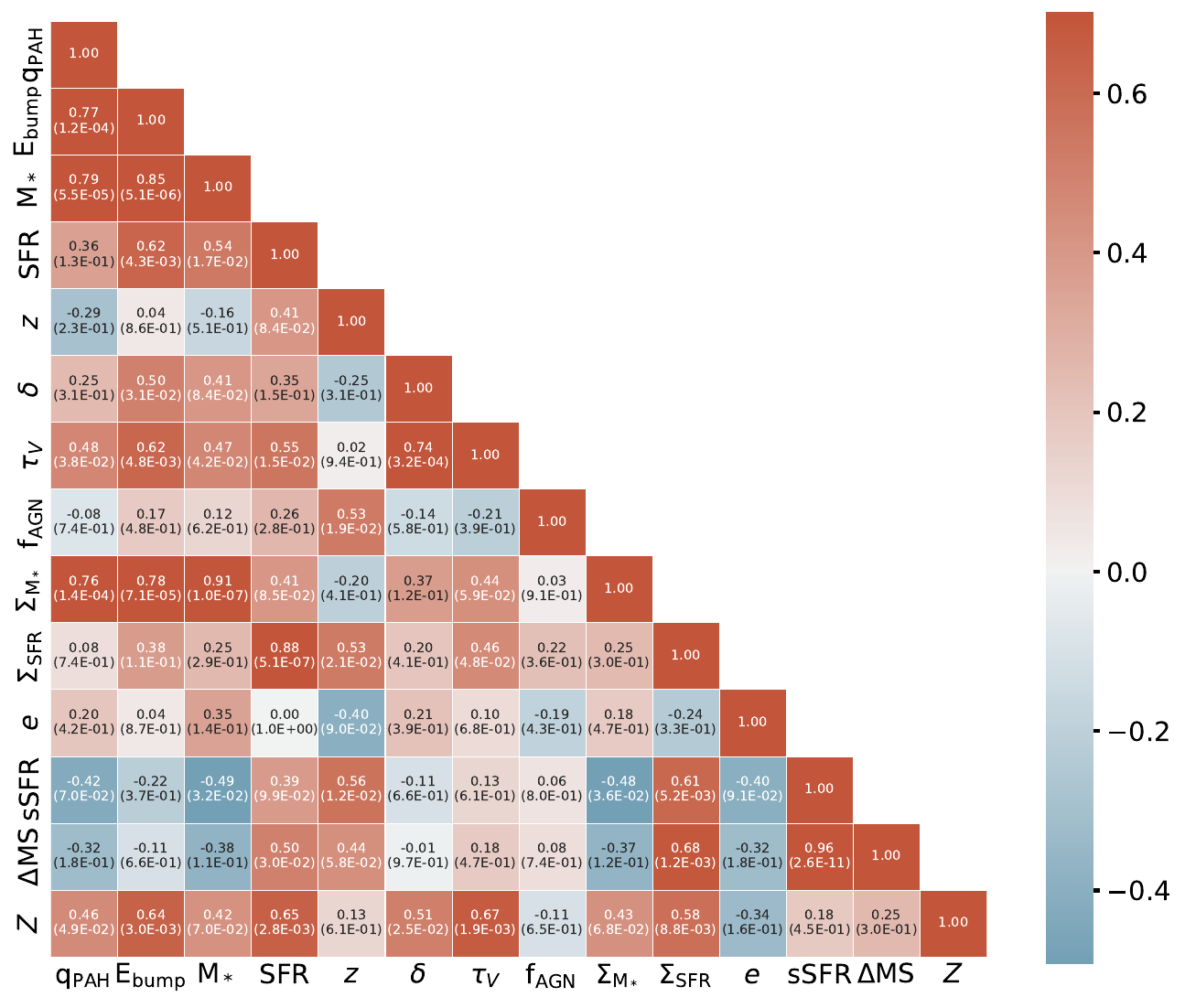}}
      \caption{Same as Fig.~\ref{app_fig01}, but for the metallicity subsample (19 galaxies; Sect.~\ref{sect02:sample}). We include oxygen abundance ($Z$).}
      \label{app_fig02}
\end{figure*}

\end{appendix}

\end{document}